\documentclass[11pt]{article}
\pdfoutput=1

\newlength{\abstractwidth}
\usepackage{changepage}
\usepackage{cite}
\usepackage{showlabels}
\usepackage{enumitem}
\usepackage{ifpdf}
 \usepackage{lmodern}
\usepackage[T1]{fontenc}
\usepackage[ansinew]{inputenc}
\usepackage{amsthm}
\usepackage{amsmath}
\usepackage{relsize}
\usepackage{amssymb}
\usepackage{comment}
\usepackage{tikz}
\usetikzlibrary{tikzmark}
\usepackage{graphicx}
\usepackage{color}
\definecolor{darkblue}{cmyk}{0.9,0.9,0,0}
\definecolor{darkgreen}{rgb}{0,0.55,0}
\definecolor{vert}{rgb}{0.1367,0.543,0.1367}
\usepackage[colorlinks=true,linkcolor=darkblue,citecolor=darkblue,urlcolor=darkblue]{hyperref}
\usepackage{epsfig}
\usepackage{epstopdf}
\usepackage{mathtools}
\usepackage{graphicx}

\usepackage{empheq} 
\usepackage{url}

\makeatletter
\long\def\@makecaption#1#2{
  \vskip\abovecaptionskip
  \sbox\@tempboxa{{\captionfonts #1: #2}}
  \ifdim \wd\@tempboxa >\hsize
    {\captionfonts #1: #2\par}
  \else
    \hbox to\hsize{\hfil\box\@tempboxa\hfil}
  \fi
  \vskip\belowcaptionskip}
\makeatother

\def\adss{AdS$_5\,\times\,$S$^5$~}
\def\fo{\mathbb{O}}

\def\llangle{\langle\!\langle}
\def\rrangle{\rangle\!\rangle}

\def\sl{SL(2,\Z)}
\def\tl{\tilde\lambda}

\def\ls{\l_{{\mathsf{S}}}}

\def\fnp{f_{\rm np}}

\def\ni{\noindent}

\def\L{\Lambda}
\def\vol{{\rm vol}}

\renewcommand{\thanks}[1]{\footnote{#1}}
\newcommand{\starttext}{
\setcounter{footnote}{0}
\renewcommand{\thefootnote}{\arabic{footnote}}}

\newcommand{\bea}{\begin{eqnarray}}
\newcommand{\eea}{\end{eqnarray}}
\newcommand{\be}{\begin{eqnarray}}
\newcommand{\ee}{\end{eqnarray}}
\newcommand{\bma}{\begin{matrix}}
\newcommand{\ema}{\cr\end{matrix}}
\newcommand{\<}{\langle}
\renewcommand{\>}{\rangle}

\def\cF{{\cal F}}
\def\cG{{\cal G}}

\def\cN{{\cal N}}
\def\cO{{\cal O}}

\def\cZ{{\cal Z}}

\def\mg{\mathfrak{g}}

\def\ZZ{{\mathbb Z}}

\def\CC{{\mathbb C}}

\def\Re{{\rm Re \,}}
\def\Im{{\rm Im \,}}

\def\det{{\rm det \,}}

\def\half{{1\over 2}}
\def\p{\partial}

\def\a{\alpha}
\def\m{\mu}
\def\b{\beta}
\def\g{\gamma}
\def\G{\Gamma}
\def\l{\lambda}

\def\no{\nonumber}

\def\({\left(}
\def\){\right)}
\def\[{\left[}
\def\]{\right]}

\def\<{\langle}
\def\>{\rangle}

\def\taub{\overline\tau}

\def\qb{\overline q}

\def\Z{\mathbb{Z}}
\def\R{\mathbb{R}}

\def\qb{\bar q}
\def\tb{\bar\tau}

\def\x{\times}

\def\bul{$\bullet$~}

\def\1{{\rm 1-loop}}

\def\cZ{\mathcal{Z}}

\def\c{\cite}

\def\cG{\mathcal{G}}
\def\cN{\mathcal{N}}

\def\G{\Gamma}
\def\p{\partial}
\def\o{\over}
\def\g{\gamma}
\def\D{\Delta}
\def\rar{\rightarrow}

\def\eqr{\eqref}

\def\O{{\cal O}}
\def\ra{\rangle}
\def\la{\langle}
\def\ssec{\subsection}
\def\sssec{\subsubsection}
\def\sec{\section}
\def\i{\infty}
\def\foot{\footnote}
\newcommand{\es}[2] {\begin{equation} \label{#1} \begin{split} #2 \end{split} \end{equation}}
\newcommand{\e}[2] {\begin{equation} \label{#1} #2 \end{equation}}

\def\t{\tau}

\def\tb{\bar\tau}

\def\Omax{\O_p^{(\text{max})}}
\def\d{\delta}

\def\m{\mu}
\def\n{\nu}
\def\){\right)}
\def\({\left( }
\def\]{\right] }
\def\[{\left[ }

\usepackage{varioref}
\usepackage{makeidx}
\makeindex

\usepackage[english]{babel}
\usepackage{caption}
\usepackage{subfig}

\usepackage{tabularx}
\renewcommand{\baselinestretch}{1.03}
 
\begin{document}
\starttext
\thispagestyle{empty}

\begin{flushright}
\end{flushright}

\vskip 1in

\begin{center}

{\Large \bf Integrated Correlators in $\mathcal{N}=4$ SYM via $SL(2,\mathbb{Z})$ Spectral Theory}

\vskip 0.3in

{Hynek Paul, Eric Perlmutter, Himanshu Raj } 
   
\vskip 0.15in

{\small Universit\'e Paris-Saclay, CNRS, CEA, Institut de Physique Th\'eorique, 91191, Gif-sur-Yvette, France}

\vskip 0.15in

{\tt \small hynek.paul,perl,himanshu.raj@ipht.fr}

\vskip 0.6in

\begin{abstract}
\vskip 0.1in

We perform a systematic study of integrated four-point functions of half-BPS operators in four-dimensional $\mathcal{N}=4$ super Yang-Mills theory with gauge group $SU(N)$. These observables, defined by a certain spacetime integral of $\<\mathcal{O}_2\mathcal{O}_2\mathcal{O}_p\mathcal{O}_p\>$ where $\mathcal{O}_p$ is a superconformal primary of charge $p$, are known to be computable by supersymmetric localization, yet are non-trivial functions of the complexified gauge coupling $\tau$. We find explicit and remarkably simple results for several classes of these observables, exactly as a function of $N$ and $\tau$. Their physical and formal properties are greatly illuminated upon employing the $SL(2,\mathbb{Z})$ spectral decomposition: in this S-duality-invariant eigenbasis, the integrated correlators are fixed simply by polynomials in the spectral parameter. These polynomials are determined recursively by linear algebraic equations relating different $N$ and $p$, such that all integrated correlators are ultimately fixed in terms of the integrated stress tensor multiplets in the $SU(2)$ theory. Our computations include the full matrix of integrated correlators at low values of $p$, and a certain infinite class involving operators of arbitrary $p$. The latter satisfy an open lattice chain equation for all $N$, reminiscent of the Toda equation obeyed by extremal correlators in $\mathcal{N}=2$ superconformal theories. We compute ensemble averages of these observables and analyze our solutions at large $N$, confirming and predicting features of semiclassical AdS$_5\, \times$ S$^5$ supergravity amplitudes. 

\end{abstract}

\end{center}
\newpage
\pagenumbering{roman}
\baselineskip .14 in
\setcounter{tocdepth}{2}
\tableofcontents
\setlength{\textheight}{9in}
\pagenumbering{arabic}
\setcounter{page}{1}
\numberwithin{equation}{section}
\baselineskip=14pt
\setcounter{equation}{0}
\setcounter{footnote}{0}

\newcommand \widebar [1] {\overline{#1}}
\def\xb{\bar{x}}

\sec{Introduction}\label{sec:intro}

It is reasonable to hope that the four-dimensional maximally-supersymmetric Yang-Mills theory may be exactly solved. In the large $N$ `t Hooft limit, the theory becomes famously integrable, and Yangian symmetries emerge, which has led to the exact determination of various planar observables. At finite $N$, these special properties are obscured; however, the theory enjoys a non-perturbative S-duality symmetry \c{Montonen:1977sn,Witten:1978mh,Osborn:1979tq,Argyres:2006qr}, which for simply-laced gauge groups is an invariance (up to global identifications \c{Aharony:2013hda}) under $\sl$ transformations of the complexified gauge coupling,
\e{}{\t = {\theta\o 2\pi} + i{4\pi\o g_{\rm YM}^2}\,.}
It seems likely that a solution of this theory (henceforth $\cN=4$ SYM) will be possible only with judicious treatment of S-duality. Even so, determining the exact $\t$-dependence of $\cN=4$ SYM observables should be very hard in general, precisely because the theory is strongly interacting over most of its parameter space, even after modding out by $\sl$. 

There is a special family of observables in $\cN=4$ SYM, introduced in \cite{Binder:2019jwn}, which sit askew of this expectation: they vary as functions of $\t$, and yet are determinable by supersymmetric localization. They are integrated four-point functions of half-BPS operators, whose definition will be recalled further below. In this paper, following \cite{Binder:2019jwn,Dorigoni:2021guq,Collier:2022emf}, we find strikingly compact solutions for an infinite class of these integrated correlators in the $SU(N)$ theory, exact in $N$ and $\tau$, and study their general systematics. Tantalizingly, they are optimally understood in a complete basis of $\sl$-invariant functions \c{Collier:2022emf}: in this basis, the integrated correlators are essentially polynomials in the relevant spectral parameter. 

To set the context for these results, let us recall the status of (unintegrated) $\cN=4$ SYM four-point functions. We henceforth specialize to the $SU(N)$ theory. Half-BPS superconformal primary operators in $\cN=4$ SYM, call them $\O_p^{(i)}$ with $p\geq 2$, are spacetime scalars in the $[0\,p\,0]$ representation of the $SU(4)$ R-symmetry with dimension $\D_p=p$. The case $p=2$ is the $\mathbf{20'}$ of the stress tensor multiplet. At $p>3$ there is degeneracy, which is captured by the label $i$. Consider the family of four-point functions $\<\O_2\O_2\O_p^{(i)}\O_p^{(j)}\>$. At finite $N$, this is known only in weak-coupling perturbation theory $(g_{\rm YM}^2\ll1)$ through two loops in general \cite{DAlessandro:2005fnh}, and three loops for $p=2$ \cite{Drummond:2013nda} (though integrands are known to four loops \cite{Eden:2012tu,Fleury:2019ydf}). In the planar limit, it is known through three loops at weak coupling \cite{Chicherin:2015edu} (with their integrands being known to five-loop order \cite{Chicherin:2018avq} and even through ten loops for $p=2$ \cite{Bourjaily:2016evz}), while at strong coupling holographic and bootstrap approaches determine the leading strong coupling limit \cite{Arutyunov:2000py,Dolan:2006ec,Uruchurtu:2008kp,Uruchurtu:2011wh,Rastelli:2016nze,Rastelli:2017udc,Arutyunov:2018tvn,Caron-Huot:2018kta}, as well as some sub-leading $1/\l$ \cite{Alday:2018pdi,Binder:2019jwn,Drummond:2019odu,Drummond:2020dwr,Abl:2020dbx,Aprile:2020mus,Alday:2022uxp} and $1/N$ \cite{Alday:2017xua,Aprile:2017bgs,Alday:2017vkk,Aprile:2017qoy,Aprile:2019rep,Alday:2019nin,Drummond:2020uni,Huang:2021xws,Drummond:2022dxw} corrections. Knowing any of the $\<\O_2\O_2\O_p^{(i)}\O_p^{(j)}\>$ exactly in the planar limit would be tantamount to knowing the holographic dual of the \adss Virasoro-Shapiro amplitude -- that is, the four-point scattering of \adss supergravitons in classical string theory -- and, upon developing its operator product expansion, would furnish a substantial part of the solution of planar $\cN=4$ SYM. Unfortunately, due to the inherent difficulty of intermediate coupling, what is known about this object is essentially perturbative around either weak or strong coupling.

And yet, rather astoundingly, if one integrates these four-point functions against a fairly vanilla spacetime measure, these correlators can be determined {\it exactly} and, as we will see, simply. In particular, the integrated correlators of interest here are defined as follows \cite{Binder:2019jwn}:
\begin{align}\label{eq:integrated_correlator_intro}
	\mathcal{G}_p^{(N|i,j)}(\tau) := -\frac{2}{\pi}\int_0^\infty dr\int_0^\pi d\theta\,\frac{r^3\sin^2\theta}{u^2}\,\mathcal{H}^{(N|i,j)}_p(u,v;\tau)\vert_{u=1+r^2-2r\cos\theta,\,v=r^2}\,,
\end{align}
where $\mathcal{H}^{(N|i,j)}_p(u,v;\tau)$ is the ``dynamical'' part of the correlator not fixed by superconformal Ward identities (recalled explicitly in Section \ref{sec:2.1} below). The magic is that $\cG_p^{(N|i,j)}(\t)$ are computable from derivatives of a free energy on $S^4$ deformed by sources, which in turn is determined by supersymmetric localization. The aim is then to find simpler expressions upon starting from the localization integrals. In \cite{Binder:2019jwn}, these were computed exactly in the `t Hooft limit for all $p$, yielding an elegant integral formula for the integrated AdS-Virasoro-Shapiro amplitude. Our goal here is to study these integrated correlators at finite $N$, for all $p$. 

For $p=2$, this was done in \cite{Dorigoni:2021guq}, who found a conjecturally exact, recursive solution at finite $N$, an elegant result subjected to many checks. (See \c{Dorigoni:2021rdo,Dorigoni:2022zcr,Wen:2022oky,Hatsuda:2022enx} for subsequent related work.) That solution was elucidated in \c{Collier:2022emf} from another point of view, one which will be crucial for this paper: namely, that of the $\sl$ spectral decomposition. To introduce this briefly, any $\sl$-invariant observable of $\cN=4$ SYM admits, by way of finiteness of free $\cN=4$ SYM and the spectral theorem for $\sl$, a decomposition into a complete, $\sl$-invariant eigenbasis of the Laplacian on the upper-half $\t$-plane. There are three branches of eigenfunctions: a continuous branch (the non-holomorphic Eisenstein series, $E_s^*(\t)$, with $\Re s=\half$), a discrete branch (the Maass cusp forms, $\phi_n(\t)$), and a constant. The overlap with the constant term is equal to the ensemble average, with respect to the Zamolodchikov metric, over the space of $\cN=4$ theories parameterized by $\t$. 

This is a powerful tool in the $\cN=4$ SYM context because the complete eigenbasis ``decouples'' the $\t$-dependence from the core information, not made redundant by S-duality, that characterizes a given observable. The entire content of an observable is characterized by its overlaps with these basis elements: $\t$-space is traded for ``spectral space.'' As in any other physical context where a complete basis is identified, it gives a systematic way to understand the possible $\t$-dependence of $\sl$-invariant $\cN=4$ SYM observables, and, therefore, to classify these dependences according to the functional complexity of the overlaps.\foot{It also makes connections, so far nascent, between $\cN=4$ SYM observables, on the one hand, and mathematical questions in arithmetic chaos and the analytic structure of the fundamental domain, on the other. See \cite{sarnak, hejrack, sarnakk} for some mathematical background.}

Applied to the $p=2$ integrated correlator, $\mathcal{G}_2^{(N)}(\tau)$, \c{Collier:2022emf} found that {\it the entire integrated correlator is determined by a single polynomial}. First, the cusp form overlap vanishes. Second, the Eisenstein overlap is given by a universal gauge-theoretic prefactor times an $N$-dependent polynomial in the spectral parameter $s$; this polynomial may be determined solely from the perturbative expansion of $\mathcal{G}_p^{(N|i,j)}(\tau)$ to finite order. These polynomials solve a short recursion relation in $N$, equivalent to the recursion originally found by \cite{Dorigoni:2021guq}. Finally, the ensemble average, which is simply (one half times) the aforementioned polynomial evaluated at $s=1$, is also just a polynomial in $N$. If one prefers to restore the $\t$-dependence, the result is
\es{g2intro}{\mathcal{G}_2^{(N)}(\tau) = {N(N-1)\o4} +\frac{1}{4\pi i}\int_{\text{Re}\,s=\frac{1}{2}}ds\,\frac{\pi}{\sin(\pi s)}s(1-s)f_2^{(N)}(s)\,E^*_s(\tau)\,,}
where the full information (including the additive constant) is contained in the polynomial $f_2^{(N)}(s)$. A consequence of this structure is that the entire $\mathcal{G}_2^{(N)}(\tau)$ is fixed by the sector of zero total instanton number. The origin of the recursion remains mysterious. However, what the $\sl$ decomposition makes clear is that the integrated correlator is about as simple of a non-trivial $\t$-dependent observable as one could hope for. 

In this paper, we carry forward this analysis to the general class of integrated correlators $\mathcal{G}_p^{(N|i,j)}(\tau)$. The results are highly uniform in $p$. We will compute the full matrix of correlators for $p\leq 5$ -- recall that for $p\geq 4$ there are multiple trace structures -- as well as a certain infinite family that exists for all $p$. The results will again be very simple in the $\sl$ spectral decomposition, with {\it all integrated correlators computed herein determined simply by polynomials in the spectral parameter $s$}. These polynomials obey fascinating recursion relations relating different values of both $N$ and $p$. They are determined by a finite number of orders in weak coupling perturbation theory. The structural rigidity of these results raises obvious questions about why, exactly, they are so, and begs for a fundamental derivation of these recursion relations. From the AdS$_5 \times$ S$^5$ point of view, the fact that these are known for all $N$ and $\tau$ means that the quantum type IIB string theory scattering amplitudes, upon suitable integration over AdS$_5$ boundary points, are known. To what extent that recasting can be made substantive, by telling us something non-trivial about type IIB string theory, is an intriguing open question for the future.  

In {\bf Section \ref{sec2}} we introduce the necessary tools and background to study the integrated correlators -- half-BPS operators, their correlators both integrated and unintegrated, and the $\sl$ spectral decomposition -- and review the $p=2$ case. 

In {\bf Section \ref{sec3}} we present our first results, namely, the integrated correlators for $p\leq 5$. At both $p=4$ and $p=5$ there are two half-BPS operators, one single-trace and one double-trace, so each case furnishes a $2\x2$ matrix. The technique is straightforward: using localization, one develops the perturbative expansion in $g_{\rm YM}^2$; infers from this the Eisenstein overlap by matching to the perturbative expansion of the spectral decomposition; and then justifies that the cusp form overlap vanishes. The result, as advertised above, is that the integrated correlators are determined in terms of a single polynomial, $f_p^{(N|i,j)}(s)$, of degree $2N+2\lfloor {p\o2}\rfloor - 4$ and symmetric under $s \rar 1-s$.\foot{In fact, we find that $f_p^{(N|i,j)}(s)\propto (2s-1)^2$, which further reduces the degree of the undetermined polynomial.} That is, $\mathcal{G}_p^{(N|i,j)}(\tau)$ takes the same functional form as \eqr{g2intro} with different polynomial and constant terms. Having enumerated the polynomials for many values of $N$, we search for, and indeed find, linear recursion relations. Let us reproduce, as an example, the recursion for $p=3$ here:
\e{}{N(N+1)\Big[f_3^{(N+1)}(s)- f_3^{(N)}(s)\Big] = 2N(N-1)\,f_2^{(N+1)}(s) + 2(N+1)(N+2)\,f_2^{(N)}(s)\,.}
This is a very tidy recursion, that relates the $p=2$ and $p=3$ polynomials $f_2^{(N)}(s)$ and $f_3^{(N)}(s)$, respectively, such that the latter are ultimately fixed by the former in the $SU(2)$ theory (plus a trivial initial condition at $N=1$). This phenomenon persists for all cases that we study. 

As for the cusp forms, in Subsection \ref{IncludingInstantons} we provide strong analytical support for their vanishing overlap from first principles, by performing explicit instanton calculations on the localization side and checking that they match the spectral result without cusp forms. These calculations use the instanton partition function in the presence of sources for half-BPS operators of higher charge \cite{Fucito:2015ofa}, generalizing the original Nekrasov partition function. All told, this means that $\mathcal{G}_p^{(N|i,j)}(\tau)$ is determined completely by the first $N+\lfloor {p\o2}\rfloor -2$ orders in perturbation theory. We do not know the fundamental reason for this remarkable fact. The recursion relations may be immediately uplifted to Laplace difference equations, which we show in Subsection \ref{ldiff}.

In {\bf Sections \ref{sec:averages} and \ref{sec:large_N}}, we further analyze the $p\leq 5$ results. Section \ref{sec:averages} collects their ensemble averages. Section \ref{sec:large_N} analyzes the integrated correlators of single-particle operators\foot{These operators, dual to single-particle states in AdS$_5 \x$ S$^5$, are linear combinations of $\O_p^{(i)}$ defined to be orthogonal to all multi-trace operators. See \eqr{spodef}.} at large $N$, developing the $1/N$ expansion of the spectral overlaps and the ensemble averages. This may be done algorithmically to arbitrary genus. We observe a match between the large $N$ ensemble averages and the strongly coupled planar results, verifying the general relation established in \c{Collier:2022emf}. Using the same correspondence between ensemble averages and bulk quantities, we also make a prediction for the one-loop AdS$_5 \times$ S$^5$ supergravity result for the integrated correlator at generic $p$ -- see \eqr{int1loop}. We explicitly develop the `t Hooft and very strongly coupled large $N$ limits from the genus expansion of the spectral overlaps, noting some interesting functional representations along the way.

In {\bf Section \ref{sec:maximal_trace}}, we present our second slate of results. We study the integrated correlator for which both operators $\O_p^{(i)}$ and $\O_p^{(j)}$ are taken to be composites of $\O_2$ with $p/2$ constituents, for arbitrary $p\in 2\Z_+$:
\e{}{\O_p^{(\rm max)} := [\O_2]^{p/2}\,.}
We call this the {\it maximal-trace family} of integrated correlators, $\cG_p^{(N|{\rm max})}(\t)$. These are a type of generalization of the extremal two-point functions studied in the context of $\cN=2$ SCFTs in four dimensions (e.g. \cite{Baggio:2014ioa,Baggio:2014sna,Baggio:2015vxa,Gerchkovitz:2016gxx,Hellerman:2017sur,Grassi:2019txd}). We again find an explicit recursive solution for all $N$ and $p$, with the same features as described earlier, but with one key difference: the recursion relation operates at {\it fixed} $N$, only shifting $p$. 
It takes the form of a semi-infinite lattice chain equation
\begin{align}\label{maxtraceintro}
\Delta_\tau \widehat{\mathcal{G}}_{2n-2}^{(N|{\rm max})}=&~2\left((n-1)^2+2nc-c \right) \widehat{\mathcal{G}}_{2n-2}^{(N|{\rm max})}-n\left(n+2c-1\right) \widehat{\mathcal{G}}_{2n}^{(N|{\rm max})} \no\\
&-(n-1)\left(n+2c-2\right) \widehat{\mathcal{G}}_{2n-4}^{(N|{\rm max})}+2c~\widehat{\mathcal{G}}_2^{(N|{\rm max})}
\end{align}
where $\widehat{\mathcal{G}}_p^{(N|{\rm max})}(\t)$ refers to an appropriately normalized ${\mathcal{G}}_p^{(N|{\rm max})}(\t)$ defined in \eqref{gHatDef}, $c=\frac14 (N^2-1)$ is the central charge and $p=2n$. This is evocative of the Toda chain equation obeyed by the extremal correlators. The ensemble average admits a closed-form expression in terms of harmonic numbers -- see \eqr{maxavg}. The large $p$ limit of this sequence is a large charge limit, the details of which will appear in future work. We also develop the $1/N$ expansion of these correlators, with more nice polynomials emerging, see e.g. \eqr{eq:maxtrace_solution} and \eqr{eq:Fp_large_N}. 

In {\bf Section \ref{sec:genpansatz}}, we make an ansatz \eqr{pansatz} for the general integrated correlator $\mathcal{G}_p^{(N|i,j)}(\tau)$ with arbitrary trace structures $i,j$.

In {\bf Section \ref{sec8}}, we conclude with some open problems and future directions. 

Several appendices round out the text with computational details, consistency checks and collected formulas. 

\sec{Setup}\label{sec2}
In this section, we introduce the half-BPS operators and their four-point functions; define the integrated correlators and how one computes them from localisation; recall the rudiments of the $\sl$ spectral decomposition, a central tool in our study of the integrated correlators; and finally, as a segue to the next sections, review the $p=2$ integrated correlator studied in \cite{Dorigoni:2021guq,Collier:2022emf}. 

\ssec{Half-BPS operators and their four-point functions}
\label{sec:2.1}
We will study four-point correlation functions of half-BPS superconformal primary operators in $\mathcal{N}=4$ SYM theory with gauge group $SU(N)$. These operators have protected scaling dimension $\Delta=p$ and transform in the $[0,p,0]$ representation of the R-symmetry group $SU(4)$. The simplest realisation of such operators is in terms of single-trace operators $T_p$, which we define by
\begin{align}\label{eq:single_trace}
	T_p(x,y) = \frac{1}{p}~ y_{I_1}\cdots y_{I_p}\text{Tr}\big(\Phi^{I_1}(x)\cdots\Phi^{I_p}(x)\big),
\end{align}
where $\Phi^I(x)$, with $I=1,\ldots6$, are the six real scalars of the theory and we introduced auxiliary $SO(6)$ null-vectors $y_I$ (obeying $y^2\equiv y_I\,y^I=0$) to project onto the symmetric traceless representation. In addition to these, we will also consider multi-trace operators, which can be obtained from products of the form 
\begin{align}\label{multiops}
	T_{p_1,\ldots,p_n}(x,y) =\frac{p_1\cdots p_n}{p}~T_{p_1}(x,y)\cdots T_{p_n}(x,y),\qquad 2\leq p_1\leq\ldots\leq p_n,
\end{align}
with $p_1+\cdots+p_n=p$. Due to this degeneracy in the space of half-BPS operators, we introduce the notation $\O_p^{(i)}$ to distinguish the different single- and multi-trace operators of a given dimension~$p$, assuming an ordering with increasing number of traces.\footnote{For example, since the single-trace operator always comes first, we simply have $\O_p^{(1)}=T_p$. Within the same number of traces we adopt lexicographic ordering, such that e.g. $T_{2,4}$ comes before $T_{3,3}$, etc.} The species of operators appearing at weight $p$ are in one-to-one correspondence with restricted integer partitions of $p$ that do not include one. In what follows we will sometimes study operators with $p\leq 5$, where we encounter at most double-trace operators:
\begin{align}\label{eq:operator_lists}
\begin{split}
	\O_2^{(i)}&\in\{T_2\},\qquad\O_3^{(i)}\in\{T_3\},\qquad \O_4^{(i)}\in\{T_4,T_{2,2}\},\qquad\O_5^{(i)}\in\{T_5,T_{2,3}\}.
\end{split}
\end{align}
Since the operators $\O_p^{(i)}$ are half-BPS, their two-point functions are protected (meaning they are independent of the coupling $\tau$) and hence can be computed within the free theory. They are of the general form
\begin{align}\label{eq:two-point_function}
	\langle \mathcal{O}_p^{(i)}(x_1,y_1) \mathcal{O}_p^{(j)}(x_2,y_2) \rangle = g_{12}^p\, R^{(i,j)}_p(N),
\end{align}
where $g_{ij} = y_{ij}^2/x_{ij}^2$ is the superpropagator and the colour-factors $R^{(i,j)}_p(N)=R^{(j,i)}_p(N)$ encode the non-trivial $N$-dependence. For the operators given in \eqref{eq:operator_lists}, we record the corresponding colour-factors in Appendix \ref{app:colour-factors}. Note that in the trace-basis the two-point functions are not diagonal.

In the context of the AdS/CFT correspondence it is also useful to consider so-called {\it single-particle operators} (SPO's), which we denote by $\O_p$ without the superscript.\footnote{In the following, any quantity without a superscript $(i)$ is evaluated in the SPO basis.} These operators are dual to single-particle states in \adss supergravity, namely the AdS$_5$ supergraviton multiplet ($p=2$) and higher Kaluza-Klein modes ($p\geq3$) on S$^5$. They are given by particular linear combinations of the $\O_p^{(i)}$ and can be elegantly defined for all $N$ by their orthogonality to all multi-trace operators \cite{Aprile:2018efk,Aprile:2019rep}: 
\e{spodef}{\langle\O_p T_{p_1,\ldots,p_n}\rangle=0\,,~~ n\geq2.}
At leading order in large $N$, $\O_p$ coincides with the single-trace operator $T_p$ but receives $1/N$ suppressed multi-trace admixtures.\footnote{The coefficient of an $n$-trace admixture is suppressed by $1/N^{n-1}$ with respect to the single-trace contribution $T_p$.} The first few SPO's are given by
\begin{align}\label{eq:SPO}
	\O_2=T_2\,,\quad \O_3=T_3\,, \quad \O_4=T_4-\frac{2N^2-3}{N(N^2+1)}T_{2,2}\,, \quad \O_5=T_5-\frac{5(N^2-2)}{N(N^2+5)}T_{2,3}\,.
\end{align}
While we do not have an explicit formula for the generic colour-factors $R_p^{(i,j)}(N)$ in \eqref{eq:two-point_function}, the $N$-dependence of the two-point function in the SPO basis is simply given by\footnote{Due to our choice of normalisation for the external operators \eqref{eq:single_trace}, the colour-factor $R_p$ used here has a factor of $p^2$ difference with respect to \cite{Aprile:2020uxk}, i.e. $R_p|_{\text{here}}=R_p|_{\text{there}}/p^2$. \label{foot4}} \cite{Aprile:2020uxk}
\begin{align}\label{eq:R_p}
	R_p(N) = (p-1)\bigg[\frac{1}{(N-p+1)_{p-1}}-\frac{1}{(N+1)_{p-1}}\bigg]^{-1},
\end{align}
For the first few SPO's given above, this yields
\begin{align}
\begin{split}
	R_2(N)&=\frac{(N^2-1)}{2},\quad R_3(N)=\frac{(N^2-1)(N^2-4)}{3N},\quad R_4(N)=\frac{(N^2-1)(N^2-4)(N^2-9)}{4(N^2+1)},\\
	R_5(N)&=\frac{(N^2-1)(N^2-4)(N^2-9)(N^2-16)}{5N(N^2+5)}\,.
\end{split}
\end{align}
Note that $R_p$ vanishes for $N=1,2,\ldots,p-1$, meaning that $\O_p$ vanishes as an operator for those values of $N$, thereby relating various elements $\cO_p^{(i)}$ in \eqref{eq:operator_lists} at fixed $p$. For instance, $2\,T_4=T_{2,2}$ for $N=2,3$ and $6\,T_5=5\,T_{2,3}$ for $N=3,4$. At large $N$, $R_p$ scales as $R_p(N) \sim N^p/p$.

Next, let us introduce four-point correlation functions of the half-BPS operators defined above. In particular, we will be interested in the class of correlators given by $\langle\O_2\O_2\O^{(i)}_p\O^{(j)}_p\rangle$. Superconformal symmetry constrains these correlators to take the form \cite{Eden:2000bk,Nirschl:2004pa}\footnote{Our conventions for the conformal and $SU(4)$ R-symmetry cross-ratios read
	\begin{align}\label{eq:crossratios}
		u= x \xb &= \frac{x_{12}^2x_{34}^2}{x_{13}^2 x_{24}^2}, &&v=(1-x)(1-\xb)=  \frac{x_{14}^2x_{23}^2}{x_{13}^2 x_{24}^2}, \notag \\
		\frac{1}{\sigma}=y \bar y &= \frac{y_{12}^2 y_{34}^2}{y_{13}^2 y_{24}^2}, &&\frac{\mu}{\sigma}=(1-y)(1-\bar y)=\frac{y_{14}^2 y_{23}^2}{y_{13}^2 y_{24}^2},
	\end{align}
We denote the second R-symmetry cross ratio	 as $\mu$ instead of the canonical $\t$, to avoid confusion with the coupling.}
\begin{align}\label{eq:superconformal_constraint}
	\langle \O_2\O_2\O^{(i)}_p\O^{(j)}_p \rangle = g_{12}^2g_{34}^{p}\,\Big[\mathcal{G}_{p,\text{free}}^{(i,j)}(u,v;\sigma,\mu)+\mathcal{I}\,\mathcal{H}^{(i,j)}_p(u,v;\tau)\Big],
\end{align}
The factor $\mathcal{H}^{(i,j)}_p$ is the only part of the correlator which depends on $\tau$, and hence contains all of the non-trivial dynamics. It is multiplied by the factor $\mathcal{I}$ which is fixed by superconformal Ward identities to take the factorised form
\begin{align}\label{eq:cal_I}
	\mathcal{I} &= \frac{(x-y)(x-\bar{y})(\xb-y)(\xb-\bar{y})}{(y\bar{y})^2}\,.
\end{align}
On the other hand, the first term in \eqref{eq:superconformal_constraint} is the correlator in the free theory which can be computed by Wick contractions. It takes a particularly nice form in the SPO basis, which in our conventions reads
\begin{align}\label{eq:freetheory}
	\mathcal{G}_{p,\text{free}}(u,v;\sigma,\mu) = R_2 R_p\left( \Big[1+\delta_{2,p}\Big(u^2\sigma^2+\frac{u^2\mu^2}{v^2}\Big)\Big] + \frac{p}{2c}\Big[u\sigma+\frac{u\mu}{v}+(p-1)\frac{u^2\sigma\mu}{v}\Big]\right),
\end{align}
where $c$ is the central charge $c=(N^2-1)/4$ of the $SU(N)$ theory. While the structure of $\mathcal{G}^{(i,j)}_{p,\text{free}}$ for generic $(i,j)$ takes a similar form in terms of the cross-ratios $(u,v)$ and $(\sigma,\mu)$, the $N$ dependence of the different $(\sigma,\mu)$ monomials is in general very non-trivial and the simplicity of the above result is a consequence of the orthogonality property of the SPO's $\O_p$ \cite{Aprile:2020uxk}.

\ssec{Integrated correlators}
\label{sec:Integrated_correlators}
Our main focus henceforth will be on a family of observables defined by a Euclidean integration of the dynamical part of $\langle \O_2\O_2\O^{(i)}_p\O^{(j)}_p \rangle$: 
\begin{align}\label{eq:integrated_correlator}
	\mathcal{G}_p^{(N|i,j)}(\tau) := -\frac{2}{\pi}\int_0^\infty dr\int_0^\pi d\theta\,\frac{r^3\sin^2\theta}{u^2}\,\mathcal{H}^{(N|i,j)}_p(u,v;\tau)\vert_{u=1+r^2-2r\cos\theta,\,v=r^2}\,.
\end{align}
It was first shown in \cite{Binder:2019jwn} that $\mathcal{G}_p^{(N|i,j)}$ is given by derivatives of $S^4$ partition function of $\mathcal{N}=2^*$ gauge theory deformed by couplings $\tau_A'$ of higher-weight chiral operators (here and below  $\tau_A'$ collectively denotes all irrelevant couplings $\tau_3' , \tau_4', ... $ which couple to chiral primary operators of dimension $\Delta>2$, whereas $\tau$ denotes the exactly marginal coupling).\footnote{	We note that \cite{Binder:2019jwn} studied the relation \eqref{eq:localisation_relation} in the 't Hooft limit where contributions from single-trace operators dominate: that is, for the case $(i,j)=(1,1)$ in our notation. However, as we will discuss in more detail below, \eqref{eq:localisation_relation} is also valid for multi-trace operators, i.e. for all pairs $(i,j)$, with the appropriate prescription for multi-trace sources.} In our conventions, this relation reads 
\begin{align}\label{eq:localisation_relation}
\mathcal{G}_p^{(N|i,j)}(\tau)=\frac{R^{(i,j)}_p}{4}~\frac{v^{i,\m}_p \bar{v}^{j,\n}_p \p_{\tau_\m'}\p_{\bar{\tau}_\n'}\p_m^2\log\cZ_N(\tau,\tau_A',m)}{v^{i,\m}_p \bar{v}^{j,\n}_p  \p_{\tau_\m'}\p_{\bar{\tau}_\n'}\log \cZ_N(\tau,\tau_A',m)}\bigg|_{\tau'_A=m=0}~.
\end{align}
The $S^4$ partition function\footnote{Strictly speaking, for non-vanishing $\tau'_A$, $\cZ_N(\tau,\tau_A',m)$ should be viewed as a generating functional for correlators of chiral primary operators.} $\cZ_N(\tau,\tau'_A,m)$ takes the following form \cite{Gerchkovitz:2016gxx,Pestun:2007rz,Nekrasov:2002qd,Fucito:2015ofa}
\begin{align}
\label{generatingfunction}
\cZ_N(\tau,\tau_A',m) =\int &d^Na  \prod_{i<j} (a_{ij})^2 \delta\big(\sum_{i=1}^N a_i\big) \bigg|\exp\big(i \pi \tau \sum_{i=1}^N a_{i}^{2} +\sum_{p=3}^\infty i\pi^{p/2}\tau_p'\sum_{i=1}^N a_i^p\big)\bigg|^2  \no\\
&\big|Z_{\text{inst}}(\tau, \tau_A', m, a)\big|^2 \frac{1}{H(m)^N}\prod_{i<j} \frac{H^2(a_{ij})}{H(a_{ij}+m)H(a_{ij}-m)}~~.
\end{align}
The derivatives in \eqr{eq:localisation_relation} ensure that $\mathcal{G}_p^{(N|i,j)}(\tau)$ are unambiguous functions of the coupling \c{Gerchkovitz:2016gxx,Bobev:2016nua,Chester:2020vyz}.

Let us unpack these expressions. The integration variables $a_i$ are over the Cartan subalgebra of the gauge group $SU(N)$ -- i.e. $a_i\in \mathbb R$ and are constrained by the condition $\sum_ia_i=0$ -- and $a_{ij} := a_i-a_j$. $m$ is the $\cN=2$-preserving mass of the adjoint hypermultiplet. The function $H(z)$ is given by product of two Barnes G-functions
\es{}{H(z)=e^{-(1+\gamma) z^{2}} G(1+i z) G(1-i z)~.}
The instanton factor $Z_{\text{inst}}(\tau, \tau_A', m, a)$ in \eqref{generatingfunction} is the localized contribution to the partition function from non-trivial instanton configurations of SYM gauge fields. It admits the decomposition
\be
\label{NekrasovPF}
Z_{\text{inst}}(\tau, \tau'_A, m, a)  = \sum_{k=0}^\infty e^{2\pi i \tau k} Z_{\text{inst}}^{(k)}(\tau'_A, m, a)~,
\ee
where $Z_{\text{inst}}^{(k)}(\tau'_A, m, a)$ denotes the contribution from the $k'$th instanton sector. The expansion of $\big|Z_{\text{inst}}\big|^2$ appearing in \eqr{generatingfunction}, in powers of the instanton counting factor $q:= e^{2\pi i \t}$, is
\es{InstantonExpansion}{\big|Z_{\text{inst}}\big|^2 = 1+ q~ Z^{(1)}_{\text{inst}}  + \bar{q}~\bar{Z}^{(1)}_{\text{inst}}  + q \bar{q}~|Z^{(1)}_{\text{inst}}|^2 +\cdots }
The explicit form of $Z_{\text{inst}}^{(k)}(\tau'_A, m, a)$ is known for generic $k$. For our purposes we will only need the expression for $k=1$ \cite{Fucito:2015ofa}:
\begin{align}
\label{1instantonPF}
Z_{\text{inst}}^{(1)}(\tau'_A, m, a)=-m^2\sum_{\ell=1}^{N} \exp\(-\sum_{p=3}^\infty i \pi^{p/2} \tau'_p  \(a_\ell^p+(a_\ell+2i)^p-2(a_\ell+i )^p\)  \) \prod_{j\neq \ell} \frac{\(a_{\ell j}+i \)^2 -m^2 }{\(a_{\ell j}+i \)^2+1} 
\end{align}
We note that this expression from \cite{Fucito:2015ofa} is for $U(N)$ gauge group.\footnote{Our notation is related to those of eq. (2.24) in \cite{Fucito:2015ofa} as follows: $a_i^{\text{there}}=-ia_i^{\text{here}},~ m^{\text{there}}=-(i m^{\text{here}}+1 )~,$ $(-i)^p 2\pi \tau_p^{\text{there}} =- \pi^{p/2} p! \tau'_p~$.} However, it is applicable for computing correlators in $\mathcal{N}=4$ SYM with $SU(N)$ gauge group by simply imposing the constraint $\sum_i a_i=0$ on the eigenvalues in the $U(N)$ matrix model.\footnote{This is unlike generic $\mathcal{N}=2$ theories. We are grateful to Silviu Pufu for a discussion.}

At the $\cN=4$ conformal point, where $m=\tau_A'=0$, $Z_{\text{inst}}(\tau, 0, 0, a) = H(0) = 1$, and $\cZ_N(\tau,0,0)$ is just a Gaussian matrix model in the (special) unitary ensemble that can be explicitly evaluated:
\be
\label{N4PartitionFSUN}
\cZ_{N}^{^{\cN=4}}(\tau) =\int d^{N} a_{i}\prod_{i<j} \(a_{i j}\)^{2} \delta\big(\sum a_i\big) \bigg|e^{i \pi \tau \sum_{i} a_{i}^{2}}\bigg|^2=\frac{G(N+2)}{(2\pi)^{N(N-1)/2}\(2 \Im\t\)^{(N^2-1)/2}\sqrt{N}}~~.
\ee

An important detail in \eqref{eq:localisation_relation} is the operator mixing problem on $S^4$. As explained in \cite{Gerchkovitz:2016gxx}, in curved space, the source for a chiral operator $\cO^{(i)}_p$ can have non-minimal couplings to lower-dimensional chiral operators $\cO^{(i)}_{p-2n}$ with $n\in\Z_+$, due to non-vanishing background fields such as the spatial curvature. Therefore, to compute correlators on $\mathbb{R}^4$ from the deformed $S^4$ partition function, one must solve this operator mixing problem. This is achieved through a Gram-Schmidt orthogonalization process which allows us to compute the complex vectors $v_p^{i,\m}$ from the matrix of $S^4$ two-point functions. These vectors encode the mixing of weight-$p$ chiral operators of species $i$ with lower-weight chiral operators. Insertion of a chiral operator $\cO_p^{(i)}$ in a correlation function on $\mathbb{R}^4$ is then obtained through the combination of derivatives $v_p^{i,\m}\p_{\tau_\m'}$, where we sum over the index $\mu$ which indexes operators in the following set:
\e{muindex}{\{\cO_p^{(i)}\}~ \cup~\{ \cO_{p-2}^{(j_1)}\} ~\cup~\{ \cO_{p-4}^{(j_2)}\} ~\cup ~\cdots~ }
where $i$ is a fixed index, whereas the $j_n$ indices run over all species of operators of indicated weight, in accord with the discussion in Subsection \ref{sec:2.1}. In a slight (but hopefully intuitive) abuse of notation, we write $\mu = p^{(j)}$ to refer to the indicated operator $\cO_p^{(j)}$ of some fixed $p$ and $j$. 

Each operator $\cO_p^{(j)}$ is associated with a source in the partition function $\cZ_N$; the meaning of $\p_{\tau_{p^{(j)}}'}$ is to differentiate with respect to that source. The derivatives are defined as follows: for a given multi-trace operator $\cO_p^{(j)}\equiv T_{p_1,\ldots,p_n}$ such that $p_1+\cdots+p_n=p$, 
\e{derivdef}{{\p\o \p{\tau_{p^{(j)}}'}} := \prod_{i=1}^n {\p\o\p_{\tau_{p_i}'}}\,,}
with the understanding that $\tau'_2\equiv\tau$. Its action on $\cZ_N$ defines differentiation with respect to the source for the composite operators.

As an example, consider the insertion of the double-trace operator $\cO_5^{(2)} \equiv T_{2,3}$ for which $\mu \in \{ 5^{(2)}, 3^{(1)}\}$. The expressions of interest will have derivatives of the form $\p/\p{\tau_{p^{(j)}}'}$ acting on $\cZ_N$. Applying \eqr{derivdef}, for $p^{(j)}=3^{(1)}$, this is a single-trace operator, so $\p \cZ_N/\p{\tau_{3^{(1)}}'} := \p_{\tau_{3}'} \cZ_N $. Likewise, for $p^{(j)}=5^{(2)}$ we have $\p \cZ_N/\p{\tau_{5^{(2)}}'} := \p_{\tau_{3}'}\p_{\tau}\cZ_N$. Appendix \ref{app:localization} contains more details on this procedure for cases up to $p=5$.


\ssec{The $SL(2,\Z)$ spectral decomposition}
\label{SpectralDecomposition}

Many observables of $\cN=4$ SYM are $\sl$-invariant. Such observables, call them $\fo(\t)$, admit a spectral decomposition into a complete $\sl$-invariant eigenbasis of the Laplacian on the upper half plane. Let us introduce the main elements of this decomposition. More complete explanations may be found in Sections 2-3 of \cite{Collier:2022emf} in the $\cN=4$ SYM context, and in \cite{Terras_2013,Iwaniec2002SpectralMO} in great mathematical detail; see also \c{DHoker:2022dxx} for a pertinent review. The fundamental domain for $\sl$ may be defined as
\begin{equation}
  \mathcal{F} = \mathbb{H}/SL(2,\mathbb{Z}) = \left\{\tau = x+i y \in\mathbb{H} \,\bigg| - \half\leq x \leq \half, |\tau| \geq 1\right\}.
\end{equation}
This space is endowed with the hyperbolic metric
\begin{equation}
  ds^2 = {dx^2+dy^2\over y^2}
\end{equation}
which defines a Laplacian
\es{}{ \Delta_\tau = -y^2(\partial_x^2+\partial_y^2)\,.}
We define inner products of square-integrable $\sl$-invariant functions via the Petersson inner product,
\begin{equation}
  (f,g) \coloneqq \int_{\mathcal{F}}{dxdy\over y^2} f(\tau)\overline{g(\tau)}.
\end{equation}
The observable $\fo(\t)$, which is square-integrable \cite{Collier:2022emf}, obeys the invariance equation
\begin{equation}
  \fo(\gamma\tau) = \fo(\tau),\quad \gamma \tau = {a\tau+b\over c\tau+d},\quad \gamma = \begin{pmatrix} a & b \\ c & d \end{pmatrix}\in SL(2,\mathbb{Z}).
\end{equation}
Restricting to real-valued $\fo(\t)$, we develop a Fourier decomposition with respect to $x$,
\e{Okfourier}{\fo(\t) = \fo_0(y) + \sum_{k=1}^\i2\cos(2\pi k x)\, \fo_k(y).}
In $\cN=4$ SYM, where $x =\theta/2\pi$, the mode number $k$ is the total instanton number. Instanton-anti-instanton pairs contribute integer powers of $q\qb = e^{-4\pi y}$, where $q \coloneqq  e^{2\pi i \t}$, with total instanton number zero. 

The Roelcke-Selberg spectral decomposition takes the following form: 
\begin{align}\label{eq:spectral_decomp}
	\fo(\tau) = \langle \fo\rangle + \frac{1}{4\pi i}\int_{\text{Re}\,s=\frac{1}{2}}ds\,\{\fo,E_s\}\,E^{*}_s(\tau)+\sum_{n=1}^\infty (\fo,\phi_n)\,\phi_n(\tau),
\end{align}
This defines a convergent decomposition for any $\tau\in\cF$. There are three branches here.

The first is the constant function, $\phi_0 = \vol(\cF)^{-1/2}$, so $\<\fo\> = \vol(\cF)^{-1}(\fo,1)$ is the normalized average of $\fo(\t)$ over $\cF$:
\es{avgdef}{\<\fo\> := \vol(\cF)^{-1}\int_\cF {dx dy\o y^2} \,\fo(\t)}
The bracket notation indicates that this is equivalent to the normalized {\it ensemble} average of $\fo(\t)$, that is, the average value of $\fo(\t)$ over the $\cN=4$ supersymmetry-preserving conformal manifold with respect to the Zamolodchikov measure. This identification between modular and ensemble averages is special to $\cN=4$ SYM, in which the Zamolodchikov metric is exactly hyperbolic due to maximal supersymmetry \cite{Papadodimas:2009eu}, unlike other sub-maximal theories whose conformal manifolds admit an $SL(2,\Z)$ action (e.g. $\cN=2$ SQCD in four dimensions \cite{Gerchkovitz:2014gta,Baggio:2014ioa}). 

The second is the continuous branch of non-holomorphic Eisenstein series, 
\begin{align}\label{eiseq}
		\Delta_\tau E^*_s(\tau) = s(1-s) E^*_s(\tau)\,,\quad \Re(s)=\frac{1}{2}.
	\end{align}
The star denotes that we use the ``completed'' Eisenstein series, whose Fourier expansion with respect to $x$ is
\begin{equation}
\label{EisensteinModeExpansion}
E^*_{s}(\tau)=\L(s)y^{s}+\L(1-s) y^{1-s}+\sum_{k=1}^{\infty} 4 \cos (2 \pi k x) \frac{\sigma_{2 s-1}(k)}{k^{s-\frac{1}{2}}} \sqrt{y} K_{s-\frac{1}{2}}(2 \pi k y),
\end{equation}
where $\L(s) := \pi^{-s} \G(s) \zeta(2s)$ is the completed Riemann zeta function and $\sigma_{2s-1}(k)$ is the divisor function. This normalization is convenient because it manifests the reflection symmetry 
\es{}{E^*_s(\t) = E^*_{1-s}(\t)\,.}
The overlap with $\fo(\t)$ is then a rescaled Petersson inner product, which may be simplified to
\es{rstrans}{\{\fo,E_s\} &:= \L(s)^{-1}(\fo,E_s)\\
&= \L(s)^{-1}\int_0^{\i} dy\, y^{-s-1}\, \fo_0(y)\,.}
The second line will be important for the physics: the Eisenstein overlap is fully determined by the zero mode of $\fo(\t)$, i.e. the projection of $\fo(\t)$ onto the sector of zero total instanton number. This sector includes the weak-coupling regime $g_{\rm YM}^2\ll1$. The reflection symmetry of $E^*_s(\t)$ implies the same property of $\{\fo,E_s\}$. Note that the ensemble average may be extracted from the $s\rar 1$ limit of the Eisenstein overlap,
\e{aveis}{\<\fo\> = \half\, \lim_{s\rar 1} \{\fo,E_s\}\,,}
which follows from $E_0(\t)=1$. 

The third is an infinite discrete branch of Maass cusp forms in the so-called $L^2$ norm, $(\phi_n,\phi_n)=1$. These functions, infinite in number, obey the eigenvalue equation
\e{}{\D_\t \phi_n(\t) = \({1\o 4} + t_n^2\) \phi_n(\t)\,,\quad n\in\Z_+\,,\quad t_n\in\R\,, \quad 0<t_1 < t_2 < \ldots}
Reality of $\fo(\t)$ restricts the basis to the even cusp forms, invariant under $x \rar -x$. The cusp forms $\phi_n(\t)$ have no zero mode, instead decaying exponentially at the cusp $y\rar\i$, hence contributing to neither the perturbative nor `t Hooft regimes. The cusp forms are rich mathematical objects exhibiting various signs of arithmetic chaos \cite{sarnak, hejrack, sarnakk}; however, quite intriguingly, the integrated correlators $\cG_p^{(N)}(\t)$ will be seen to have {\it vanishing} cusp form overlap, so we will not say anything further about the cusp forms here. 

In summary, the entire content of $\fo(\t)$ is stored in the spectral overlaps $\{\fo,E_s\}$ and $(\fo,\phi_n)$.

Looking ahead slightly, our computations will begin from the zero mode, $\fo_0(y)$:
\es{fmode}{\fo_{0}(y) =\langle\fo\rangle+\frac{1}{2 \pi i} \int_{\operatorname{Re} s=\frac{1}{2}} d s\,\{\fo, E_{s}\} \Lambda(s) y^{s}}
As shown in \cite{Collier:2022emf}, $\cN=4$ SYM observables with a consistent perturbative expansion in $g_{\rm YM}^2$ admit a rather constrained analytic form for their Eisenstein overlaps:
\es{EisensteinOverlap}{	\{\fo,E_s\} = \frac{\pi}{\sin(\pi s)} s(1-s)f_{\text{p}}(s) + f_{\text{np}}(s)}
The function $f_{\text{p}}(s)$ encodes the complete perturbative part of $\fo(\tau)$, its residues giving the weak coupling ($y \gg 1$) data. The function $f_{\text{np}}(s)$ encodes non-perturbative, instanton-anti-instanton corrections. Both functions must be invariant under $s\mapsto1-s$, and are regular for $s\in\CC$ away from $s=0$ and its reflection. For the observables we consider, it will turn out that $ f_{\text{np}}(s)=0$, a fact whose origin lies in the Borel summability of their perturbative expansions; this further implies \cite{Collier:2022emf} that $f_{\text{p}}(s)$ must be regular at $s=0$ and is thus entire. 

\ssec{Prelude: Review of $p=2$ result}\label{sec:summary_2222}

To illustrate the spectral decomposition \eqref{eq:spectral_decomp} at work -- and to set the stage for our computations ahead -- let us consider the example of the $p=2$ integrated correlator, $\mathcal{G}^{(N)}_2(\tau)$. The result is
\begin{align}\label{eq:spectral_decomp_G2}
	\mathcal{G}^{(N)}_2(\tau)=\frac{N(N-1)}{4}+\frac{1}{4\pi i}\int_{\text{Re}\,s=\frac{1}{2}}ds\,\frac{\pi}{\sin(\pi s)}s(1-s)f^{(N)}_2(s)\,E^*_s(\tau).
\end{align}
That is, the ensemble average, and Eisenstein and cusp form overlaps, are
\begin{align}\label{g2overlaps}
	\langle\mathcal{G}^{(N)}_2\rangle=\frac{N(N-1)}{4},\quad \{\mathcal{G}^{(N)}_2,E_s\}=\frac{\pi}{\sin(\pi s)} s(1-s)f^{(N)}_{2}(s),\quad (\mathcal{G}^{(N)}_2,\phi_n)=0.
\end{align}
For $N\leq 2$,
\e{eq:f_2_coefficients}{f_2^{(1)}(s)=0,\quad f_2^{(2)}(s)=(2s-1)^2.}
For $N>2$, it turns out that the spectral overlaps obey a powerful three-term recursion relation \cite{Dorigoni:2021guq}
\begin{align}\label{eq:recursion_2222}
	N(N-1)f^{(N+1)}_2(s) = \Big[2(N^2-1)-s(1-s)\Big] f^{(N)}_2(s)-N(N+1) f^{(N-1)}_2(s),
\end{align}
This fully determines $\mathcal{G}^{(N)}_2(\t)$ for any $SU(N)$ in terms of the $SU(2)$ result alone (and its triviality at $N=1$). 

To reiterate, the complete content of $\mathcal{G}^{(N)}_2(\t)$ is contained in \eqr{g2overlaps}-\eqr{eq:recursion_2222}. 

This observable also admits a (formally equivalent) lattice-integral representation, as originally seen in \cite{Dorigoni:2021guq}, which may be derived by performing an $SL(2,\Z)$ Poincar\'e sum of the zero mode of (a Borel resummation of) $\mathcal{G}^{(N)}_2(\t)$. The same will be true of the $p>2$ integrated correlators.\foot{It is straightforward to prove \c{Collier:2022emf} that any $SL(2,\Z)$-invariant, square-integrable observable $\fo(\t)$ with $(\fo,\phi_n)=0$ admits a lattice-integral representation of the form admitted by $\mathcal{G}^{(N)}_2(\t)$. On the other hand, a generic $SL(2,\Z)$-invariant observable, with nonzero cusp form overlap, need not.}
 As discussed in the Introduction and hopefully made clear by our presentation, instead writing $\mathcal{G}^{(N)}_2(\t)$ in a complete, $\sl$-invariant basis illuminates just how simple it is in the space of possible $\cN=4$ observables:

\bul The cusp form overlap vanishes for all $N$.

\bul The non-perturbative part of the Eisenstein overlap vanishes for all $N$.

\bul The perturbative part of the Eisenstein overlap, $f^{(N)}_2(s)$, is an even polynomial of order $2N-2$. 

\ni The passage between bases also helps to clarify the physical meaning of some of the mathematical aspects of the lattice-integral representation itself, such as the identification of the lattice-integral kernel as an $\sl$ Borel transform and the observed properties of $\mathcal{G}^{(N)}_2(\t)$ under integration and inversion.

Let us make a few further comments. First, the $SU(2)$ result is, rather literally, the simplest result consistent with the analyticity properties of the overlap: in particular, $f_p^{(N)}(s)$ must be an entire, reflection-symmetric (i.e. even under $s \rar 1-s$), non-constant function of $s$, the simplest example of which is the degree-two monomial shown above. Second, since the cusp form overlap vanishes, the above recursion relation may be uplifted to a `Laplace difference equation' for the integrated correlator \cite{Dorigoni:2021guq}, which reads 
\begin{align}\label{LDEp2}
	N(N-1)\mathcal{G}^{(N+1)}_2(\tau) = \Big[2(N^2-1)-\D_\t\Big]\mathcal{G}^{(N)}_2(\tau)-N(N+1) \mathcal{G}^{(N-1)}_2(\tau),
\end{align}
where we have used \eqr{eiseq}. This differential relation is equivalent to the algebraic relation \eqr{eq:recursion_2222}. Finally, the large $N$ expansion of $\mathcal{G}^{(N)}_2(\t)$ was analyzed in \cite{Dorigoni:2021guq, Collier:2022emf}, in various limits and including non-perturbative corrections in both $1/\l$ \cite{Dorigoni:2021guq} and $1/N$ \c{Collier:2022emf}, where $\l:=g_{\rm YM}^2 N$ is the `t Hooft coupling. We refer the reader to those papers for details. Here we just highlight one result, to be generalized to $p>2$ later in this work: at $\l \gg 1$, 
\e{}{\mathcal{G}^{(N)}_2(\l) \approx {N^2\o 4}\(1-24 \zeta(3)\l^{-3/2} + O(\l^{-5/2})\) + O(N^0)}
The leading term, which gives the value of the integrated correlator in tree-level \adss supergravity in our normalization, manifestly equals the ensemble average in \eqr{g2overlaps} at large $N$ \c{Collier:2022emf}.

With these observations in place, we now proceed to study the more general class of integrated correlators $\mathcal{G}_p^{(N|i,j)}(\tau)$. The $\sl$ spectral decomposition will greatly simplify the physical and technical analysis.

\sec{Results I: Integrated correlators for $p\leq 5$}\label{sec3}
We now derive the integrated correlators $\cG_p^{(N|i,j)}(\t)$ for $p\leq 5$ in the $\sl$ spectral decomposition. As we will see, their general structure is uniform in $p$, essentially identical to that of the $p=2$ case presented in \eqr{g2overlaps}-\eqr{eq:recursion_2222}. The derivation is a three-step process:

\begin{enumerate}[label=\textbf{\arabic*})]
\item In Subsection \ref{ss31}, we reconstruct the full zero(-instanton) mode of the integrated correlator from its weak-coupling expansion. The latter is computed from localization as outlined in Section \ref{sec:Integrated_correlators} and Appendix \ref{app:localization}. 

\item In Subsection \ref{sec:spectral_overlaps}, we leverage the zero mode to deduce the Eisenstein overlap \eqref{EisensteinOverlap} by matching to the spectral decomposition. Just as for $p=2$, these will have vanishing non-perturbative part, $f_{\rm np}(s)=0$, and the remaining piece will be fixed by polynomials obeying powerful recursion relations in $N$. 

\item Finally, in Subsection \ref{IncludingInstantons}, we upgrade this to the result for the full integrated correlator by arguing that the cusp form overlap vanishes. We do so with several explicit computations. This then implies that the aforementioned recursion relations are relations for the full integrated correlator. We note that they may be recast as Laplace difference equations, an exercise we carry out in Subsection \ref{ldiff}.

\end{enumerate}

\ssec{Perturbative expansion}\label{ss31}
The weak-coupling expansion of $\mathcal{G}_p^{(N|i,j)}$ (around a zero-instanton background), may be extracted from the localization expression \eqr{eq:localisation_relation}-\eqr{generatingfunction} by expanding the latter at $y\gg1$ (recall that $y := \Im \t$). Details of this computation may be found in Appendix \ref{app:localization} for all $p\leq 5$. The final result may be written as
\be
\label{PerturbativeResults}
\mathcal{G}_p^{(N|i,j)}(\t)\Big|_{\text{0-inst}}=\frac{R^{(i,j)}_p}{4} \frac{C_p^{(N|i,j)}}{D_p^{(N|i,j)}}\,,
\ee
where $C_p^{(N|i,j)}$ and $D_p^{(N|i,j)}$ are given in \eqref{CValues} and \eqref{DValues} respectively. Let us copy the result from Appendix \ref{app:localization} for $p=2,3$ below:
\begin{align}
\mathcal{G}_2^{(N)}(\t)\Big|_{\text{0-inst}}= &\frac{R_2}{4}  \bigg( \frac{12 N \zeta (3)}{y\pi} -\frac{75 N^2 \zeta (5)}{y^2 \pi^2} +\frac{735 N^3 \zeta (7)}{2 y^3 \pi^3}-\frac{945 N^2 \left(7 N^2+2\right) \zeta (9)}{4 y^4 \pi^4}\no\\
&+\frac{114345 N^3 \left(N^2+1\right) \zeta (11)}{16 y^5 \pi^5}-\frac{351351 N^2 \left(11 N^4+25 N^2+4\right) \zeta (13)}{128 y^6 \pi^6}+\ldots\bigg),
\end{align}
\begin{align}
\mathcal{G}_3^{(N)}(\t)\Big|_{\text{0-inst}} = &\frac{R_3}{4}  \bigg(\frac{18 N \zeta (3)}{y\pi}-\frac{90 N^2 \zeta (5)}{y^2 \pi^2}+\frac{735 N^3 \zeta (7)}{2 y^3 \pi^3}-\frac{2835 N^2 \left(2 N^2+1\right) \zeta (9)}{4 y^4 \pi^4}\no\\
&+\frac{114345 N^3 \left(3 N^2+5\right) \zeta (11)}{64 y^5 \pi^5}-\frac{117117 N^2 \left(11 N^4+40 N^2+9\right) \zeta (13)}{64 y^6 \pi^6}+\ldots\bigg).
\end{align}
The results for $p=4,5$ may similarly be assembled from Appendix \ref{app:localization}.

Before extracting the spectral overlaps from these, we note that these results are in agreement with previous results from two-loop perturbation theory: in particular, following the strategy of \cite{Dorigoni:2021guq}, in Appendix \ref{app:2LoopPerturbation} we take the two-loop results for the {\it unintegrated} $\<\O_2\O_2\O_p^{(i)}\O_p^{(j)}\>$ correlators from \cite{DAlessandro:2005fnh}, integrate them against the measure \eqr{eq:integrated_correlator}, and compare to the localization result shown above. The results match, as seen in Appendix \ref{app:2LoopPerturbation}. This provides a non-trivial check of the method used here. It also probes a novel aspect of the $p=4,5$ cases, in which there is a matrix of integrated correlators involving {\it double-trace} composite operators; the agreement described above verifies that the localization method works even when the external operators are composites.\foot{This is also a further verification that the Konishi operator does not contribute to $\mathcal{G}_p^{(N|i,j)}(\tau)$, as was discussed in \cite{Binder:2019jwn} in the 't Hooft limit and \cite{Dorigoni:2021guq} for finite $N$.}
\ssec{Spectral overlaps and recursion relations}\label{sec:spectral_overlaps}

To deduce the Eisenstein overlap from the perturbative expansion, one must develop the latter from the spectral decomposition. This is straightforwardly done by closing the integration contour in \eqref{fmode} to the left, yielding\foot{As we will point out momentarily, the non-perturbative part of the overlap vanishes for all $p$. This allows us to employ an amusing abuse of notation, here and henceforth, in which we refer to the ``$p$'' subscript on $f_p(s)$ as the label for the $p$'th integrated correlator.}
\begin{align}\label{eq:perturbative_expansion}
\mathcal{G}_p^{(N|i,j)}(\tau)\Big|_{\text{0-inst}}=-\sum_{n=1}^\infty(-1)^n n (n+1)\L(n+\tfrac{1}{2})f^{(N|i,j)}_p(n+1)\, y^{-n}~.
\end{align}
where we recall the definition of the completed Riemann zeta function $\L(n+\tfrac{1}{2})$ below \eqr{EisensteinModeExpansion}. This is then compared to the weak-coupling expansion of $\mathcal{G}_p^{(N|i,j)}$, from which $f^{(N|i,j)}_p(n+1)$ is extracted. Using the perturbative expansions derived in the previous subsection and Appendix \ref{app:localization}, we find that in all cases we considered, 
\begin{align}\label{eq:f_p_ansatz}
{	f_p^{(N|i,j)}(s)=(2s-1)^2\,g_p^{(N|i,j)}(s),}
\end{align}
where $g_p^{(N|i,j)}(s)$ are reflection-symmetric polynomials of  degree $2N+2\lfloor{p\o 2}\rfloor -6$, for all labels $(i,j)$. 

The perturbative expansions are Borel summable. This typically (but not always \c{Grassi:2014cla,Dorigoni:2017smz}) implies, in the context of resurgence, that non-perturbative corrections in $g_{\rm YM}^2$ are not present. This was confirmed in \c{Dorigoni:2021guq} for $p=2$, implying that the non-perturbative part of the Eisenstein overlap \eqref{EisensteinOverlap} vanishes \cite{Collier:2022emf}. One can confirm this property for the $p>2$ cases considered here as well:\footnote{One can see this from the localisation point of view as follows. From the expansion \eqref{InstantonExpansion} and the partition function \eqref{1instantonPF} it can be observed that the instanton-anti-instanton term $|Z_{1\text{-inst}} |^2$ goes like $m^4$, and therefore does not contribute to the integrated correlators \eqref{eq:localisation_relation}. Persistence of this property to higher orders leads to the above conclusion.} 
\e{}{{f_{\rm np}^{(N|i,j)}(s) = 0}}
Therefore, \eqr{eq:f_p_ansatz} gives the {\it full} Eisenstein overlap. Summarizing so far, the Eisenstein overlaps of $\cG_p^{(N|i,j)}$ for $p=3,4,5$ take the form
\e{EisensteinOverlapGp}{\boxed{\{\cG_p^{(N|i,j)},E_s\} = \frac{\pi}{\sin(\pi s)} s(1-s)(2s-1)^2 g_{p}^{(N|i,j)}(s)\,, \quad \text{where}\quad \text{deg}\[g_p^{(N|i,j)}(s)\] = 2N+2\left\lfloor\frac{p}{2}\right\rfloor-6\,. }}
This implies that $\{\cG_p^{(N|i,j)},E_s\}$ may be completely fixed by the first $N+\left\lfloor\frac{p}{2}\right\rfloor-2$ orders in weak-coupling perturbation theory. 

Given this simplicity, and inspired by the $p=2$ case \eqref{eq:recursion_2222}, we are encouraged to seek recursive relations for $p>2$. To do so, we use the above observations on the general structure of the spectral overlaps to construct the most restrictive polynomial ansatz, allowing us to compute the overlaps $f_p^{(N|i,j)}(s)$ for $p\leq5$ and $N\leq 20$ from the perturbative expansions to order $y^{-20}$. As we will discuss in detail in the following, we are then able to find many interesting relations generalising the $p=2$ recursion to relations not only among theories with different values of $N$, but also among integrated correlators of different values of $p$.

\sssec{$p=3$}
In the $p=3$ case, the perturbative expansion vanishes for $N=1,2$ (due to vanishing of the operator $\O_3$ itself) and hence we have $f^{(1)}_3(s)=f^{(2)}_3(s)=0$. For concreteness, we display the expressions for the next few values of $N$ which read
\begin{align}\label{eq:f_3_coefficients}
\begin{split}
	f_3^{(3)}(s)&=\tfrac{1}{3}(2 s-1)^2(s^2-s+18),\\
	f_3^{(4)}(s)&=\tfrac{1}{12}(2 s-1)^2(s^4-2s^3+47s^2-46s+264),\\
	f_3^{(5)}(s)&=\tfrac{1}{120}(2s-1)^2(s^6-3s^5+95s^4-185s^3+1824s^2-1732s+6240).
\end{split}
\end{align}
Considering many more cases up to $N=20$, we find that the above overlaps satisfy the following striking recursion relation:
\es{eq:recursion_2233_1}{\boxed{	N(N+1)\Big[f_3^{(N+1)}(s)- f_3^{(N)}(s)\Big] = 2N(N-1)\,f_2^{(N+1)}(s) + 2(N+1)(N+2)\,f_2^{(N)}(s)\,.}}
This is a remarkable formula: it allows one to recursively compute the $p=3$ overlaps for generic $N$ from knowledge of the $p=2$ overlaps $f^{(N)}_2(s)$, which in turn are fully determined by the $SU(2)$ theory alone, together with the trivial initial condition $f^{(1)}_3(s)=0$! Let us note that this recursion is consistent starting from the value $N=1$. In particular, the fact that $f_3^{(2)}(s)$ vanishes is compatible with $f_2^{(2)}(s)\neq0$ in the first term on the RHS thanks to the presence of an $(N-1)$ factor. 

An interesting feature of the above relation is that it has no explicit dependence on $s$, in contrast with the $p=2$ recursion relation \eqref{eq:recursion_2222}. This is concordant with the fact that the spectral overlaps $f_p^{(N)}(s)$ are polynomials of the same degree for $p=2$ and $p=3$.

This is not the only interesting relation the spectral overlaps $f^{(N)}_3(s)$ obey. At the cost of introducing some $s$-dependent terms, we find a recursion relation involving only the $p=3$ spectral overlaps:
\begin{align}
\begin{split}\label{eq:recursion_2233_2}
	(N-2)(N-3)(N+1)f_3^{(N+1)}(s) &=~~ (N-3)\Big[(N^2+2N-2)-s(1-s)\Big]\,f_3^{(N)}(s)\\
	 &\,~~+(N+2)\Big[(N^2-4N+1)-s(1-s)\Big]\,f_3^{(N-1)}(s)\\
	 &\,~~-(N-2)(N+2)(N+1)\,f_3^{(N-2)}(s)\,.
\end{split}
\end{align}
Note that due to the factors of $(N-2)(N-3)$ on the RHS, the recursion starts at $N=4$, and as a consequence of being a four-term recursion relation one needs to supply $f^{(N)}_3(s)$ with $N=2,3,4$ as initial data, making it a somewhat less powerful statement than the previous relation \eqref{eq:recursion_2233_1}.

\sssec{$p=4$}
Next, $p=4$ is the first case with a degeneracy in the spectrum of half-BPS operators: namely, the $T_4$ and $T_{2,2}$ operators. As such, we have three cases to consider: $(i,j)=(1,1),\,(1,2),\,(2,2)$. We will also consider the case of single-particle operators, $(i,j) \rar \text{(SPO, SPO)}$. 

Interestingly, while for $N=1$ the perturbative expansions vanish, we find that for $N=2,3$ the three expansions degenerate and are simply proportional to each other termwise in their $1/y$ expansions. In terms of spectral overlaps, this yields
\begin{align}\label{eq:f4_low_N_degeneracy}
\begin{split}
	N=1:\qquad\, f_4^{(1|i,j)}(s)&=0\,,\\
	N=2:\qquad f_4^{(2|2,2)}(s)&=2\,f_4^{(2|1,2)}(s)=4\,f_4^{(2|1,1)}(s)= (2s-1)^2(s^2-s+8)\,,\\
	N=3:\qquad f_4^{(3|2,2)}(s)&=2\,f_4^{(3|1,2)}(s)=4\,f_4^{(3|1,1)}(s)= \tfrac{1}{2}(2s-1)^2(s^2-s+6)(s^2-s+18)\,.
\end{split}
\end{align}
This phenomenon is a non-trivial consequence of considering the $SU(N)$ theory and is explained by the fact that for low values of $N$ the single-trace operator $T_4$ is no longer linearly independent of the double-trace operator $T_{2,2}$: for $N=2,3$ one has $T_4\propto T_{2,2}$ and therefore also their correlators become proportional to each other. In view of the different intricate $N$-dependence of the weak-coupling expansions \eqref{eq:c_p=4_1}-\eqref{eq:c_p=4_3}, the degenerations for $N=2,3$ seem highly non-trivial (and further support our claim that the localisation integrals correctly compute multi-trace insertions).

For $N\geq4$, the spectral overlaps $f_4^{(N|i,j)}(s)$ start to differ for different $(i,j)$. For example, the case $(i,j)=(2,2)$, corresponding to the correlator with two double-trace operators, is the one with simplest $s$-dependence:
\begin{align}
\begin{split}
	f_4^{(4|2,2)}(s)&=\tfrac{1}{12}(2s-1)^2(s^2-s+4)(s^2-s+18)(s^2-s+32)\,,\\
	f_4^{(5|2,2)}(s)&=\tfrac{1}{144}(2s-1)^2(s^2-s+50)(s^6-3 s^5+55 s^4-105 s^3+664 s^2-612 s+1440)\,.
\end{split}
\end{align}
We have obtained analogous expressions for the $(i,j)=(1,1)$ and $(1,2)$ cases. However, we refrain from presenting more explicit data since, by considering many cases of different $N$, we discover that the $p=4$ spectral overlaps likewise obey interesting recursion relations, which we present now.

\paragraph{$\mathbf{(i,j)=(2,2)}$:} Let us start with the simplest case given by the correlator involving two double-trace operators. In this case we find the very direct relation
\begin{align}\label{eq:recursion_2244|2,2}
	\boxed{f^{(N|2,2)}_4(s) = \Big[2N^2-s(1-s)\Big]\,f_2^{(N)}(s)\,.}
\end{align}
The overlaps $f_4^{(N|2,2)}(s)$ are fully determined by the $p=2$ overlaps $f_2^{(N)}(s)$ only. This will be elaborated upon in Section \ref{sec:maximal_trace}.

\paragraph{$\mathbf{(i,j)=(1,2)}$:} For the mixed correlator with one single-trace and one double-trace operator,
\begin{align}\label{eq:recursion_2244|1,2}
\begin{split}
	f_4^{(N+1|1,2)}(s)-f_4^{(N|1,2)}(s) &= \frac{1}{2}\bigg[\frac{N^3+4N^2+13N-2}{N+1}-s(1-s)\bigg]\,f_2^{(N+1)}(s)\\
	&~~~~-\frac{N^3+2N^2-4N-6}{N}\,f_2^{(N)}(s)\\		&~~~~+\frac{N(N+1)}{2}\,f_2^{(N-1)}(s)\,,
\end{split}
\end{align}
This, too, determines $f_4^{(N|1,2)}(s)$ from $f^{(N)}_2(s)$ alone.\foot{Here and in what follows, we leave implicit the trivial vanishing conditions of the spectral overlaps at $N=0,1$.} 

\paragraph{$\mathbf{(i,j)=(1,1)}$:} Recursion relations also exist for the overlaps of the correlator involving two single-trace operators albeit in somewhat more complicated form. For this case, we find
\begin{align}\label{eq:recursion_2244|1,1}
\begin{split}
	f_4^{(N+1|1,1)}(s)-f_4^{(N|1,1)}(s) &=
	\frac{1}{4}\bigg[\frac{3N^4+3N^3+23N^2-23N+26}{(N+1)^2}-s (1-s)\bigg]\,f_2^{(N+1)}(s)\\
	&~~~~+\frac{3(2N^2+2N+3)}{N(N+1)}\,f_3^{(N)}(s)\\
	&~~~~-\frac{N^5+7N^4-18N^2-12N-36}{2N^2(N+1)}\,f_2^{(N)}(s)\\
	&~~~~-\frac{N(N+1)}{4}\,f_2^{(N-1)}(s)\,,
\end{split}
\end{align}
The new feature relative to the previous relations is the presence of $f_3^{(N)}(s)$ on the RHS. However, since the latter is fixed by $f_2^{(N)}(s)$ via \eqr{eq:recursion_2233_1}, so, ultimately, is $f_4^{(N|1,1)}(s)$.\foot{ We have also found a recursion relation which does not involve the $p=3$ overlaps, but its structure as well as the $N$-dependence of the coefficients is more complicated and we refrain from giving it here. Notably, the RHS of that relation has a pole at $N=3$ and hence the recursion can be used only for $N\geq4$.}

\paragraph{\textbf{(SPO, SPO):}} Lastly, we can assemble the spectral overlap $f_4^{(N)}(s)$ of the correlator of two SPO's $\O_4$, defined as the linear combination in \eqref{eq:SPO}. From that definition, one has
\begin{align}\label{eq:f4_spo}
	f_4^{(N)}(s) = f_4^{(N|1,1)}(s)+2\beta f_4^{(N|1,2)}(s)+\beta^2 f_4^{(N|2,2)}(s)\,,
\end{align}
with mixing coefficient $\beta=-\frac{2N^2-3}{N(N^2+1)}$. Note that this particular linear combination is precisely such that $f_4^{(N)}(s)=0$ for $N=1,2,3$ since, by construction, the operator $\O_4$ itself vanishes for those values of $N$.\footnote{Because the overlaps $f_4^{(N|i,j)}(s)$ become proportional to each other for $N=2,3$ (cf.  \eqref{eq:f4_low_N_degeneracy}), the mixing coefficient $\beta$ must take the same value for $N=2$ and $N=3$ in order to be consistent with $f_4^{(N)}(s)=0$ for those $N$. Indeed, one has $\beta|_{N=2}=\beta|_{N=3}=-\frac{1}{2}$, which is a somewhat non-trivial property of $\beta$.}
By combining the recursion relations for $f_p^{(N|i,j)}(s)$ given in equations \eqref{eq:recursion_2244|2,2}-\eqref{eq:recursion_2244|1,1} one can derive a similar recursion for the overlaps of $f_p^{(N)}(s)$, whose precise form we record in Appendix \ref{app:spo_recursions}.

\sssec{$p=5$}
Similarly to $p=4$, the $p=5$ case presents three cases to consider: $(i,j)=(1,1),\,(1,2),\,(2,2)$, as well as the case of single-particle operators, $(i,j) \rar \text{(SPO, SPO)}$. 

As for $p=4$, we observe that the perturbative expansions degenerate for low values of $N$: as a consequence of the two half-BPS operators becoming proportional to each other, i.e. $T_5\propto T_{2,3}$ for $N\leq4$, we have that the weak-coupling expansions vanish for $N=1,2$ (due to $\O_3$ which ceases to exist for $N=1,2$), and that for $N=3,4$ the expansions degenerate. As  mentioned for $p=4$, the latter property is not at all obvious from the different $N$-dependent coefficients in the perturbative expansions, see equations \eqref{eq:c_p=5_1}-\eqref{eq:c_p=5_3}. For the spectral overlaps we therefore have
\begin{align}\label{eq:f5_low_N_degeneracy}
\begin{split}
	N=1,2:\qquad &f_5^{(N|i,j)}(s)=0\,,\\[3pt]
	N=3:\qquad &f_5^{(3|2,2)}(s)=\tfrac{6}{5}\,f_5^{(3|1,2)}(s)=\tfrac{36}{25}\,f_5^{(3|1,1)}(s)\\
		&~=\tfrac{4}{25}(2s-1)^2(3 s^4-6 s^3+98 s^2-95 s+498)\,,\\[3pt]
	N=4:\qquad &f_5^{(4|2,2)}(s)=\tfrac{6}{5}\,f_5^{(4|1,2)}(s)=\tfrac{36}{25}\,f_5^{(4|1,1)}(s)\\
		&~=\tfrac{3}{50}(2 s-1)^2(2 s^6-6 s^5+149 s^4-288 s^3+2297 s^2-2154 s+7704)\,,
\end{split}
\end{align}
and only from $N\geq5$ onwards their $s$-dependence differs. Comparing again cases with varying $N$, we are able to find recursion relations for the $p=5$ spectral overlaps, which we list below:

\paragraph{$\mathbf{(i,j)=(2,2)}$:}
\begin{align}\label{eq:recursion_2255|2,2}
\begin{split}
	f_5^{(N+1|2,2)}(s) &= \frac{12 (N-1)}{25 (N+1)}\bigg[(N^3+20N^2+24N+18)-6s(1-s)\bigg]\,f_2^{(N+1)}(s)\\
	&~~~~+\frac{18(N^2+2N+18)}{25}\,f_3^{(N)}(s)-\frac{36(N+2)(3N^2+2N-18)}{25N}\, f_2^{(N)}(s)\,.
\end{split}
\end{align}

\paragraph{$\mathbf{(i,j)=(1,2)}$:}
\begin{align}\label{eq:recursion_2255|1,2}
\begin{split}
	f_5^{(N+1|1,2)}(s)-f_5^{(N|1,2)}(s) &= \frac{12 (N-1)}{5 (N+1)}\bigg[\frac{2 N^3+3 N^2+8 N-3}{N+1}-s(1-s)\bigg]\,f_2^{(N+1)}(s)\\
	&~~~~+\frac{24 (N+2)}{5 N}\,f_4^{(N|1,2)}(s)+\frac{18 (N-1) (3 N-2)}{5 N (N+1)}\,f_3^{(N)}(s)\\
	&~~~~-\frac{12 (N+2)(2 N^4+N^3-12 N^2+11 N-8)}{5 N^2 (N+1)}\,f_2^{(N)}(s)\,.
\end{split}
\end{align}

\paragraph{$\mathbf{(i,j)=(1,1)}$:}
\begin{align}\label{eq:recursion_2255|1,1}
	f_5^{(N+1|1,1)}(s)-f_5^{(N|1,1)}(s) &= \frac{2(N-1)}{(N+1)}\bigg[\frac{3 N^4+4 N^3+14 N^2+4 N+15}{(N+1)^2}-s(1-s)\bigg]\,f_2^{(N+1)}(s)\qquad\nonumber\\
	&~~~~+\frac{8(N^2+N+2)}{N (N+1)}\,f_4^{(N|1,1)}(s)\nonumber\\
	&~~~~-\frac{3(2 N^5-3 N^4-4 N^3-23 N^2-24 N-24)}{N^2 (N+1)^2}\,f_3^{(N)}(s)\\
	&~~~~-\frac{2(N^7+8 N^6-18 N^4-5 N^3-42 N^2-24 N-48)}{N^3 (N+1)^2}\,f_2^{(N)}(s)\nonumber\\
	&~~~~+4N\,f_2^{(N-1)}(s)\,,\nonumber
\end{align}
where the above recursion relations start from $N=1$ with initial values $f_p^{(N|i,j)}(s)=0$ for $N\leq1$. As noted earlier in the $p=3$ case, the factors of $(N-1)$ in front of the $f_2^{(N+1)}(s)$ terms on the RHS are required by consistency of these formulae for $N=1$.

\paragraph{\textbf{(SPO, SPO):}} Lastly, the $p=5$ spectral overlaps of the correlator of SPO's are given by an analogous equation as given in \eqref{eq:f4_spo}, 
\begin{align}\label{eq:f4_spo}
	f_5^{(N)}(s) = f_5^{(N|1,1)}(s)+2\beta f_5^{(N|1,2)}(s)+\beta^2 f_5^{(N|2,2)}(s)\,,
\end{align}
where the mixing coefficient now reads $\beta=-\frac{5(N^2-2)}{N(N^2+5)}$. This is the right linear combination which yields $f_5^{(N)}(s)=0$ for $N=1,2,3,4$.\footnote{Consistency with the proportionality relations \eqref{eq:f5_low_N_degeneracy} demands that for $N=3,4$ the $p=5$ mixing coefficient takes the same value. This is indeed the case and one finds $\beta|_{N=3}=\beta|_{N=4}=-\frac{5}{6}$.} We can then again use the recursion relations for $f_5^{(N|i,j)}(s)$ to assemble a similar recursion for $f_5^{(N)}(s)$ which we give in Appendix \ref{app:spo_recursions}.

\ssec{Including instantons: completion to the full correlator}
\label{IncludingInstantons}

Having computed the zero mode of $\mathcal{G}^{(N)}_p(\t)$, and therefore the Eisenstein spectral overlap, what remains is to determine the cusp form overlap. We claim that it vanishes:
\e{nocusp}{\boxed{\(\mathcal{G}_p^{(N|i,j)}, \phi_n\)=0 \quad \forall~n>0\,.}}
That is, the full integrated correlator is 
\begin{align}\label{eq:spectral_decomp_Gp_ansatz}
	\boxed{{\mathcal{G}_p^{(N|i,j)}(\tau) = \langle\mathcal{G}_p^{(N|i,j)}\rangle +\frac{1}{4\pi i}\int_{\text{Re}\,s=\frac{1}{2}}ds\,\frac{\pi}{\sin(\pi s)}s(1-s)(2s-1)^2g_p^{(N|i,j)}(s)\,E^*_s(\tau)\,,}}
\end{align}
where recall that the average is $\langle\mathcal{G}_p^{(N|i,j)}\rangle = \half g_p^{(N|i,j)}(0)$. Note how strongly constraining this is: because $g_p^{(N|i,j)}(s)$ is an even polynomial of degree given in \eqr{EisensteinOverlapGp}, the integrated correlator may be completely fixed by the first $N+\left\lfloor\frac{p}{2}\right\rfloor-2$ orders in weak-coupling perturbation theory. 

We will provide strong support for \eqr{nocusp} with explicit instanton computations below. The basic idea behind the check is as follows. First, one can compute the perturbative expansion of the integrated correlators in nonzero instanton sectors as predicted by \eqr{eq:spectral_decomp_Gp_ansatz}. This result is then to be compared with a direct computation from localisation formula \eqref{eq:localisation_relation} using the relevant instanton partition function. For $p=2$, the absence of cusp forms was extensively checked in this manner in \cite{Collier:2022emf,Dorigoni:2021guq}. In this subsection, we carry out the check for several $p>2$ integrated correlators using the instanton partition function \eqref{1instantonPF}.\foot{Note that it is sufficient to check this in the one-instanton sector: as argued in \c{Collier:2022emf} using properties of the $\sl$ eigenbasis, all $k>1$ sectors are fully determined by the $k=0,1$ sectors.} Happily, the two results agree.

\sssec{Calculation}

The first step of extracting the perturbative expansion around $k$ instantons from \eqref{eq:spectral_decomp_Gp_ansatz} is straightforward. (The anti-instanton result, proportional to $e^{-2\pi i k \tb}$, is identical.) One just inserts the $k$'th Fourier mode of $E_s^*(\t)$ from \eqref{EisensteinModeExpansion} and the $y\gg1$ expansion
\es{}{\sqrt{y} K_{s-\frac{1}{2}}(2 \pi k y)=\frac{1}{2 \sqrt{k}} e^{-2 \pi k y} \sum_{n=0}^{\infty} \frac{a_{n}(s)}{(2 \pi k y)^{n}}~,\qquad a_{n}(s)=\frac{(s)_{n}(1-s)_{n}}{(-2)^{n} n !}~ .}
Borel summability allows us to swap the $s$ integral in \eqref{eq:spectral_decomp_Gp_ansatz} with the sum over $n$ \cite{Collier:2022emf}. Closing the contour towards the left, and taking $k=1$, we find 
\begin{align}
\label{instfromfps}
\mathcal{G}_p^{(N|i,j)}(\tau)\Big|_{\text{1-inst}}={1\o 2} e^{2\pi i \t} \sum_{n=0}^{\infty} \frac{1}{(2 \pi y)^{n}}\(\sum_{s=1}^\infty (-1)^s s(s-1)a_{n}(s)  f_p^{(N|i,j)}(s)\)~,
\end{align}
where we have used the symmetry property $a_{n}(-m)=a_{n}(1+m)$. For fixed $n$, the sum in parentheses is divergent, since both $f_p^{(N|i,j)}(s)$ and $a_n(s)$ are polynomials in $s$. The sum can be regulated in a standard manner using an exponential or zeta function regulator. 

Let us first apply this strategy to the $p=3$ integrated correlator, $\cG_3^{(N)}(\t)$. Plugging its spectral overlap into \eqr{instfromfps} gives the prediction
\begin{align}\label{oneinstanton2233}
\mathcal{G}_3^{(N)}(\tau)\Big|_{\text{1-inst}}= e^{2 \pi i \tau} \frac{R_3}{4} \sum_{n=0}^\infty \frac{\Gamma \left(N-\frac{1}{2}-n\right)}{\sqrt{\pi }  \Gamma (N+3)(\pi  y)^n}\, h_{3,n}(N)\,,
\end{align}
where $h_{3,n}(N)$ are polynomials in $N$ of degree $n+2$. These may be obtained algorithmically for any $n$ via recursion. For first few values of $n$,
\begin{alignat}{2}
h_{3,0} (N) & = -6 N (4 N+3)\,, &\qquad h_{3,1} (N) & =-\frac{9}{8} N (4 N^2-33 N-12)\,,\no\\
h_{3,2} (N) & =\frac{45}{256} N^2 (4 N^2+47N-237)\,, &\qquad h_{3,3} (N) & =-\frac{315}{4096} N^2 (N+1) (N+2) (4 N-9)\,.\
\end{alignat}
We now compare \eqr{oneinstanton2233} to the exact localization calculation of the same quantity via the definition \eqref{eq:localisation_relation}. To proceed, we use the expression \eqref{1instantonPF} for the one-instanton partition function and set the higher-weight couplings $\t_{p>3}=0$, for which we have
\be\label{1instantonPFp3}
Z^{(1)}_{\text{inst}}(\tau_3,m,a) = -m^2 \sum_{\ell=1}^{N} e^{6\pi^{3/2} \tau_3 (i a_\ell-1)} \prod_{j\neq \ell} \frac{\(a_{\ell j}+i\)^2 -m^2 }{\(a_{\ell j}+i\)^2+1}~.
\ee
Plugging this expression into \eqref{generatingfunction}, we develop a weak-coupling (large $y$) expansion for the $p=3$ integrated correlator in the one-instanton sector (exactly in the same manner as we did for the zero-instanton computations) using the definition \eqref{eq:localisation_relation}. A finite generic $N$ computation is technically involved due to the form of the instanton partition function. But for low values of $N$, the computation is straightforward. Performing this calculation for $N=3,4,5$ through the first several orders in $1/y$, we find precise agreement with \eqref{oneinstanton2233} as obtained from the spectral decomposition.\footnote{The authors of \c{Chester:2021aun} have, in work to appear, independently studied the $p=3$ integrated correlator at low values of $N$ and performed numerical checks of the vanishing cusp form overlap at finite $y$, using the convergent expansion of the deformed instanton partition function. This further complements our conclusion. We thank Shai Chester for discussions.}

The above strategy applies for any integrated correlators $\cG_p^{(N|i,j)}(\t)$. Analogous results for the full matrix of $p=4,5$ correlators are deferred to Appendix \ref{app:Instantons} for which we have also checked in this way that the cusp form overlaps vanish at low values of $N$.\foot{It would be interesting to reproduce and independently confirm these perturbative results by explicitly integrating the one-instanton correction to the unintegrated $\<\cO_2\cO_2\cO_p^{(i)}\cO_p^{(j)}\>$ correlators against the measure \eqr{eq:integrated_correlator}. The latter may be computed in principle using methods based on the ADHM construction of instantons \cite{Bianchi:1998nk, Bianchi:1999ge, Alday:2016tll,Alday:2016jeo,Alday:2016bkq}, though very few explicit results for instanton corrections to the unintegrated correlator exist in the literature \cite{Bianchi:1998nk,Green:2002vf,Kovacs:2003rt}.} 

There is yet another class of integrated correlators -- to be introduced more thoroughly in Section \ref{sec:maximal_trace} -- which is especially simple. This is what we call the {\it maximal-trace family} of integrated correlators. For $p\in 2\Z_+$, these are defined by taking both operators $\O_p^{(i)}$ and $\O_p^{(j)}$ to be multi-trace composites of $\O_2$ with $p/2$ constituents,
\es{opmax2}{\O_p^{(\rm max)} := T_{2,\ldots,2}\,.}
The lowest of these operators besides $T_2$ is $\O_4^{(\rm max)} = T_{2,2}$, whose corresponding integrated correlator $\cG_4^{(N|2,2)}(\t)$ was studied in Section \ref{sec:spectral_overlaps} above. Its Eisenstein overlap $f_4^{(N|2,2)}(s)$ was computed in \eqref{eq:recursion_2244|2,2}. Inserting this into \eqref{instfromfps}, we find
\begin{align}
\label{oneinstanton2244}
\mathcal{G}_4^{(N|2,2)}(\tau)\Big|_{\text{1-inst}}= e^{2 \pi i \tau}\, \frac{R_4^{(2,2)}}{4} \frac{1}{N^2+1}\sum_{n=0}^\infty  \frac{\Gamma \left(N-\frac{3}{2}-n\right)}{\sqrt{\pi }  \Gamma (N+2)(\pi y)^n}h_{4,n}^{(2,2)}(N)\,,
\end{align}
where $h_{4,n}^{(2,2)}(N)$ are polynomials in $N$ of degree $n+4$. For first few values of $n$ these are
\begin{align}
\begin{split}
h_{4,0}^{(2,2)}(N) & =-\frac{3N}{2}\( 8 N^3-12 N^2+3 N-12 \),\\
h_{4,1}^{(2,2)}(N) & =-\frac{9N}{32}\(8 N^4-52 N^3+83 N^2-57 N+80 \),\\
h_{4,2}^{(2,2)}(N) & =\frac{45}{2^{10}}(N-2) N^2 (N+1) \left(8 N^2-12 N-21\right),\\
h_{4,3}^{(2,2)}(N) & =-\frac{315}{2^{14}} N^2(N+1) \left( 8 N^4-84 N^3+219 N^2-9 N-414 \right).
\end{split}
\end{align}
We now compare \eqr{oneinstanton2244} to the exact localization calculation of the same quantity using the instanton partition function with vanishing higher-weight sources ($\t_{p\geq 3}=0$),
 \be\label{1instantonPFp3}
Z^{(1)}_{\text{inst}}(m,a) = -m^2  \sum_{\ell=1}^{N} \prod_{j\neq \ell} \frac{\(a_{\ell j}+i\)^2 -m^2 }{\(a_{\ell j}+i\)^2+1}~.
\ee
It is convenient to first compute $\<\p_m^2 Z^{(1)}_{\text{inst}}(0,a) \>$ in a weak-coupling expansion (the expectation value is in the ensemble \eqref{N4PartitionFSUN}). This result was obtained for finite $N$ in \cite{Chester:2019jas, Dorigoni:2021guq}, with the first few orders being
\begin{align}
\<\p_m^2 Z^{(1)}_{\text{inst}}(0,a) \> &= \bigg(-\frac{4 \Gamma \left(N+\frac{1}{2}\right)}{\pi^{1/2} \Gamma (N)}-{1\o y}\frac{3 \Gamma \left(N-\frac{1}{2}\right)}{4 \pi ^{3/2} \Gamma (N-1)}+{1\o y^2}\frac{15 N \Gamma \left(N-\frac{3}{2}\right)}{128 \pi ^{5/2}\Gamma (N-1)}\no\\[5pt]
&-{1\o y^3}\frac{105 (N-3) N \Gamma \left(N-\frac{5}{2}\right)}{2048 \pi ^{7/2} \Gamma (N-1)}+{1\o y^4}\frac{945 N (N (5 N-33)+58) \Gamma \left(N-\frac{7}{2}\right)}{131072 \pi ^{9/2} \Gamma (N-1)}+\ldots\bigg).
\end{align}
Plugging the above into \eqref{eq:localisation_relation} (using \eqref{combo} for solving the operator mixing problem) we precisely find \eqref{oneinstanton2244}. The same calculation may be repeated for higher-$p$ maximal-trace integrated correlators, studied further in Section \ref{sec:maximal_trace}; having done so for many $p>4$ again yields perfect agreement.

Altogether, these results provide substantial evidence of vanishing cusp form overlap for integrated correlators $\cG_p^{(N|i,j)}(\t)$. We leave a deeper physical understanding of this intriguing mathematical property for the future. 

\ssec{Laplace difference equations}\label{ldiff}
Having given evidence for vanishing cusp form overlap and thereby completion to the full correlator, the various recursion relations from Section \ref{sec:spectral_overlaps} should be regarded as non-trivial identities between the {\it full} integrated correlators. Said another way, relations for $f_p^{(N|i,j)}(s)$ simply `uplift' to relations for $\cG_p^{(N)}(\t)$. Starting from a given recursion relation for the spectral overlap $f_p^{(N|i,j)}(s)$, this is achieved by applying the replacement rules
\begin{align}\label{eq:replacement_rules}
	f_p^{(N|i,j)}(s) ~\mapsto~ \mathcal{G}_p^{(N|i,j)}(\tau)-\langle\mathcal{G}_p^{(N|i,j)}\rangle\,, \qquad s(1-s)f_p^{(N|i,j)}(s) ~\mapsto~ \Delta_\tau\,\mathcal{G}_p^{(N|i,j)}(\tau)\,.
\end{align}
Prime among these relations is the $p=3$ recursion relation \eqref{eq:recursion_2233_1}, which uplifts to the following powerful difference equation obeyed by the integrated correlator $\mathcal{G}_3^{(N)}(\tau)$:
\begin{align}
\label{Beautifulp3}
	{N(N+1)\Big[\mathcal{G}_3^{(N+1)}(\tau)-\mathcal{G}_3^{(N)}(\tau)\Big] = 2N(N-1)\,\mathcal{G}_2^{(N+1)}(\tau)+2(N+2)(N+1)\,\mathcal{G}_2^{(N)}(\tau)}\,.
\end{align}
The absence of an inhomogeneous term coming from the ensemble averages $\langle\mathcal{G}_p^{(N|i,j)}\rangle$ follows from consistency of the recursion relations evaluated at $s=1$.

One may, if desired, recast all of the previously derived recursion formulas as Laplace difference equations. Let us just give one more example explicitly, namely, the very simple relation \eqr{eq:recursion_2244|2,2} between $\mathcal{G}_4^{(N|2,2)}(\tau)$ and $\mathcal{G}_2^{(N)}(\tau)$:
\begin{align}
	\mathcal{G}_4^{(N|2,2)}(\tau) = -(\Delta_\tau-2N^2)\,\mathcal{G}_2^{(N)}(\tau)\,.
\end{align}
We leave other applications of \eqr{eq:replacement_rules} as an exercise for the enthusiastic reader.

\sec{Ensemble averages}\label{sec:averages}

We now consider $\langle\mathcal{G}_p^{(N|i,j)}\rangle$, the ensemble averages of the integrated correlators with respect to the Zamolodchikov measure, cf. \eqr{avgdef}. As can be seen from equation \eqr{aveis}, these are obtained by simply evaluating the Eisenstein spectral overlaps $f^{(N|i,j)}_p(s)$ at $s=1$.

For $p=2$, this yields the result already given in \cite{Collier:2022emf}, which for convenience we reproduce here:
\begin{align}\label{eq:ensemble_average_2222}
	\langle\mathcal{G}_2^{(N)}\rangle = \frac{N(N-1)}{4}\,.
\end{align}
We find that for higher $p$ the ensemble averages continue to be given by rational functions of $N$. Explicitly, we find (in slight abuse of matrix notation)
\begin{align}\label{eq:ensemble_average_2233}
	\langle\mathcal{G}_3^{(N)}\rangle &= \frac{1}{6}(N-1)(N-2)(2N+3)\,,\\[3pt]
	\langle\mathcal{G}_4^{(N|i,j)}\rangle &= 
		\begin{pmatrix}
		\frac{(N-1)(3 N^4-3 N^3-17 N^2+12 N+36)}{8 N} & \frac{(N-1)(4 N^2-N-6)}{4} \\[5pt]
		\frac{(N-1)(4 N^2-N-6)}{4} & \frac{(N-1) N^3}{2} 
		\end{pmatrix},\label{eq:ensemble_averages_2244}\\[3pt]
	\langle\mathcal{G}_5^{(N|i,j)}\rangle &= 
		\begin{pmatrix}
		\frac{(N-2) (N-1)(4 N^5+2 N^4-17 N^3+10 N^2+120 N+120)}{10 N^2} & \frac{3 (N-2) (N-1)(3 N^3+4 N^2-8 N-10)}{5 N}\\[5pt]
		\frac{3 (N-2) (N-1)(3 N^3+4 N^2-8 N-10)}{5 N} & \frac{3(N-2)(N-1)(3 N^3+5 N^2+9 N+13)}{25}
		\end{pmatrix}.\label{eq:ensemble_averages_2255}
\end{align}

We observe that at leading order in large $N$ the diagonal entries scale as $N^p$, while the off-diagonal ones scale as $N^{p-1}$:
\begin{align}\label{g4g5}
	\langle\mathcal{G}_4^{(N|i,j)}\rangle ~\longrightarrow~ \begin{pmatrix}
	\frac{3N^4}{8} & N^3\\[3pt]
	N^3 & \frac{N^4}{2}
	\end{pmatrix}, \qquad 
	\langle\mathcal{G}_5^{(N|i,j)}\rangle ~\longrightarrow~ \begin{pmatrix}
		\frac{2N^5}{5} & \frac{9N^4}{5}\\[3pt]
		\frac{9N^4}{5} & \frac{9N^5}{25}
		\end{pmatrix}.
\end{align}
Furthermore, note that the ensemble averages of correlators with two single-trace operators (i.e. the $(i,j)=(1,1)$ components) at large $N$ exhibit $p$-dependence given by 
\begin{align}\label{eq:ensemble_average_large_N}
	\langle\mathcal{G}_p^{(N|1,1)}\rangle~ \xrightarrow[]{~N\to\infty~} ~\frac{p-1}{2p}~N^p+\ldots\,,
\end{align}
For the correlators with one single-trace and one double-trace operator, the available data for the ensemble averages at large $N$ is consistent with the formula
\begin{align}\label{stdt}
	\langle\mathcal{G}_p^{(N|1,j)}\rangle~ \xrightarrow[]{~N\to\infty~} ~\frac{p_1p_2(p-2)}{2p}~N^{p-1}+\ldots\,,
\end{align}
where the label $j$ denotes a specific double-trace operator $T_{p_1,p_2}$ with $p_1+p_2=p$. 

\ssec*{Ensemble averages for SPO's}
To make contact with correlators describing scattering amplitudes in AdS$_5\times$S$^5$, let us also consider the ensemble averages of integrated correlators involving SPO's. For $p=2,3$ these are equal to the averages for single-trace operators already given in equations \eqref{eq:ensemble_average_2222} and \eqref{eq:ensemble_average_2233} above. On the other hand, for $p=4,5$ we find
\begin{align}\label{eq:ensemble_average_2244_2255_SPO}
\begin{split}
	\langle \mathcal{G}_4^{(N)} \rangle &= {\G(N)\o \G(N-3)}\,\(\frac{3 N^6+12 N^5+15 N^4+17 N^3+16 N^2-5 N-6}{8 N(N^2+1)^2}\),\\[3pt]
	\langle \mathcal{G}_5^{(N)} \rangle &= {\G(N)\o \G(N-4)}\,\(\frac{4 N^7+30 N^6+95 N^5+245 N^4+475 N^3+325 N^2-130 N-120}{10 N^2 (N^2+5)^2}\).
\end{split}
\end{align}
Due to the gamma factors, these have the expected property that they vanish for $N=1,2,\ldots,p-1$. While at first sight these expressions look rather messy, a surprising simplification occurs upon pulling out a factor of the SPO two-point function normalisation $R_p$, given in \eqr{eq:R_p}, and performing a decomposition into partial fractions:
\begin{align}\label{eq:ensemble_averages_partial_fractions}
\begin{split}
	\langle \mathcal{G}_2^{(N)} \rangle &= \frac{R_2}{2}\, \Big(1-\frac{1}{(N+1)}\Big)\,,\\
	\langle \mathcal{G}_3^{(N)} \rangle &= \frac{R_3}{2}\, \Big(2-\frac{1}{(N+1)}-\frac{2}{(N+2)}\Big)\,, \\
	\langle \mathcal{G}_4^{(N)} \rangle &= \frac{R_4}{2}\, \Big(3-\frac{1}{(N+1)}-\frac{2}{(N+2)}-\frac{3}{(N+3)}+\frac{(N+1)}{(N^2+1)}-\frac{1}{N}\Big)\,, \\
	\langle \mathcal{G}_5^{(N)} \rangle &= \frac{R_5}{2}\, \Big(4-\frac{1}{(N+1)}-\frac{2}{(N+2)}-\frac{3}{(N+3)}-\frac{4}{(N+4)}+\frac{(N+5)}{(N^2+5)}-\frac{1}{N}\Big)\,. 
\end{split}
\end{align}
Recalling that at large $N$ we have $R_p\sim N^p/p$, one recognises that the first term in the brackets is again consistent with the leading contribution \eqref{eq:ensemble_average_large_N}, which is then followed by a string of $1/N$ suppressed terms. Moreover, a clear pattern is visible in the above expressions, namely
\begin{align}\label{eq:ensemble_averages_guess}
	\langle \mathcal{G}_p^{(N)} \rangle = \frac{R_p}{2}\, \bigg((p-1)-\sum_{m=1}^{p-1}\frac{m}{(N+m)}+\frac{(N+x_p)}{(N^2+x_p)}-\frac{1}{N}\bigg)\,,
\end{align}
with the variable $x_p$ taking the values $x_p=0,0,1,5$ for $p=2,3,4,5$, respectively. Let us emphasise that the above pattern is based on data up to $p=5$ only. It would be interesting to consider higher values of $p$ to either confirm that this nice structure persists or to find a more general structure of $1/N$ corrections.

We will further analyze the large $N$ structure of $\langle\mathcal{G}_p^{(N|i,j)}\rangle$ in Section \ref{LargeNAverage}. 

\sec{Large $N$}
\label{sec:large_N}

In this section we study the large $N$ limit of the integrated correlators $\mathcal{G}_p^{(N)}(\tau)$. In order to make contact with the dual \adss supergravity, we restrict ourselves in this Section to correlators of single-particle operators (SPOs) $\O_p$. As such, we drop the $(i,j)$ labels.

All of the $N$-dependence of a given observable $\fo(\t)$ lies purely in its spectral overlaps, so the large $N$ expansion of the former follows from that of the latter. The large $N$ expansion of $\mathcal{G}_p^{(N)}(\tau)$, therefore, follows from that of $f_p^{(N)}(s)$, the Eisenstein overlaps. It was shown in \cite{Collier:2022emf} that the $1/N$ expansion takes the form\foot{The canonical development of the genus expansion of a general observable $\fo(\t)$ starts at order $N^2$. While a given observable may have vanishing contributions at low genus -- for example, a conformal dimension which starts at order $N^0$ -- it is nevertheless convenient to refer to its leading contribution as genus zero, instead adjusting the leading power of $N$ accordingly. Our conventions for $\cG_p^{(N)}(\t)$ are, for various reasons of clarity, unnormalized, which gives them a leading power $N^p$. Upon normalizing by the two-point functions of $\O_2$ and $\O_p$ using the colour factors $R_2$ and $R_p$ given in \eqr{eq:R_p}, their leading power would be $1/N^2$, which is the correct scaling of a normalized connected four-point function in a CFT obeying large $N$ factorization. \label{footy}}
\begin{align}\label{eq:f_p_large_N}
	f_p^{(N)}(s) = \sum_{g=0}^\infty N^{p-2g}\Big[N^{s-1}f_p^{(g)}(s)+(s \rar 1-s)\Big]\,.
\end{align}
Anticipating the genus expansion of the 't Hooft limit in \eqr{genusexp}, we call $f_p^{(g)}(s)$ the genus-$g$ spectral overlap. 

Developing the genus expansion of $f_p^{(N)}(s)$ is straightforward because of the recursion relations of Section \ref{sec:spectral_overlaps}. Solving the recursion in a $1/N$ expansion yields $f_p^{(g>0)}(s)$ in terms of $f_p^{(0)}(s)$. The latter may be computed in various ways: for example, by studying the leading-order $N$-dependence order-by-order in the $1/y$ weak-coupling expansion and reconstructing the $s$-dependence; or, by using the previously known results for $p=2$ to seed the recursion. 

In Subsection \ref{GenusExpfps} we present results for $f_p^{(g)}(s)$ through $g=2$, though it is straightforward to compute to arbitrary genus using the recursion formulas. Taking their $s\rar 1$ limit, it is then trivial to compute ensemble averages at large $N$ (cf. \eqr{aveis}) and make contact with \adss supergravity. We do this in Subsection \ref{LargeNAverage}. Finally, plugging in the large $N$ expansion \eqref{eq:f_p_large_N} of the spectral overlaps into the spectral decomposition \eqref{eq:spectral_decomp} allows us to make manifest two different large $N$ limits of $\cG_p^{(N)}(\t)$: the 't Hooft limit in Subsection  \ref{'tHooftLimit} and the very strongly coupled (VSC) limit in Subsection \ref{VSCLimit}.

\ssec{Genus expansion of spectral overlaps}
\label{GenusExpfps}
At genus zero, we find the following result for $p=2$
\begin{align}\label{eq:genus_0_p=2}
	f_2^{(0)}(s) = \frac{2^{2 s-1} (2 s-1) \Gamma \left(s+\frac{1}{2}\right)}{\sqrt{\pi } \Gamma (s+1) \Gamma (s+2)}\,,
\end{align}
which reproduces the result of \cite{Dorigoni:2021guq,Collier:2022emf}. The arbitrary $p$ generalization of this result is given by 
\begin{align}\label{eq:genus_0_overlaps}
{	f_p^{(0)}(s) = \frac{s+1}{2 s-1}\(1+(-1)^p\frac{\Gamma (s-1) \Gamma (s+1)}{\Gamma (s-p) \Gamma (s+p)}\)~f_2^{(0)}(s)\,.}
\end{align}
For a derivation of the equation we refer the reader to Appendix \ref{BesselSpectralEquivalence}. The presence of simple poles at $s=\frac12-m$ for $m\in \mathbb{Z}_+$ (in addition to zeros at $s\in \ZZ_-$) is consistent with the allowed general properties of spectral overlaps in the genus expansion.\foot{We recall from \cite{Collier:2022emf} that whereas the finite $N$ overlap $f_p^{(N)}(s)$ must be regular for all complex $s\neq 0,1$, the large $N$ expansion allows $f_p^{(g)}(s)$ to have poles at $s\in\Z/2$ at fixed genus $g$.} 

Let us make a curious observation. If we write \eqr{eq:genus_0_overlaps} in the form
\begin{align}
	f_p^{(0)}(s) = \frac{n^{(0)}_p(s)}{(s+2)_{p-2}}~f_2^{(0)}(s)\,,
\end{align}
we find that the numerator polynomials $n^{(0)}_p(s)$, of degree $2\lfloor {p-2\o 2}\rfloor$, obey a reflection symmetry for all $p\geq 2$: that is, 
\e{}{n_p^{(0)}(1-s) = n_p^{(0)}(s)\,.}
This symmetry is not required by $SL(2,\mathbb{Z})$-invariance of the spectral decomposition.

Having computed the leading order result, we then solve the recursion relations for the spectral overlaps for $p=2,3,4,5$ in a $1/N$ expansion and find that the higher-genus overlaps are related to the genus-zero $p=2$ overlap in the following simple manner:
\begin{align}\label{eq:higher_genus_overlaps}
	f_p^{(g)}(s) = \frac{n^{(g)}_p(s)}{(-2)^g\,(4g)!\,\left(\frac{3}{2}-s\right)_g (s+2)_{p-2}}~f_2^{(0)}(s)~.
\end{align}
Here we note that the higher-genus overlaps have additional poles from the $(\frac{3}{2}-s)_g$ factor, which includes a finite number of positive values, as opposed to genus zero (this point will be important later when we discuss the 't Hooft limit).  The polynomials $n^{(g)}_p(s)$ are of degree $2\lfloor\frac{p-2}{2}+2g\rfloor$; unlike the genus-zero case, the $n^{(g>0)}_p(s)$ are not reflection-symmetric. To give some explicit examples, the polynomials $n_p^{(1)}(s)$ of the genus-one overlaps read
\begin{align}\label{eq:genus_1_data}
\begin{split}
	n_2^{(1)}(s) &= (s-6) (s-1) s (s+1)\,,\\
	n_3^{(1)}(s) &= 4(s+2) (s^3-2 s^2-51 s+72)\,,\\
	n_4^{(1)}(s) &= (s+3) (s^5+2 s^4-181 s^3+622 s^2-3180 s+3816)\,,\\
	n_5^{(1)}(s) &= 12(s+4) (s^5+11 s^4-387 s^3+1125 s^2-3270 s+3528)\,,
\end{split}
\end{align}
whereas at genus two we find
\begin{align}\label{eq:genus_2_data}
\begin{split}
	n_2^{(2)}(s) &= 7(s-4) (s-3) (s-2) (s-1) s (s+1) (5 s^2-47 s+30)\,,\\
	n_3^{(2)}(s) &= 28(s-1) s (s+1) (s+2) (5 s^4-42 s^3-473 s^2+4374 s-7560)\,,\\
	n_4^{(2)}(s) &= 7(s+3) (5 s^9+8 s^8-2084 s^7+8138 s^6+62635 s^5\\
		&\qquad\qquad\quad -13978 s^4-601996 s^3+2937672 s^2-9717120 s+8985600)\,,\\
	n_5^{(2)}(s) &= 84 (s+4) (5 s^9+103 s^8-4022 s^7-16146 s^6+485093 s^5\\
		&\qquad\qquad\quad~ +40775 s^4-7260020 s^3+24585444 s^2-47580480 s+35942400)\,.
\end{split}
\end{align}
Let us point out the occurrence of certain zeroes in these polynomials, which have non-trivial consequences for the large $N$ expansion of the ensemble averages discussed in the next section. In particular, note that for $p=2$ the spectral overlap $f_p^{(g)}(s)$ vanishes at $s=0,1$ for $g\geq1$, while for $p=3$ the same is the case for $g\geq2$. On the other hand, the $p=4,5$ overlaps do not vanish at $s=0,1$, and the reason for this will be explained next.

\ssec{Ensemble averages at large $N$ and supergravity}
\label{LargeNAverage}

We now make contact with semiclassical \adss supergravity. 

It was shown in \cite{Collier:2022emf} that for any $SL(2,\Z)$-invariant observable $\fo(\t)$ that admits a genus expansion in the `t Hooft limit, the ensemble average $\<\fo\>$ and the large `t Hooft coupling limit of $\fo(\t)$ are equal at leading order in large $N$:
\e{infavg2}{\langle \fo \rangle = \fo(\l\rar\i)\,.}
This correspondence extends to all genera: defining the leading-order term at genus-$g$ as
\es{}{\llangle\fo^{(g)}\rrangle &\coloneqq  \lim_{N\rar\i} N^{2g-2} \<\fo^{(g)}\>\,,}
there is an equivalence
\e{infavg2}{\llangle\fo^{(g)}\rrangle = \fo^{(g)}(\l\rar\i)~~ \forall~g\,.}
Translated into bulk terms, the genus-zero relation implies an equivalence between the large $N$ ensemble average of $\fo(\t)$ and its value in tree-level \adss supergravity, while the $g>0$ relation relates the higher-genera averages to loop-level \adss supergravity after string theory regularization of divergences.\foot{We note that one can perform a mutual consistency check of this dictionary and our results: namely, matching the large $N$ expansion of the finite $N$ ensemble averages $\langle\mathcal{G}_p^{(N)}\rangle$ to the genus expansion of the spectral overlaps $f_p(s)$. We do so in Appendix \ref{appcheck}.} (The $g>0$ statement of \eqr{infavg2} will be elaborated upon below.)

At $g=0$, our results confirm this equivalence. As foreshadowed in \eqref{eq:ensemble_average_large_N}, we find
\e{gpg0}{\llangle\cG_p^{(g=0)}\rrangle=\frac{p-1}{2p}~.}
After accounting for differences in normalisations, this agrees with the $\l\rar\i$ limit at $g=0$ \cite{Binder:2019jwn}. 

At $g=1$, the results \eqref{eq:ensemble_averages_partial_fractions} for $p\leq 5$ are consistent with the following pattern:
\e{}{\llangle\cG_p^{(g=1)}\rrangle=-\frac{(p^2-1)(p^2-4)}{48}~.}
Via \eqr{infavg2}, this makes a prediction for the finite term in the $\l\rar\i$ limit at genus one. This amounts to a holographic computation of the 1-loop supergravity computation of $\cG_p^{(N)}(\t)$. To put this in conventional supergravity language, we note that the bulk loop expansion proceeds not in powers of $1/N$ but in powers of $1/c=4/(N^2-1)$. In addition, let us divide by the product of colour factors $R_2 R_p$ to normalize the correlator,\foot{In this convention, the tree-level supergravity result is $\widetilde\cG^{(N)}_{p}(\t)\approx (p-1)/4c$.} 
\e{}{\widetilde \cG_p^{(N)}(\t) := \cG_p^{(N)}(\t)(R_2 R_p)^{-1} \,.}
Putting things together leads to
\e{int1loop}{\widetilde\cG^{(N)}_{p}(\t)\Big|_{\text{1-loop sugra}}={1\o c^2}\frac{(p-1)p(p+1)(p+2)}{384}~.}
This sets a benchmark for a direct bulk computation of the integrated correlators in regularized 1-loop \adss supergravity.\foot{The results of \c{Aprile:2019rep} imply that \eqr{int1loop}, in conjunction with the results of \cite{Aprile:2017qoy,Alday:2019nin} and the flat space limit, can be used to determine the {\it unintegrated} 1-loop $\< \O_2\O_2\O_p\O_p\>$ correlator, as anticipated in \cite{Chester:2019pvm}.\label{f22}} We reiterate that this is the {\it finite} result that remains after the (unambiguous) string theory regularization of 1-loop supergravity divergences. It is notable that the ensemble average is sensitive to the UV details of this regularization; in particular, string theory regularizes divergences unambiguously, choosing a specific renormalization scheme, and the $\sl$ average identifies precisely this scheme.

Finally, let us make the side remark that \eqr{stdt} gives a prediction for tree-level supergravity, this time for the integrated correlator involving one single-trace operator $T_p$ and one double-trace operator $T_{p_1,p_2}$ with $p_1+p_2=p$.

\ssec{Integrated correlators at strong coupling}\label{sec:strong_coupling_expansions}

As emphasized throughout this section, the spectral overlaps in the $1/N$ expansion contain complete information about the large $N$ integrated correlators themselves: simply plug the formulas for the overlaps $f_p^{(g)}(s)$ into the spectral representation \eqref{eq:spectral_decomp}, and expand in the desired limit. 

Having said that, for convenience of future study, we assemble these ingredients into explicit expressions for $\cG_p^{(N)}(\t)$ in two strongly coupled limits of interest: the `t Hooft limit, and the very strongly coupled (VSC) limit. We will briefly review salient aspects of these limits in the spectral language; a systematic treatment of these limits for general $\sl$-invariant observables can be found in \cite{Collier:2022emf}.

\sssec{'t Hooft limit}
\label{'tHooftLimit}

In the 't Hooft limit, where $g_{\text{YM}}\to 0$ and $N\to \infty$ with $\lambda:=g^2_{\text{YM}}N$ fixed, the $1/N$ expansion organises into a genus expansion, up to non-perturbative corrections in $1/N$. Because instantons and anti-instantons are non-perturbatively suppressed in $1/N$, the genus expansion is an expansion of the zero-instanton mode:
\es{genusexp}{
	\mathcal{G}_{p}^{(N)}(\tau)= \sum_{\mathfrak{g}=0}^\infty N^{p-2\mathfrak{g}}\,\mathcal{G}_{p,0}^{(\mathfrak{g})}(\lambda) + (\text{non-perturbative})~.}
We henceforth drop the ``0'' subscript with the understanding that we are working perturbatively in $1/N$. The parameter $\mathfrak{g}$ differs from the previous definition of genus in \eqref{eq:f_p_large_N} because of a non-trivial repackaging of the $1/N$ expansion by the $\sl$ spectral decomposition, a fact we return to momentarily.

We want to perform a spectral decomposition of this quantity. The zero mode of completed Eisenstein series may be written in terms of a `t Hooft coupling as
\es{t'HooftEisenstein}{E^*_{s,0}(y) = \Lambda(s) N^s\tilde\lambda ^{-s}+ \Lambda(1-s) N^{1-s}\tilde\lambda ^{s-1}~,}
where 
\e{}{\tilde{\lambda}\coloneqq {\l\o4\pi}~.}
Upon inserting \eqref{t'HooftEisenstein} and the large $N$ expansion of the perturbative overlaps %
\eqref{eq:f_p_large_N} into the spectral decomposition \eqref{eq:spectral_decomp}, one finds
\begin{align}\label{eq:spectral_decomp_large_N}
	\mathcal{G}_{p}^{(N)}(\lambda) =&~ \langle \mathcal{G}_p^{(N)} \rangle\\
	 +& \frac{1}{2\pi i}\int_{\text{Re}\,s=\frac{1}{2}}ds \frac{\pi}{\sin(\pi s)}s(1-s)\sum_{g=0}^\infty N^{p-2g}\Big[\Lambda(1-s)\tilde{\lambda}^{s-1}+\Lambda(s)N^{2s-1}\tilde{\lambda}^{-s}\Big]f_p^{(g)}(s)~,\no
\end{align}
As the constant term in the spectral decomposition, the ensemble average $\langle \mathcal{G}_p^{(N)} \rangle$ does not depend on $\lambda$. There are two terms in brackets, each of a different nature. Large $N$ forces us to close the contour of the second term in brackets, carrying an $N^{2s-1}$, to the left; this generates terms with positive powers of $\lambda$, regardless of the value of $\l$. These terms were denoted `renormalization terms' in \cite{Collier:2022emf}, due to their strong coupling interpretation as  bulk UV divergences regularized by the string scale cutoff.\foot{At strong coupling, these terms lead to super-leading powers of $\l$ with respect to the supergravity term; as argued in \cite{Aharony:2016dwx}, they can be seen as finite string theory counterterms which regulate the UV divergences of the \adss supergravity loop expansion. Their presence is in fact crucial to restoring the $SL(2,\Z)$ invariance (S-duality) of the strong coupling expansion. At weak coupling, their cancellation -- required so that the perturbative expansion proceeds only in integer powers of $\l$ --  implies intriguing inter-genera relations among the residues, see eq. (7.9) of \cite{Collier:2022emf}. (See also \c{Hatsuda:2022enx}.)} The integration contour of the first term in brackets may be closed either to the left to develop the strong coupling expansion (in negative powers of $\tilde{\lambda}$); to the right to develop the weak-coupling expansion (in positive powers of $\tilde{\lambda}$); or not deformed at all, thus giving an expression which is exact in $\l$. 

For example, at $\mathfrak{g}=0,1$ one finds
\begin{align}
\label{eq:genus_0_correlator}
	\mathcal{G}_{p}^{(\mathfrak{g}=0)}(\lambda) =&~ \frac{p-1}{2p}+\frac{1}{2\pi i}\int_{\text{Re}\,s=\frac{1}{2}} ds~\frac{\pi}{\sin(\pi s)}\,s(1-s)\,\Lambda(1-s)\,\tl^{s-1}\,f_p^{(0)}(s)~,\\[5pt]
\label{eq:genus_1_correlator}
	\mathcal{G}_{p}^{(\mathfrak{g}=1)}(\lambda) =&-\frac{p^2-1}{48}\sqrt{\l}-\frac{(p^2-1)(p^2-4)}{48}\no\\
&+\frac{1}{2\pi i}\int_{\text{Re}\,s=\frac{1}{2}} ds~\frac{\pi}{\sin(\pi s)}\,s(1-s)\,\Lambda(1-s)\,\tl^{s-1}\,f_p^{(1)}(s)~.
\end{align}
where $f_p^{(0)}(s)$ and $f_p^{(1)}(s)$ were given in \eqr{eq:genus_0_overlaps} and \eqr{eq:higher_genus_overlaps}, respectively. The first term in \eqref{eq:genus_1_correlator} is the `renormalization term' discussed above, and descends from the second term in brackets in \eqref{eq:spectral_decomp_large_N} for $g=0$. At risk of repetition, we stress anew that this spectral representation is {\it exact} in $\lambda$: no expansion has been made. 

$\cG_p^{(N)}(\t)$ contains non-perturbative effects at both large $\l$ and large $N$. This follows from the analysis in \cite{Collier:2022emf}, where they were determined for $p=2$, and the fact that the $p$-dependence of $f_p^{(g)}(s)$ in \eqr{eq:genus_0_overlaps} and \eqr{eq:higher_genus_overlaps} does not enter the large-order asymptotics of the relevant series (for all genera). The effects at large $\l$ and large $N$ will therefore come in positive even powers of the scales \cite{Collier:2022emf}
\e{npscales}{\Lambda_{\l} \coloneqq  \exp\(-\sqrt{\l}\)\qquad \text{and} \qquad \Lambda_{\ls} \coloneqq  \exp\(-\sqrt{\ls}\)}
respectively, where $\ls := {16\pi^2 N^2/\l}$ is an S-dual `t Hooft coupling. Both corrections are present in the `t Hooft limit. The $p$-dependence of the non-perturbative series, entering only in the expansion coefficients, would be nice to examine in future work.

It is formally interesting to derive two alternative presentations of these same objects $\mathcal{G}_{p}^{(\mathfrak{g})}(\lambda)$. 

First, it is possible to show (see Appendix \ref{App:F1}) that the genus-$\mathfrak{g}$ correlator can be written entirely in terms of genus-$\mathfrak{g}$ spectral overlaps as follows:
\begin{align}
\label{eq:genus_g_correlator1}
\mathcal{G}_{p}^{(\mathfrak{g})}(\lambda)= &~\frac{1}{2\pi i}\int_{\text{Re}\,s=\mg+\frac12+\epsilon}ds \frac{\pi}{\sin(\pi s)}s(1-s)\Lambda(1-s)\tilde{\lambda}^{s-1}f_p^{(\mg)}(s) \no \\[5pt]
	 &- \sum_{k=2}^{\mg}(-1)^{k+1}\tilde{\lambda}^{k-1}k(k-1) \Lambda(1-k) f_{{p}}^{(\mg)}(k)
\end{align}
This rewriting involves shifting the contour of \eqr{eq:spectral_decomp_large_N} and making use of identities relating different genera -- namely, eq. (7.9) of \cite{Collier:2022emf} -- that are required by consistency of the weak-coupling expansion in integer powers of $\l$. The novelty of this expression is that the integrand involves only the genus-$\mathfrak{g}$ overlap, unlike what \eqr{eq:spectral_decomp_large_N} seems to give. Note the absence of the second line for $\mg<2$. 

Secondly, in \c{Binder:2019jwn} the genus-zero result $\mathcal{G}_{p}^{(\mathfrak{g}=0)}(\lambda)$ was derived for all $\l$ as an integral expression involving squared Bessel functions. In Appendix \ref{BesselSpectralEquivalence}, we prove its equivalence to \eqref{eq:genus_0_correlator}.  Indeed, a ``Bessel representation'' at all genera maybe derived by deforming the contour in \eqref{eq:genus_g_correlator1} to the right, yielding
\begin{align}
\label{eq:genus_g_correlator2}
\mathcal{G}_{p}^{(\mathfrak{g})}(\lambda)=- \sum_{k=2}^{\infty}(-1)^{k+1}\tilde{\lambda}^{k-1}k(k-1) \Lambda(1-k) f_{{p}}^{(\mg)}(k)\,.
\end{align}
This is convergent for $|\lambda|\leq\pi^2$ for all finite $\mg$ and $p$. The expansion may be resummed using the integral identity
\be
\label{ZetaIdentity}
\frac{2^{2 s-2}}{\Gamma(2 s)} \int_{0}^{\infty} d w \frac{w^{2 s-1}}{\sinh ^{2}(w)}=\zeta(2 s-1)\,.
\ee
The resulting expression is a one-dimensional integral of products of Bessel functions. For example, we present such a result for $\mathcal{G}_{3}^{(\mathfrak{g}=1)}(\lambda)$:
\begin{align}
\label{p3g1overlap}
\mathcal{G}_{3}^{(\mg=1)}(\lambda)=&-\int_{0}^{\infty} \frac{d \omega}{12 \pi ^2 \lambda  \omega \sinh^2\omega} \bigg[ \left(7 \lambda ^2 \omega ^4+48 \pi ^2 \lambda  \omega ^2-192 \pi ^4\right) J_1(\sqrt{\lambda } \omega /\pi )^2\\
+&\lambda  \omega ^2 \left(5 \lambda  \omega ^2-48 \pi ^2\right) J_0(\sqrt{\lambda } \omega /\pi )^2+4 \pi  \sqrt{\lambda } \omega  \left(48 \pi ^2-\lambda  \omega ^2\right) J_1(\sqrt{\lambda } \omega /\pi ) J_0(\sqrt{\lambda } \omega /\pi ) \bigg].\no
\end{align}
For the sake of completeness, we provide the results at genus one for $p=2,4,5$ in Appendix \ref{BesselSpectralEquivalence}. Similar expressions for $\cG_p^{(\mg)}(\lambda)$ for various $\mg$ and $p$ are easily obtained from \eqref{eq:genus_g_correlator2}.

\sssec{Very strongly coupled limit}
\label{VSCLimit}

The VSC limit is the limit of $N\rar\i$ with $\t$ fixed. In contrast to the 't Hooft limit, instanton contributions are not suppressed. This limit is trivially extracted from the spectral decomposition \eqref{eq:spectral_decomp} upon inserting the large $N$ expansion of the overlaps \eqref{eq:f_p_large_N} into \eqref{eq:spectral_decomp}, and deforming the integration contour to develop the $1/N$ expansion. 
Summing over residues in this way gives \cite{Collier:2022emf}
\begin{equation}
\label{VSCDecomp2}
\mathcal{G}_p^{(N)}(\tau)=\sum_{g=0}^{\infty} N^{p-2g}\(\llangle \mathcal{G}_p^{(g)} \rrangle -\sum_{m=1}^{\infty} N^{-m-\frac{1}{2}} \mathsf{R}_{m}^{(g)} E_{\frac{1}{2}+m}(\tau)\),
\end{equation}
with residue function $\mathsf{R}_{m}^{(g)}$ defined as
\begin{equation}
\mathsf{R}_{m}^{(g)}:=\underset{s=\frac{1}{2}+m}{\operatorname{Res}}\left[\frac{\pi}{\sin \pi s} s(1-s) \Lambda(s) f_{{p}}^{(g)}(1-s)\right].
\end{equation}
The $\t$-dependence appears only via Eisenstein series, and only down by odd-half-integer powers of $1/N$.\foot{There can be terms down by powers of $1/N^2$ as well, due to the ensemble average terms $\llangle \mathcal{G}_p^{(g)} \rrangle$, but these are $\t$-independent.} Both of these features, like many others in this work, follow from the relations $\fnp^{(N)}(s) = (\cG_p^{(N)},\phi_n)=0$.

For example, applying the above formulae to the $p=3$ case using the genus expansion  \eqref{eq:higher_genus_overlaps} of its spectral overlap through $g=2$ yields the following VSC expansion:
\begin{align}
\label{VSCp3}
&\cG_3^{(N)}(\tau)=\frac{N^3}{3}-\frac{1}{2} N^{\frac32} \widetilde{E}_{\frac{3}{2}}(\tau)-\frac{5N}{6}+\frac{45}{2^5}N^{\frac12}\widetilde{E}_{\frac{5}{2}}(\tau)+25N^{-\frac12} \left(\frac{31 \widetilde{E}_{\frac{3}{2}}(\tau)}{2^{10}}-\frac{189 \widetilde{E}_{\frac{7}{2}}(\tau)}{2^{12}}\right) \no\\
-&225 N^{-\frac32}\left(\frac{25 \widetilde{E}_{\frac{5}{2}}(\tau)}{2^{13}}+\frac{147 \widetilde{E}_{\frac{9}{2}}(\tau)}{2^{15}}\right)+N^{-\frac52}\left(\frac{10517 \widetilde{E}_{\frac{3}{2}}(\tau)}{2^{19}}-\frac{2159325 \widetilde{E}_{\frac{7}{2}}(\tau)}{2^{22}}-\frac{49116375 \widetilde{E}_{\frac{11}{2}}(\tau)}{2^{24}}\right)\no\\
+&45 N^{-\frac72}\left(\frac{26699 \widetilde{E}_{\frac{5}{2}}(\tau)}{2^{24}}-\frac{215355 \widetilde{E}_{\frac{9}{2}}(\tau)}{2^{23}}-\frac{93648555 \widetilde{E}_{\frac{13}{2}}(\tau)}{2^{28}}\right)+O(N^{-\frac{9}{2}})\,,
\end{align}
where 
\begin{align}
	\widetilde{E}_{s}(\tau):=\frac{2 \Lambda(s)}{\Gamma(s)}\, E_{s}(\tau)\,.
\end{align}

\sec{Results II: Integrated maximal-trace correlators for all $p$}\label{sec:maximal_trace}

So far, we have mostly restricted ourselves to integrated correlators with $p\leq5$. Recall that our approach was to obtain the weak-coupling expansion of $\cG_{p}^{(N|i,j)}(\tau)$ using the localisation relation \eqref{eq:localisation_relation}, from which we then inferred the spectral overlaps. When trying to generalise the localisation computation to higher values of $p$ for most choices of $(i,j)$, one is met with some computational expense, to be elaborated upon in Section \ref{sec:genpansatz}. 

However, there exists a special class of integrated correlators, defined for arbitrary $p$, which we now determine exactly: the so-called {\it maximal-trace correlators}. 

The key point is that, as shown in \cite{Gerchkovitz:2016gxx}, one can recursively construct a basis of operators on the sphere such that certain families of operators become orthogonal to each other. The simplest such family of operators with this property is given by maximal-trace operators, introduced earlier in \eqr{opmax2}, which for even $p$ are given by
\e{Omax}{\Omax\equiv T_{2,\ldots,2}\,.}
They mix only among themselves, leading to an effective reduction in the size of mixing matrices.\footnote{	The same reasoning applies also to the odd $p$ family of maximal-trace operators, $\{T_3,\,T_{2,3},\,T_{2,2,3},\,T_{2,2,2,3},\,\ldots\}$, but in what follows we restrict ourselves to the even $p$ case.} The solution to the sphere mixing is given in terms of the vectors $v_p^j$, recall the discussion around \eqref{muindex}. For the first few cases of $p$ we find
\begin{align}
\begin{split}
	v_2^j &= \Big\{1\quad 0\quad 0\quad 0\quad 0~~\cdots\Big\}\,,\\
	v_4^j &= \Big\{-i\tfrac{N^2+1}{2y}\quad 1\quad 0\quad 0\quad 0~~\cdots\Big\}\,,\\
	v_6^j &= \Big\{-\tfrac{3(N^2+1)(N^2+3)}{16y}\quad -i\tfrac{3(N^2+3)}{4y}\quad 1\quad 0\quad 0~~\cdots\Big\}\,.
\end{split}
\end{align}
The above vectors admit the closed form expression
\begin{align}
	v_p^j = \binom{\frac{p}{2}}{j}\,\frac{\Big(\frac{N^2-1}{2}+j\Big){}_{\frac{p}{2}-j}}{(2i y)^{\frac{p}{2}-j}}\,,
\end{align}
where the label $j=1,2,3\ldots$ now runs only over the subset of maximal-trace operators $\Omax\in\{T_2,\,T_{2,2},\,T_{2,2,2},\ldots\}$ due to decoupling from all other operators on the sphere.

The other simplifying feature is that, since $\Omax$ is built from products of $T_2$ only, one only needs to act with repeated $\partial_{\tau}$ and $\partial_{\bar{\tau}}$ derivatives in order to bring down insertions of these operators in the localisation computation. As such, no new matrix integrals have to be performed when considering correlators of these maximal-trace operators. 

In what follows, we will denote the integrated correlators $\langle\O_2\O_2\Omax\Omax\rangle$ for $p$ even by $\cG_p^{(N|\text{max})}(\tau)$:
\begin{align}\label{eq:integrated_correlator_max}
	\mathcal{G}_p^{(N|\text{max})}(\tau) :=-\frac{2}{\pi}\int_0^\infty dr\int_0^\pi d\theta\,\frac{r^3\sin^2\theta}{u^2}\,\mathcal{H}^{(N|\text{max})}_p(u,v;\tau)\Big\vert_{u=1+r^2-2r\cos\theta,\,v=r^2}\,.
\end{align}
following the notation of \eqr{eq:superconformal_constraint} and \eqr{eq:integrated_correlator}.\footnote{	Note that for $p=2,4$, $\cG_p^{(N|\text{max})}(\tau)$ reduces to the previously discussed cases $\cG_2^{(N)}(\tau)$ and $\cG_4^{(N|2,2)}(\tau)$, respectively.} 
The corresponding colour-factor appearing in the localisation relation \eqref{eq:localisation_relation} is recorded in \eqr{rpmax}.
We also introduce a hatted notation to denote normalization by $R_p^{(\text{max})}(N)$, e.g.
\e{gHatDef}{\widehat\cG_p^{(N|\text{max})}(\t) := \cG_p^{(N|\text{max})}(\t) \,R_p^{(\text{max})}(N)^{-1}\,,}
and likewise for other quantities.

Let us present, then elaborate upon, the result:
\e{maxresult}{{\widehat\cG_p^{(N|\text{max})}(\t) = \< \widehat\cG_p^{(N|\text{max})}\> + {1\o 4\pi i} \int_{\Re s =\half} ds \,{\pi\o \sin(\pi s)}s(1-s)(2s-1)^2\,\widehat g_p^{(N|{\rm max})}(s)\, E^*_s(\t)}}
The ensemble average is
\e{maxavg}{\boxed{\langle\widehat\cG_p^{(N|\text{max})}\rangle = \<\cG_2^{(N)}\>\, \Big(H_{\frac{N^2+p-3}{2}}-H_{\frac{N^2-3}{2}}\Big)}}
where $H_n$ denotes the harmonic number, and the overlaps obey the polynomial recursion
\begin{empheq}[box=\fbox]{equation}\label{maxrecursion}
\begin{aligned}
p (N^2+p-3)\,\widehat g_p^{(N|\text{max})}(s) &= 2\Big[\big((p-1)\,N^2 +p^2-5 p+5\big)-2\,s(1-s)\Big]\, \widehat g_{p-2}^{(N|\text{max})}(s) \\&~~~~-(p-2) (N^2+p-5)\, \widehat g_{p-4}^{(N|\text{max})}(s) +4g_{2}^{(N)}(s)\,,
\end{aligned}
\end{empheq}
Recall that $\<\cG_2^{(N)}\> = N(N-1)/4$.

We first address the spectral overlaps. These are deduced as usual from the weak-coupling expansion obtained from localisation.\footnote{
	As mentioned earlier, one can verify that the weak-coupling expansion agrees with the explicit results from perturbation theory upon integration. For the maximal-trace correlators, we have performed this check to two-loop order and up to $p=6$, see Appendix \ref{app:2LoopPerturbation}.
	}
The usual function $f_p^{(N|\text{max})}(s)$ parameterizing the Eisenstein overlap \`a la \eqref{EisensteinOverlap} is found to be of the now-familiar form
\begin{align}
	f_p^{(N|\text{max})}(s) = (2s-1)^2\,g_p^{(N|\text{max})}(s)\,,\quad \text{where}\quad \text{deg}\[g_p^{(N|\text{max})}(s)\] = 2N+p-6
\end{align}
where the polynomials $g_p^{(N|\text{max})}(s)$ are even and of indicated degree. Considering many cases of different $N$ and $p$ leads to \eqr{maxrecursion}.\foot{As we have observed in previous cases, for low values of $p$ the above formula neatly degenerates in a consistent way. This can be most easily seen when considering the recursion relation for the unnormalised overlaps $g_p^{(N|\text{max})}(s)$, i.e. after reinstating the normalisation factors $R_p^{(max)}(N)$ in the recursion \eqref{maxrecursion}: for $p=2$, the equation is trivially satisfied. For $p=4$, the first term in the second line vanishes, as it should since $g_{0}^{(N|\text{max})}(s)=0$: $p=0$ corresponds to an external identity operator, for which the connected correlator vanishes. Moreover, the other coefficients combine to correctly recover relation \eqref{eq:recursion_2244|2,2}. Finally, for $p=6$, the two terms in the second line of \eqref{maxrecursion} degenerate and in particular their coefficients combine to give the correct recursion relation consistent with the data.} The central property here is that there are no shifts in $N$: in contrast to the previously-studied cases, this is a {\it recursion relation in $p$ alone}, with $g_2^{(N)}(s)$ as the initial condition (together with the trivial condition $g_0^{(N|\text{max})}(s)=0$). Indeed, the solution of the recursion relation for general $p$ and finite $N$ is simply {\it proportional} to the $p=2$ solution:
\begin{align}\label{eq:maxtrace_solution}
	{\widehat g_p^{(N|\text{max})}(s) = F_p(N,s)\,\widehat g_2^{(N)}(s)\,.}
\end{align}
The functions $F_p(N,s)$ have the structure
\begin{align}\label{eq:Fp_structure}
	F_p(N,s)=\frac{2^{-\frac{p}{2}}}{\Big(\frac{N^2+1}{2} \Big){}_{\frac{p}{2}-1}}\,\sum _{m=0}^{\infty} \frac{2^{m+1}}{\Gamma (m+2)^2}\,\Big(\frac{p-2 m}{2}\Big)_{m+1}\,{h}_p^{(m)}(s)\,N^{p-2-2m}\,,
\end{align}
where the complete $s$-dependence is captured by the $h_p^{(m)}(s)$, which turn out to be polynomials of order $2m$. Since $p\in 2\Z_+$, the infinite sum over $m$ truncates at $m=p/2$. For some low values of $m$, the polynomials $h_p^{(m)}(s)$ read
\begin{align}\label{eq:h_p_data}
\begin{split}
	h_p^{(0)}(s) &= 1\,,\\
	h_p^{(1)}(s) &=s^2-s+p-4\,,\\
	h_p^{(2)}(s) &= s^4-2 s^3+\tfrac{1}{4} (9 p-28) s^2-\tfrac{1}{4}(9p-32)s+\tfrac{3}{8}(3 p^2-26 p+44)\,,\\
	h_p^{(3)}(s) &= s^6-3 s^5+(4p-9) s^4-(8p-23) s^3+\tfrac{1}{2}(9 p^2-62 p+76) s^2\\
	&\quad-\tfrac{1}{2}(9p^2-70p+100) s+\tfrac{3}{2} (p-4)(p^2-10 p+12)\,.
\end{split}
\end{align}
More data can be generated straightforwardly from the recursion relation \eqref{maxrecursion}. The reflection symmetry of $\widehat g_p^{(N|\text{max})}(s)$ implies that $F_p(N,s)$ and the 
individual polynomials $h_p^{(m)}(s)$ also have this symmetry:
\begin{align}\label{eq:Fp_reflection}
	F_p(N,s) = F_p(N,1-s)\,, \qquad h_p^{(m)}(s) = h_p^{(m)}(1-s)\,.
\end{align}

As is evident from \eqr{maxresult}, the cusp form overlap vanishes:
\e{maxnocusp}{(\cG_p^{(N|\text{max})},\phi_n)=0\quad \forall~n>0\,.}
This assertion follows from the computations discussed in Section \ref{IncludingInstantons}: for many values of $p$, we have directly checked to the first few perturbative orders in $1/y$ around one instanton that \eqr{maxresult} agrees with the first-principles localisation computation.

Finally, we turn to the ensemble averages $\langle\widehat\cG_p^{(N|\text{max})}\rangle$. For general $p$ we have found an explicit formula \eqr{maxavg}, not just a recursion relation. Since $p\in2\Z_+$, an equivalent representation of \eqref{maxavg} for the ensemble averages is given by\foot{It may be worth noting that the unnormalised average, $\langle\mathcal{G}_p^{(N|\text{max})}\rangle$, is actually a polynomial in $N$ of total degree $p$. This follows from the explicit formula \eqref{rpmax} for $R_p^{(\text{max})}(N)$, whereupon the Pochhammer factor and the sum combine to give rise to non-trivial polynomials in $N$ of degree $p-2$ containing only even powers of $N$.} 
	\begin{align}\label{normavgmaxtrace}
		\langle \widehat \cG_p^{(N|\text{max})}\rangle &= {N(N-1)\o 2}~\sum_{m=0}^{\frac{p-2}{2}}\frac{1}{N^2+2m-1}\,.
	\end{align}
	At leading order in large $N$ (with finite $p$), 
\e{eq:large_N_average_maxtrace}{\langle\widehat\cG_p^{(N|\text{max})}\rangle ~\xrightarrow[]{~N\rightarrow\infty~}~{p\o 4}\,.}
For $p=2$ and $p=4$, we recover equations \eqref{eq:ensemble_average_large_N} and the lower-right entry of \eqr{g4g5}, respectively, after multiplying by $R_p^{(\rm max)}(N)$ at large $N$.\foot{We also observe that the linear scaling in $p$ matches the large $p$ limit of the classical, {\it un}integrated $\< pppp \>$ correlator studied in \cite{Aprile:2020luw}, a different, but related, observable.}

\ssec{Comments and connections to $\cN=2$ extremal correlators}
\label{sec:6.1}

The maximal-trace family of integrated correlators, $\widehat\cG_p^{(N|\text{max})}(\t)$, has been determined above for all $N$ and arbitrary $p$. Compared to the general trace structures studied earlier in this work, its solution is especially clean. It also has interesting connections to various other quantities studied in the literature. 

First, note that $\cG_p^{(N|\text{max})}(\t)$ may be thought of as a generalization of the so-called ``extremal correlators'' studied in 4d $\cN=2$ SCFTs. Let us summarize the latter. Extremal correlators involve several chiral operators $O_m$ and a single anti-chiral operator $\overline{O}_n$. The most well-studied extremal correlators are two-point functions, 
\e{extcorr}{ G_{2n}(\t) := \<O_n(0)\overline{O}_n(\i)\>_{\R^4}}
These depend non-holomorphically on the complexified gauge coupling $\t$, and may be determined by localisation. In an $\cN=2$ SCFT of arbitrary rank, one may take $O_n$ to be $n$-fold composites of the weight-two chiral primary, what we called $\O_2$. Such an $O_n$ is precisely our family $\O_p^{(\rm max)}$ defined in \eqref{Omax}, with the identification $2n=p$. The correlators $G_{2n}(\t)$ solve differential recursion relations in $n$ \cite{Papadodimas:2009eu, Baggio:2014ioa}. In rank-one $\cN=2$ SCFTs, such as $SU(2)$ SQCD with four flavors, the $G_{2n}(\t)$ obey semi-infinite Toda equations relating different values of $n$:\foot{In the special case of $\cN=4$ SYM, this may be solved for $G_{2n}(\t)$ \cite{Gerchkovitz:2016gxx}, and matches the color factor $R_p^{(\rm max)}(N)$ defined in \eqr{rpmax}, up to a $p^2$ factor due to normalization differences (see footnote \ref{foot4}). In this case, $G_{2n}(\t) \propto y^{-2n}$, so the RHS is manifestly $\t$-independent.}
\e{toda}{\D_\t \log G_{2n}(\t) = -4y^2\({G_{2n+2}(\t)\o G_{2n}(\t)}-{G_{2n}(\t)\o G_{2n-2}(\t)}-{G_{2}(\t)}\)}
At higher rank, differential relations are believed to exist, but to take a more involved form that couples different families of operators besides merely composites of $\O_2$ \c{Gerchkovitz:2016gxx}. 

We can now state the connections between $G_{2n}(\t)$ and $\cG_p^{(N|\text{max})}(\t)$ with $2n=p$. Both are correlators involving the composite operators $\O_p^{(\rm max)}$. While the former are two-point functions, the latter are {\it four}-point functions, but integrated over positions, leaving only $\t$-dependence. Strikingly, both obey a similar form of differential recursion: as a consequence of \eqr{maxnocusp}, the recursion \eqr{maxrecursion} uplifts to a differential relation for $\widehat\cG_p^{(N|\text{max})}(\t)$,
\es{maxcorrelator}{\,\Delta_\tau\, \widehat \cG_{p-2}^{(N|\text{max})}(\tau) =&-\frac{p}{4} (N^2+p-3)\,\widehat \cG_p^{(N|\text{max})}(\tau)+ \frac12\big((p-1)\,N^2 +p^2-5 p+5\big)\widehat \cG_{p-2}^{(N|\text{max})}(\tau)  \\&-\frac14(p-2) (N^2+p-5)\, \widehat \cG_{p-4}^{(N|\text{max})}(\tau) + \cG_{2}^{(N)}(\tau)\,. }
This was written in terms of the central charge and $n=p/2$ in \eqr{maxtraceintro}. Despite obvious differences, this equation is intriguingly similar to the decoupled semi-infinite Toda chain equations obeyed by the extremal two-point functions $G_{2n}(\t)$ in the $SU(2)$ theory \cite{Baggio:2014ioa,Baggio:2014sna,Baggio:2015vxa,Gerchkovitz:2016gxx}.\foot{Dividing both sides by $G_2(\t)$, the RHS becomes reminiscent of a connected four-point function.} Note, though, that \eqr{maxcorrelator} holds for {\it all} $N$! Evidently, the extra power of $\cN=4$ supersymmetry makes the integrated four-point function even more soluble than its $\cN=2$ extremal counterparts. Altogether, this suggests that one may think of the Laplace difference equations for integrated $\cN=4$ SYM correlators in general, and \eqr{maxcorrelator} in particular, in the same spirit as the Toda equations \eqr{toda}, and that they may be derivable by some generalization of the $tt^*$ method that led to \eqr{toda}. 

The other point we wish to make is the following: {\it the large $p$ limit of $\cG_p^{(N|\text{max})}(\t)$ is a large charge limit} \c{Hellerman:2015nra,Hellerman:2017sur}. Since $\cG_p^{(N|\text{max})}(\t)$ is determined explicitly for all $p$, this regime is easy to access, given the simplicity of the recursion \eqr{maxrecursion}. For example, together with the spectral decomposition \eqr{maxresult}, it is quick to see that $\cG_{p\gg 1}^{(N|\text{max})}(\t)$ limit admits a `t Hooft-like expansion with $g_{\rm YM}^2\, p$ fixed, for {\it any} finite $N$. This generalizes a main result of \cite{Grassi:2019txd}, which established the `t Hooft-like limit of the extremal correlators \eqr{extcorr} in rank-one SCFTs at large charge (also previously studied in \cite{Bourget:2018obm,Beccaria:2018xxl,Beccaria:2018owt}). Non-perturbative effects in large $p$ or large $\l$ are also straightforward to determine along the lines of \cite{Collier:2022emf}, again using the explicit recursion above. Moreover, there are several interesting regimes of $p$ and $N$ scaling together. Work in these directions is ongoing \cite{Paul}. 
 
\subsection{Large $N$ expansion}
Let us proceed to develop the large $N$, fixed $p$ expansion of the (normalised) maximal-trace overlaps $\widehat g_p^{(N|\text{max})}(s)$. 

In light of \eqref{eq:maxtrace_solution}, the large $N$ expansion follows from that of the functions $F_p(N,s)$ and $\widehat g_2^{(N)}$. The large $N$ expansion of the latter\foot{Recall that $\widehat g_2^{(N)}(s)=2/\big((2s-1)^2(N^2-1)\big)\,f_2^{(N)}(s)$.}
 has been discussed in \cite{Dorigoni:2021guq,Collier:2022emf}, with the genus-$g$ overlaps $f_2^{(g)}(s)$ being of the general form depicted in \eqref{eq:higher_genus_overlaps}.

On the other hand, the genuinely new information about the $p$-dependence of the maximal-trace overlaps lies entirely in the functions $F_p(N,s)$. For some fixed value of $p$, we note that both the sum in the numerator as well as the Pochhammer factor in the denominator of \eqref{eq:Fp_structure} give rise to even polynomials in $N$ of degree $p-2$. Therefore, the large $N$ expansion of $F_p(N,s)$ starts at order $N^0$ and proceeds in even powers of $1/N$. Explicitly, for the first few orders we find the expansion
\begin{align}\label{eq:Fp_large_N}
\begin{split}
	F_p(N,s) &= \frac{p}{2} + \frac{p(p-2)}{8}(s+1)(s-2)\,N^{-2}\\
	&\quad+\frac{p (p-2)}{72} (s+1)(s-2)\big[(p-4)s(s-1)-3(2p-5)\big]\,N^{-4}\\
	&\quad+\frac{p(p-2)}{1152}(s+1) (s-2)\big[(p-4)(p-6)s^3(s-2)-(p-4)(17p-38)s^2\\
	&\quad\qquad\qquad\qquad\qquad\qquad\quad\quad+2(p-4)(9p-22)s+24(3p^2-14 p+14)\big]\,N^{-6}\\
	&\quad+O(N^{-8})\,.
\end{split}
\end{align}
At order $N^{-2m}$ (with $m\geq1$) we find an overall factor $(s+1)(s-2)$ times a degree $2m-2$ polynomial in $s$, whose coefficients depend only polynomially on $p$. We have verified that this simple structure persists to high order in $m$. Moreover, as a consequence of \eqref{eq:Fp_reflection}, every term in this expansion is invariant under $s\mapsto1-s$.

With this expansion of $F_p(N,s)$ at hand, it is clear that the normalised overlaps $\widehat g_p^{(N|\text{max})}(s)$ for general $p$ admit the large $N$ expansion
\begin{align}\label{eq:large_N_ghat}
	\widehat g_p^{(N|\text{max})}(s) = \sum_{g=0}^\infty N^{-2g}\Big[N^{s-1}\,\widehat g_p^{(g|\text{max})}(s)+(s \rar 1-s)\Big]\,.
\end{align}
Let us reiterate that the genus-$g$ overlaps $\widehat g_p^{(g|\text{max})}(s)$ can be computed using \eqref{eq:maxtrace_solution} from the expansion \eqref{eq:Fp_large_N} of $F_p(N,s)$ together with the known genus-$g$ overlaps $f_2^{(g)}(s)$ studied previously in \cite{Dorigoni:2021guq,Collier:2022emf}. Importantly, since the $F_p(N,s)$ are polynomial in $s$, the pole structure of $\widehat g_p^{(g|\text{max})}(s)$ is identical to their $p=2$ counterparts, which we discussed in Section \ref{GenusExpfps}. As such, the general structure of the expansion in the 't Hooft or the very strongly coupled limit remains unchanged, the only effect being that the coefficients acquire some $p$-dependence.
Let us illustrate this by explicitly developing the corresponding large $N$ expansions.

\subsubsection{'t Hooft and very strongly coupled limits}
We start by considering the 't Hooft limit of the normalised correlators $\widehat{\mathcal{G}}_p^{(N|\text{max})}(\tau)$, which admit the genus expansion
\begin{align}\label{eq:maxtrace_genusexp_normalised}
	\widehat{\mathcal{G}}_{p}^{(N|\text{max})}(\tau)= \sum_{\mathfrak{g}=0}^\infty N^{-2\mathfrak{g}}\,\mathcal{G}_{p,0}^{(\mathfrak{g}|\text{max})}(\lambda) + (\text{non-perturbative in $N$})\,.
\end{align}
As mentioned previously, (anti-)instanton contributions are non-perturbatively suppressed in $1/N$ and hence this is an expansion in the zero-instanton sector only.\footnote{Again, we will drop the `0' subscript and it is understood that we work perturbatively in $1/N$.} Note that since we consider the \textit{normalised} correlator, there is a difference of $N^p$ in the overall power of $N$ compared to \eqref{genusexp}. Since the genus-zero overlaps are directly related to the $p=2$ overlaps by an overall factor of $p$, i.e. $\widehat{f}_p^{(0|\text{max})}(s)=p\,f_2^{(0)}(s)$, it follows that the $\mathfrak{g}=0$ term is simply given by
\begin{align}
	\widehat{\mathcal{G}}_{p}^{(\mathfrak{g}=0|\text{max})}(\lambda)=p\,\mathcal{G}_2^{(\mathfrak{g}=0)}(\lambda)\,,
\end{align}
where $\mathcal{G}_2^{(\mathfrak{g}=0)}(\lambda)$ was given in Section \ref{'tHooftLimit}.  
At higher orders in $1/N$ such a factorisation no longer happens and some polynomial dependence on $p$ will appear.
For example at the next order, for $\mathfrak{g}=1$, one has
\begin{align}
\begin{split}\label{eq:genus_1_correlator_normalised}
	\widehat{\mathcal{G}}_{p}^{(\mathfrak{g}=1|\text{max})}(\lambda) &=-\frac{p}{16}\,\sqrt{\lambda }-\frac{p(p-4)}{8}+\frac{1}{2\pi i}\int_{\text{Re}\,s=\frac{1}{2}} ds~\frac{\pi}{\sin(\pi s)}\,s(1-s)\,\Lambda(1-s)\,\tilde{\lambda}^{s-1}\,\widehat{f}_p^{(1|\text{max})}(s)\,.
\end{split}
\end{align}
Like the SPO integrated correlators considered earlier, this admits a Bessel representation, which we provide in Appendix \ref{appf2}. Its strong coupling expansion is
\es{}{
	\widehat\cG_{p}^{(\mathfrak{g}=1|\text{max})}(\lambda \gg1) & = \frac{p}{16}\bigg[-\sqrt{\lambda}-2(p-4)+\frac{15(32p-169)\,\zeta(3)}{32\lambda^{\frac{3}{2}}}+\frac{315(16p+1)\,\zeta(5)}{64\lambda^{\frac{5}{2}}}\\
	&\qquad\quad+\frac{14175(576 p-1009)\,\zeta(7)}{8192\lambda^{\frac{7}{2}}}+\frac{1091475(80p-173)\,\zeta(9)}{4096\lambda^{\frac{9}{2}}}+O(\lambda^{-\frac{11}{2}})\bigg].}

Lastly, let us also consider the large $N$ expansion in the VSC limit. Following the same steps as described in Section \ref{VSCLimit}, we arrive at
\begin{align}
\begin{split}
	\widehat{\mathcal{G}}_p^{(N|\text{max})}(\tau) &= \frac{p}{4} - \frac{3p}{16}N^{-\frac{3}{2}}\widetilde{E}_{\frac{3}{2}}(\tau)- \frac{p(p-4)}{8}  N^{-2} + \frac{45p}{2^8} N^{-\frac{5}{2}} \widetilde{E}_{\frac{5}{2}}(\tau)\\[3pt]
	&\quad+\frac{15 p}{2^{15}} N^{-\frac{7}{2}} \Big((128p-676)\, \widetilde{E}_{\frac{3}{2}}(\tau)+315 \widetilde{E}_{\frac{7}{2}}(\tau)\Big)  + \frac{p(p^2-6 p+11)}{12} N^{-4}\\[3pt]
	&\quad+\frac{315 p}{2^{18}} N^{-\frac{9}{2}}\Big((64p+4)\,\widetilde{E}_{\frac{5}{2}}(\tau)+315 \widetilde{E}_{\frac{9}{2}}(\tau)\Big)\\[3pt]
	&\quad-\frac{p}{2^{27}}N^{-\frac{11}{2}}\Big(32(143360 p^2-866400 p+1961161)\,\widetilde{E}_{\frac{3}{2}}(\tau)\\[3pt]
	&\quad\qquad\qquad-56700(576 p-1009)\,\widetilde{E}_{\frac{7}{2}}(\tau)-245581875 \widetilde{E}_{\frac{11}{2}}(\tau)\Big)+ O(N^{-6})\,.
\end{split}
\end{align}
This takes a remarkably simple form considering that this describes an infinite family of integrated correlators. Compared to the corresponding expansion of the (unnormalised) $p=2$ correlator, as given in e.g. eq. (5.67) of \cite{Dorigoni:2021guq}, one notes that also higher-order \textit{even} powers of $1/N$ are present, which comes from the non-trivial large $N$ expansion of the ensemble averages $\langle \widehat{\mathcal{G}}_p^{(N|\text{max})}\rangle$, cf. \eqref{maxavg}.

\sec{General $p$ ansatz}\label{sec:genpansatz}
There is unlikely to be a simple closed-form expression for generic integrated correlators for arbitrary multi-trace insertions $(i,j)$ at arbitrary $p$, unlike for the maximal-trace correlators considered in Section \ref{sec:maximal_trace}. However, our computations provide a good deal of insight into their general structure. In particular, the difficulties in obtaining general higher $p$ results appear to be of computational nature only.\foot{To be clear, generic integrated correlator computations encounter two main technical complications. The first is the degeneracy in the space of half-BPS operators and the associated mixing of chiral operators on the sphere: as explained in Section \ref{sec:Integrated_correlators}, to obtain correlators on $\R^4$ from the localisation relation \eqref{eq:localisation_relation} one first needs to solve the mixing problem on $S^4$. As a consequence of the growing number of multi-trace operators, the size of the corresponding mixing matrices increases rapidly with $p$. The second is the presence of multi-trace operators which leads to higher-point insertions in the matrix-model integrals. This complicates the evaluation of their expectation values considerably. As described in Appendix \ref{sec:EV} this amounts to the computation of higher-index $N$-dependent coefficients, which one has to perform on a case-by-case basis.} As such, we do not expect any qualitative changes or new features to arise when considering correlators beyond the explicit cases considered so far. 

With some confidence afforded by the structural uniformity of our results, we now make some conjectural assertions about the form of generic integrated correlators. The ansatz is
\e{pansatz}{\boxed{\cG_p^{(N|i,j)}(\t) = \half g_p^{(N|i,j)}(0) + {1\o 4\pi i} \int_{\Re s =\half} ds \,{\pi\o \sin(\pi s)}s(1-s)(2s-1)^2\,g_p^{(N|i,j)}(s)\, E^*_s(\t)\,,}}
where the function $g_p^{(N|i,j)}(s)$ is polynomial in $s$ with rational coefficients. Let us explicate this ansatz in further detail:

\begin{itemize}
	\item We expect that the Eisenstein overlaps continue to be of the form
	\e{}{\{\cG_p^{(N|i,j)},E_s\} = {\pi\o \sin(\pi s)}s(1-s) f_p^{(N|i,j)}(s)\,,
}
with the structure
	\begin{align}\label{eq:f_p_ansatz_2}
	f_p^{(N|i,j)}(s)=(2s-1)^2\,g_p^{(N|i,j)}(s)\,, \quad \text{where}\quad \text{deg}\[g_p^{(N|i,j)}(s)\] = 2N+2\left\lfloor\frac{p}{2}\right\rfloor-6\,.
	\end{align}
The reflection-symmetric polynomials $g_p^{(N|i,j)}(s)$ have the indicated degree, with expansion coefficients rational in $N$ and $p$. Consequently, the integrated correlator may be completely fixed by the first $N+\left\lfloor\frac{p}{2}\right\rfloor-2$ orders in weak-coupling perturbation theory.

	\item 
	In Section \ref{IncludingInstantons}, we gave robust evidence that the cusp form overlaps vanish for $\cG_{p}^{(N|i,j)}(\tau)$ by matching the form \eqr{pansatz} to explicit calculations of instanton effects using localisation. We conjecture that this remarkable property extends to general $p$ and all trace-structures $(i,j)$: 
	\e{}{(\cG_{p}^{(N|i,j)},\phi_n)=0 \quad \forall~ (N|i,j),\,p,\,n>0\,.}

\item In accord with our remarks above, the normalised ensemble averages will be {\it rational} functions of $N$ and $p$. This is clearly visible in the maximal-trace case, see \eqr{normavgmaxtrace}.

The individual averages computed in Section \ref{sec:averages} do not seem to exhibit any particular pattern, beyond their rationality and some uniformity in the large $N$ limit. On the other hand, the ensemble averages $\langle\cG_p^{(N)}\rangle$ of correlators of SPO's {\it do} exhibit nice patterns, leading us to conjecture their general form
\begin{align}\label{eq:average_structure}
	\langle\mathcal{G}_p^{(N)}\rangle =Q_p(N)\,(N-p+1)_{p-1}\,,\qquad (\text{for SPO's})\,,
\end{align}
for some rational function $Q_p(N)$.\footnote{The fact that $Q_p(N)$ is rational essentially follows from the rational $N$-dependence of the coefficients which appear in the weak-coupling expansion obtained from the matrix model integrals. See Appendix \ref{app:localization}.}  The Pochhammer factor ensures the vanishing of $\langle\mathcal{G}_p^{(N)}\rangle$ for $N=1,2,\ldots,p-1$. By considering the large $N$ expansion of \eqref{eq:average_structure}, we can further constrain the function $Q_p(N)$: 
	\begin{enumerate}[label=(\roman*)]
	\item At leading order in large $N$, the SPO's reduce to single-trace operators and hence we should recover the large $N$ limit of $\langle\cG_p^{(N|1,1)}\rangle$ given in \eqref{eq:ensemble_average_large_N}. Therefore, we have
	\begin{align}
			Q_p(N)~ \xrightarrow[]{~N\to\infty~} ~ \frac{p-1}{2p}\,N + \ldots.
	\end{align}
	\item To be consistent with the general large $N$ expansion of the ensemble average, cf. \eqref{eq:large_N_average}, $Q_p(N)$ must admit an expansion containing only integer powers of $1/N$, i.e.
	\begin{align}
		Q_p(N) = \frac{p-1}{2p}\,N + \sum_{k=0}^{\infty}\alpha_k(p) \, N^{-k}\,,
	\end{align}
	with coefficients $\alpha_k(p)$ which we expect to be rational functions of $p$.
	\end{enumerate}
	
	\item All recursion relations derived herein determine $f_p^{(N|i,j)}(s)$ for all $N$ in terms of lower $p$ spectral overlaps, together with the trivial initial condition $f_p^{(N=1|i,j)}(s)=0$. Ultimately, everything is recursively determined from the $p=2$ spectral overlap at $N=2$, given by $f_2^{(2)}(s)=(2s-1)^2$. As such, the $p=2$ integrated correlator in the $SU(2)$ theory (together with the knowledge of the recursion relations themselves!) determines the higher-charge correlators. We expect that this determinism will extend to all $p$. 

\item The functional form of the recursion relations themselves is less clear. For $p=4,5$, the recursion had the general structure
	\begin{align}\label{eq:recursion_structure}
	\begin{split}
	f_p^{(N+1|i,j)}(s)-f_p^{(N|i,j)}(s) &= \Big[W_p^{(0)}(N)-W_p^{(1)}(N)\,s(1-s)\Big]\,f_2^{(N+1)}(s) \\
	&~~~~+ \sum_{q=2}^{p-1}W_p^{(q)}(N)\, f_q^{(N|i,j)}(s) + W_p^{(p)}(N)\,f_2^{(N-1)}(s) \,,
	\end{split}
	\end{align}	
	for some functions $W_p^{(n)}(N)$, rational in $N$, which also depend on the precise trace-structure $(i,j)$.\footnote{As we have pointed out a few times by now, in cases where $f_p^{(N|i,j)}(s)$ vanishes for $N=2$, the consistency of such a recursion relation requires the functions $W_p^{(0)}(N)$ and $W_p^{(1)}(N)$ to vanish at $N=1$.} However, in light of the growing degree with $p$ of the overlaps $f_p^{(N|i,j)}(s)$ -- see equation \eqref{eq:f_p_ansatz_2} -- the above structure cannot hold for $p\geq6$. To account for the higher degree polynomials requires some additional terms on the RHS of \eqref{eq:recursion_structure}. This could potentially include terms with higher powers of $s(1-s)$, or higher integer shifts of $N$ of the form $f_2^{(N+n)}(s)$. As such, it would be necessary to consider further explicit examples with $p\geq6$ to determine the most general structure of these powerful recursion relations.

\end{itemize}

\sec{Future directions}\label{sec8}

The integrated correlators $\cG_p^{(N|i,j)}(\t)$ are remarkable observables: they are non-trivial functions of the complexified gauge coupling $\t$, and yet tractable enough not only to be exactly determined, but to be determined simply as polynomials in an appropriate functional basis. This basis is, moreover, one in which S-duality invariance is manifest. This link suggests that there may be lessons in the offing for how to better understand $\t$-dependence of $\cN=4$ SYM observables at large.

These properties raise many conceptual and fundamental questions. We close by discussing some of these, but we first describe some more concrete open problems. There are several computations that are ripe for completion, which include the following:

\begin{itemize}
\item Our formulas allow the explicit determination of non-perturbative corrections, both in $1/\l$ and $1/N$, to $\cG_p^{(N|i,j)}(\t)$ in the large $N$ limit. This was analyzed previously for the $p=2$ case, where corrections in powers of the two scales \eqr{npscales} -- identified with fundamental and D-string worldsheet instantons in AdS$_5\,\times\,$S$^5$, respectively -- are easily identified from the $\sl$ spectral integral, the two being related by S-duality. It would be interesting to understand the $p$-dependence of these corrections. We also note that the analysis of \cite{Collier:2022emf} did not compute {\it all} non-perturbative corrections in $N$, in particular leaving the sum over genera to future study; this is an interesting problem because the genus sum can in principle generate higher-order (e.g. ``doubly'') non-perturbative effects. 

\item Another set of computations pertains to the genus-one correlators $\<\O_2\O_2\O_p\O_p\>$, both integrated and unintegrated. On the integrated side, we have made a prediction in \eqr{int1loop} for a bulk computation of the finite one-loop correction to $\cG_p^{(N)}(\t)$ in \adss supergravity. This would be satisfying to see verified in the bulk. Moreover, as noted in footnote \ref{f22}, this prediction (in combination with the flat space limit) can be used to fix the undetermined ambiguities in the {\it unintegrated} $\<\O_2\O_2\O_p\O_p\>$ correlator at one-loop level from \cite{Aprile:2019rep,Alday:2019nin}, another worthwhile computation.

\item The ensemble averages of the (SPO|SPO) correlators through $p=5$ were seen in \eqr{eq:ensemble_averages_partial_fractions}-\eqr{eq:ensemble_averages_guess} to obey an intriguing pattern. Can a closed-form expression be derived, or perhaps inferred from more data points, for generic $p$?

\item The $\sl$ spectral decomposition faciliates the study of the ensemble statistics of observables over the $\cN=4$ conformal manifold $\mathcal{M}$. The variance of the integrated correlators over $\mathcal{M}$ follows immediately from their explicit spectral overlaps and a formula derived in \cite{Collier:2022emf}:
\e{}{\mathcal{V}\(\cG_p^{(N|i,j)}\) = {\rm{vol}}^{-1}(\cF)\({1\o 4\pi i} \int_{\Re s=\half} ds\,\left|{\pi \o \sin\pi s} s(1-s)\L(s)f_p^{(N|i,j)}(s)\right|^2\)}
This was studied at both finite and large $N$ in \cite{Collier:2022emf} for the $p=2$ case. It would be interesting to ask how the general $p$ integrated correlator statistics behave as well.

\end{itemize}

There are various generalizations of our work -- e.g. extension to $\cN=4$ SYM with other gauge groups, analysis of other families of integrated correlators $\cG_p^{(N|i,j)}(\t)$ that exist for all $p$, integration of $\la \O_2\O_2\O_2\O_2\ra$ against other measures \cite{Chester:2020dja,Chester:2020vyz} -- which may be worth pursuing. Slightly further afield, it remains unknown whether there is an analogous supersymmetric construction of integrated $\<\O_p\O_p\O_p\O_p\>$, or for that matter even $\<\O_p\O_q\O_r\O_s\>$, correlators. If so, new ideas are needed to determine the supersymmetric integration measure and to relate it to some as-yet-unknown generalization of the localization methods based on $\cN=2^*$ partition functions. It should also be possible to extend the integrated correlator construction to SCFTs in two spacetime dimensions using localisation techniques.

Most fascinating are various foundational questions about these observables, centered around the question of why they exhibit such simplicity, and the relation to manifest S-duality invariance.

Our results for $\cG_p^{(N|i,j)}(\t)$ are characterized by strong uniformity in the half-BPS charge $p$. This calls for a deeper explanation. At large $N$, an obvious possible touchstone is the 10-dimensional ``hidden conformal symmetry'' that is known to govern at least some aspects of planar correlators at weak and strong coupling \cite{Caron-Huot:2018kta,Caron-Huot:2021usw}. The origin and regime of applicability of the hidden conformal symmetry itself are still unknown. It is tempting to ask whether, and how, the integrated correlators $\cG_p^{(N|i,j)}(\t)$ are organized in some way by this symmetry, perhaps at large $N$ only. If so, one may wonder whether the integrated correlators are subject to stronger constraints than the unintegrated ones. 

Is there a natural bulk description of these objects {\it besides} as integrals over boundary points? That point of view is rather clumsy from the bulk spacetime perspective. The integrated correlators have an alternative formulation as derivatives of the free energy deformed by sources, but that quantity breaks the $\cN=4$ superconformal symmetry. Perhaps there is a nicer picture which preserves the full symmetries without using the unintegrated correlator or free energy as an intermediary. 

Of course, since the $\cG_p^{(N|i,j)}(\t)$ are soluble for all $N$ and $\tau$, we are really after their bulk {\it string theory} description. The formulas herein for $\cG_p^{(N|i,j)}(\t)$ may be regarded as the exact \adss quantum string theory answers. As emphasized in the introduction, the integrated correlators seem to be picking out some privileged subsector of the \adss four-point string amplitude. Whether this is a useful perspective that can be leveraged to teach us something more general about string theory, either in \adss or in 10-dimensional flat space, is an open question. It would obviously be very interesting to make this more explicit from the worldsheet point of view, perhaps first in various simplifying limits. 

The central role of $\sl$ in our results underscores not only the power, but also the practical utility, of S-duality. That the integrated correlators are so streamlined in the $\sl$ eigenbasis suggests that the $\sl$ spectral decomposition may be generally useful in studying unprotected observables of $\cN=4$ SYM such as operator dimensions or OPE coefficents. This deserves further investigation in the context of the superconformal bootstrap and integrability (and their combination \c{Cavaglia:2022qpg,Alday:2022uxp,Caron-Huot:2022sdy}). In the present case, that the $\cG_p^{(N|i,j)}(\t)$ can be written down in such simple form for finite $N$ and $\tau$ is partly\foot{Note that there do exist non-perturbative physical observables that have computable nonzero cusp form overlaps. An example is the torus partition function of two free bosons on a Narain lattice, for all values of the moduli \c{Benjamin:2021ygh}.} due to the absence of Maass cusp forms in their $\sl$ spectral decomposition. {\it Why} do the cusp form overlaps vanish? What is the {\it physical} principle that tells us when an $\sl$-invariant observable in $\cN=4$ SYM has vanishing cusp form overlap? Observables with this property are ``non-chaotic'' in a precise mathematical, if not (yet) physical, sense. It seems important to bridge this gap. This may help understand whether the mathematical notion of {\it arithmeticity} is a useful physical criterion in characterizing CFT observables.\foot{We point out the recent work \cite{Kravchuk:2021akc,Bonifacio:2021aqf} as another, possibly related, arena for this question. For earlier work in a slightly different context, see \c{Moore:1998zu}.} 

Indeed, the radical simplicity of the integrated correlators --  exact for all $N$ and $\tau$, with ingredients no more complicated than rational functions of $N$ and $p$ -- deserves to be understood in the deepest terms possible. The fundamental origin of the recursion relations derived herein, generalizing the $\cG_2^{(N)}(\t)$ recursion relation of \cite{Dorigoni:2021guq}, is a major open question. We made some suggestions about their provenance in Section \ref{sec:6.1}. Can they be derived from a 6d (2,0) theory point of view? Can we bootstrap them directly without the input of localization? And finally, what are these equations telling us about {\it unintegrated} CFT data at finite $N$ and $\tau$, and can we somehow extract these data by cleverly combining the whole family of integrated correlators indexed by $p$? 

The integrated correlators seem to be beautiful and rich observables, standing out even in the wonderful world of $\cN=4$ SYM. They are, in our view, well-deserving of intensive further study. 


\section*{Acknowledgments}
We thank Shai Chester, Scott Collier, Gabriele di Ubaldo, Bertrand Eynard and Silviu Pufu for discussions. HR would like to  acknowledge useful discussions with participants of the Simons Summer Workshop organized at the Simons Center for Geometry and Physics, Stony Brook. This research was supported by ERC Starting Grants 853507 and 679278. 

\appendix
\sec{Colour factors}\label{app:colour-factors}
Here we list the colour factors $R_p^{(i,j)}(N)$ used in the main text. For the cases $p=2,3$, these are just given by the general formula \eqref{eq:R_p}. For $p=4,5$ we have
\begin{alignat}{2}
\label{combo}
	R_4^{(1,1)}(N)&=\frac{(N^2-1)(N^4-6 N^2+18)}{4N^2} ~, & \qquad 	R_5^{(1,1)}(N)&=\frac{(N^2-1)(N^2-4)(N^4+24)}{5N^3}~,\no\\
	R_4^{(1,2)}(N)&=\frac{(N^2-1)(2 N^2-3)}{2N} ~, & \qquad 	R_5^{(1,2)}(N)&=\frac{6(N^2-1)(N^2-4)(N^2-2)}{5N^2}~,\\
	R_4^{(2,2)}(N)&=\frac{(N^4-1)}{2} ~, & \qquad 	R_5^{(2,2)}(N)&=\frac{6 (N^2-1)(N^2-4)(N^2+5)}{25N}~.\no
\end{alignat}
The colour-factor of the two-point function of maximal-trace operators $\O_p^{(\text{max})}=T_{2,\ldots,2}$ has been derived in \cite{Gerchkovitz:2016gxx}. In our conventions, for even $p$ it reads
\begin{align}\label{rpmax}
	R_p^{(\text{max})}(N)=\frac{2^p\,\big(\frac{p}{2}\big)!}{p^2}\,\bigg(\frac{N^2-1}{2} \bigg)_{\frac{p}{2}}\,.
\end{align}

At large $N$, the counting simplifies. In particular, computation of colour factors may be computed by large $N$ factorization. Single-trace colour factors become
\e{}{R_p^{(1,1)}(N) \approx {N^p\o p}\,.}
This is especially useful for computation of colour factors involving multi-trace operators $T_{p_1,\cdots,p_n}$, defined in \eqr{multiops}. The large $N$ colour factor of $T_{p_1,\ldots,p_n}$ with $n$ distinct indices is
\e{}{\({p_1\cdots p_n\o p}\)^2 R_{p_1}(N)\ldots R_{p_n}(N)\Big|_{N\rar\i} \approx {p_1\cdots p_n\o p^2}N^p \qquad (\text{for}~p_1\neq p_2 \neq \cdots \neq p_n)\,,}
where $p = p_1+ \cdots + p_n = p$. For example, taking $p_1=2$ and $p_2=3$ reproduces the large $N$ asymptotic of $R_5^{(2,2)}(N)$. For coincident indices, extra combinatoric factors emerge. For example, the large $N$ colour factor of $T_{q,\ldots, q}$ with $n$ identical indices is
\e{}{\({q^n\o n q}\)^2 n! \,(R_{q}(N))^n \Big|_{N\rar\i} \approx{q^n\o (n q)^2}n!\,N^{n q} \,.}
Taking $q=2$ and $n=p/2$ reproduces the large $N$ asymptotic of $R_p^{(\text{max})}(N)$. 

\sec{Localization details}
\label{app:localization}
\subsection{Solving operator mixing on $S^4$ and integrated two-point functions}
\label{denominator}
In this Section we focus on computing the denominator of \eqref{eq:localisation_relation} for $p=2,3,4,5$. Let us define
\be
\label{DenFactor}
D_p^{(N|i,j)} \equiv \sum_{\m,\n}v^{i,\m}_p \bar{v}^{j,\n}_p  \p_{\tau_\m'}\p_{\bar{\tau}_\n'}\log \cZ_N(\tau,\tau_A',m)\bigg|_{\tau'_A=m=0}\,,
\ee
where we recall that $\mu$ belongs to the index set \eqref{muindex}. One of the main problems is to determine the vectors $v^{i,\m}_p$ which encode the mixing of weight-$p$ chiral operators of species $i$, denoted by $\cO_p^{(i)}$, with lower-weight chiral operators. This is done through the algorithm devised in \cite{Gerchkovitz:2016gxx}. On $S^4$, a chiral operator of weight $p$ will generically mix with all other chiral operators of weight $p-2,\, p-4,\,\ldots$ through the non-trivial background fields. For a given $p$, we begin by listing all such operators of dimension $\geq 2$.\footnote{On $S^4$, operators of even conformal dimension also mix with the identity operator. However, there is no need for us to consider this in the unmixing problem because the final results differ only by one-point functions on $\R^4$, which vanish.} The operators $\cO_2$ and $\cO_3$ do not mix with any other operator on the sphere. The two chiral operators $\cO_4^{(i)}$ mix with $\cO_2$ and likewise the two chiral operators $\cO_5^{(i)}$ mix with $\cO_3$ (recall the operator definitions in \eqref{eq:operator_lists}).

Next, we construct the matrix of {\it connected} two-point functions on $S^4$ for a given chiral operator of weight $p$ with every other chiral operator of weight $p-2,\, p-4,\,\ldots$ with which it mixes. We will list below all such matrices for operators up to $p=5$. First, we introduce some short-hand notation which will be convenient. In what follows we restrict our treatment and notation to single- and double-trace operators only, but everything can be straightforwardly generalized to arbitrary multi-trace insertions following the prescription of \cite{Gerchkovitz:2016gxx}. 

Let us introduce the notation $O_p$ and $O_{p,q}$ for single- and double-trace chiral operators, respectively, which couple to respective sources $\t'_p$ and $\t'_{p,q}$. Two-point functions on $S^4$ are constructed via
\es{defnC}{\<O_p\bar{O}_q\>:= \frac{\p_{\tau_p'}\p_{\taub_q'}\cZ_N}{\cZ_N}  \bigg|_{\tau'_A=m=0},~ 
\<O_{p,q}\bar{O}_r\>:= \frac{\p_{\tau'_{p,q}}\p_{\taub_r'}\cZ_N}{\cZ_N}  \bigg|_{\tau'_A=m=0},~ 
\<O_{p,q}\bar{O}_{r,s}\>:= \frac{\p_{\tau'_{p,q}}\p_{\taub'_{r,s}}\cZ_N}{\cZ_N}  \bigg|_{\tau'_A=m=0}.}
Note that $O_p = \cO_p^{(1)}$ is the weight-$p$ single-trace chiral operator introduced in the main text, while $O_{p,q}$ is a single representative of the set of weight-$(p+q)$ double-trace chiral operators $\cO_{p+q}^{(j)}$ indexed by $j$. Let us also introduce a standard notation for the connected two-point functions on $S^4$ (one-point functions do not vanish on $S^4$), obtained by differentiation with respect to the free energy:
\es{}{\<O_p \bar{O}_q \>_c &:= \p_{\tau'_p}\p_{\bar \tau'_q} \log\cZ_N\big|_{\tau'_A=m=0} \\&= \<O_p \bar{O}_q \> -\<O_p \>\<\bar{O}_q \>\,,}
and likewise for the other two-point functions in \eqr{defnC}. 

Through $p=5$, we thus have the following mixing matrices:

\begin{itemize}

\item $p=2$: There is just one single-trace operator which does not mix with any other operator. So there is nothing to unmix here and we have $M_2 = \<O_2 \bar{O}_2 \>_c$.

\item $p=3$: Again there is nothing to unmix, and we have $M_3= \<O_3 \bar{O}_3 \>_c.$

\item $p=4$: There are two mixing matrices to consider here since operators of weight four have degeneracy two. For the single-trace weight-four operator $\cO_4^{(1)}$ we denote the mixing matrix by $M_4^{(1)}$ whereas for the double-trace weight-four operator $\cO_4^{(2)}$, we denote the corresponding mixing matrix by $M_4^{(2)}$. These are 
\begin{equation}
M_4^{(1)} =\left(\begin{array}{cc}
\<O_2 \bar{O}_2 \>_c &\<O_2 \bar{O}_4 \>_c   \\[5pt]
\<O_4 \bar{O}_2 \>_c & \<O_4 \bar{O}_4 \>_c
\end{array}\right),\qquad
M_4^{(2)} =\left(\begin{array}{cc}
\<O_2 \bar{O}_2 \>_c &\<O_2 \bar{O}_{2,2} \>_c   \\[5pt]
\<O_{2,2} \bar{O}_2 \>_c & \<O_{2,2} \bar{O}_{2,2} \>_c
\end{array}\right).
\end{equation}

\item $p=5$: Again, there are two mixing matrices to consider here since operators of weight five have degeneracy two. For the single-trace weight-five operator $\cO_5^{(1)}$ we denote the mixing matrix by $M_5^{(1)}$ whereas for the double-trace weight-five operator $\cO_5^{(2)}$, we denote the corresponding mixing matrix by $M_5^{(2)}$. These are 
\begin{equation}
M_5^{(1)} =\left(\begin{array}{cc}
\<O_3 \bar{O}_3 \>_c &\<O_3 \bar{O}_5 \>_c   \\[5pt]
\<O_5 \bar{O}_3 \>_c & \<O_5 \bar{O}_5 \>_c
\end{array}\right),\qquad 
M_5^{(2)} =\left(\begin{array}{cc}
\<O_3 \bar{O}_3 \>_c &\<O_3 \bar{O}_{2,3} \>_c   \\[5pt]
\<O_{2,3} \bar{O}_3 \>_c & \<O_{2,3} \bar{O}_{2,3} \>_c
\end{array}\right).
\end{equation}

\end{itemize}
We explictly compute $M_2,M_3,M_4^{(1)},M_4^{(2)},M_5^{(1)},M_5^{(2)}$ as a function of $N$ and $\tau$ by evaluating the matrix integrals using methods outlined in Section \ref{sec:EV}.

The next step is Gram-Schmidt orthogonalization of the matrices $M_p^{(i)}$. As mentioned above there is nothing to unmix in the $p=2,3$ cases. In particular this gives $\p_{\tau_3} = \p_{\tau_3'}$. For $p=4,5$, the Gram-Schmidt process gives the new orthogonal vectors $\p/\p{\tau_{4^{(j)}}},\p/\p{\tau_{5^{(j)}}}$ (which brings insertions of chiral operators $\cO_4^{(j)}$ and $\cO_5^{(j)}$, respectively, in correlation functions on $\mathbb{R}^4$):
\begin{align}
&\frac{\p}{\p\tau_{4^{(j)}}} = \frac{\p}{\p\tau'_{4^{(j)}}} + v_4^{j,2}\frac{\p}{\p{\tau}}\,,\qquad \frac{\p}{\p\tau_{5^{(j)}}} = \frac{\p}{\p\tau'_{5^{(j)}}} + v_5^{j,3}\frac{\p}{\p \tau_3}\,,
\end{align}
where $j=1,2$ corresponds to single- and double-trace operators, respectively. We find the following results for the $v$-vectors:
\begin{alignat}{2}
\label{combo}
v_4^{1,2} &= -\frac{ \<O_4 \bar{O}_2 \>_c}{ \<O_2 \bar{O}_2 \>_c}=-\frac{ \left(2 N^2-3\right)}{2yN }\,, & \qquad v_5^{1,3} &= -\frac{ \<O_5 \bar{O}_3 \>_c}{ \<O_3 \bar{O}_3 \>_c}=-\frac{5\left(N^2-2\right)}{4yN}\,, \\
v_4^{2,2} &= -\frac{ \<O_{2,2} \bar{O}_2 \>_c}{ \<O_2 \bar{O}_2 \>_c}=-i\frac{\left(N^2+1\right)}{2y}\,, & \qquad v_5^{2,3} &= -\frac{ \<O_{2,3} \bar{O}_3 \>_c}{ \<O_3 \bar{O}_3 \>_c}=-i\frac{N^2+5}{4y}\,.\no
\end{alignat}
This data is sufficient to solve the sphere mixing problem for $p=2,3,4,5$ and therefore allows to compute both the numerator and denominator of \eqref{eq:localisation_relation}. 

As for the denominator, assembling everything, we find the following results for the matrix of integrated two-point functions $D_p^{(N|i,j)}$ on $\mathbb{R}^4$ as defined in \eqref{DenFactor}:
\begin{align}
\begin{split}
\label{DValues}
D_2^{(N)}  & = \frac{N^2-1}{2^3y^2}\,,\qquad\qquad D_3^{(N)} = \frac{3  (N^2-4) (N^2-1)}{2^6y^3 N }\,,\\[5pt]
D_4^{(N)}  & = \left(
\begin{array}{cc}
 \frac{\left(N^2-1\right) \left(N^4-6 N^2+18\right)}{2^6y^4 N^2} & -i\frac{\left(N^2-1\right) \left(2 N^2-3\right)}{2^5y^4N} \\[10pt]
i \frac{ \left(N^2-1\right) \left(2 N^2-3\right)}{2^5y^4 N} & \frac{N^4-1}{2^5y^4} \\
\end{array}
\right),\\[5pt]
D_5^{(N)} & = \left(
\begin{array}{cc}
 \frac{5  \left(N^2-4\right) \left(N^2-1\right) \left(N^4+24\right)}{2^{10}y^5 N^3 } & -i \frac{15 \left(N^2-4\right) \left(N^2-2\right) \left(N^2-1\right)}{2^9y^5 N^2} \\[10pt]
i \frac{15 \left(N^2-4\right) \left(N^2-2\right) \left(N^2-1\right)}{2^9y^5 N^2} & \frac{3 \left(N^2-4\right) \left(N^2-1\right) \left(N^2+5\right)}{2^9y^5 N } \\
\end{array}
\right).
\end{split}
\end{align}
Note that the matrices $D_p^{(N|i,j)}$ are all positive definite for physical values of $N$, as they should be for two-point functions. We can further diagonalize $D_4^{(N|i,j)}$ and $D_5^{(N|i,j)}$ in a $\tau,\bar{\tau}$ independent single-particle operator basis (discussed in Section \ref{sec:2.1}) but for our current purpose we do not need such an expression. It is straightforward to check that the $N$-dependence of these results are precisely those given by computing the flat space two-point function using Wick contractions in the free $\cN=4$ SYM with gauge group $SU(N)$.

\subsection{Integrated four-point functions}
In this Section, we compute the numerator of \eqref{eq:localisation_relation} for $p=2,3,4,5$ in the zero instanton sector. 
\be
\label{NumFactor}
C_p^{(N|i,j)} \equiv \sum_{\m,\n}v^{i,\m}_p \bar{v}^{j,\n}_p \p_{\tau_\m'}\p_{\bar{\tau}_\n'}\p_m^2\log\cZ_N(\tau,\tau_A',m)\Big|_{\tau'_A=m=0}~,
\ee
where we recall that $\mu$ belongs to the index set \eqref{muindex}. We first note that $\p_{\tau_\m'}\p_{\bar{\tau}_\n'}\p_m^2\log\cZ_N$ gives the following five terms that compute the {\it connected} integrated correlator
\begin{align}\label{5terms}
&\p_{\tau_\m'}\p_{\bar{\tau}_\n'}\p_m^2\log\cZ=\no\\[5pt]
&\frac{\p_{\tau_\m'}\p_{\bar{\tau}_\n'}\p_m^2 \cZ}{\cZ} - \(\frac{\p_{\tau_\m'}\p_{\bar{\tau}_\n'}\cZ}{\cZ}\frac{\p_m^2\cZ}{\cZ} + \frac{\p_{\tau_\m'}\cZ}{\cZ}\frac{\p_{\bar{\tau}_\n'}\p_m^2\cZ}{\cZ}+\frac{\p_{\bar{\tau}_\n'}\cZ}{\cZ}\frac{\p_{\tau_\m'}\p_m^2\cZ}{\cZ}\) + 2 \frac{\p_{\tau_\m'}\cZ}{\cZ}\frac{\p_{\bar{\tau}_\n'}\cZ}{\cZ}\frac{\p_m^2\cZ}{\cZ}
\end{align}
Single mass derivatives of the $\cN=2^*$ partition function do not appear above since those terms vanish for vanishing mass. At vanishing sources and with $\m,\n$ refering to single-trace species $p_1,q_1$ respectively, the above derivatives give the following insertions in the matrix model integral:
\begin{align}
\label{insertions}
\begin{split}
\p_{\tau_p'} &\to i\pi^{p/2} \sum_{i=1}^N a_i^{p}~,\qquad \p_{\bar{\tau}_q'} \to -i\pi^{q/2} \sum_{i=1}^N a_i^{q}\\
\p^2_m &\to -4 \sum_{i<j}\sum_{\ell=1}^\infty (-1)^\ell (2 \ell+1) \(a_{ij}\)^{2\ell} \zeta (2 \ell+1)
\end{split}
\end{align}
For $\m,\n$ refering to multi-trace species, as explained in Section \ref{sec:Integrated_correlators}, the precise insertion depends upon the species under consideration. The first line in \eqref{insertions} is easy to see from \eqref{generatingfunction}. On the other hand, insertions from mass derivatives resulting from \eqref{generatingfunction} are complicated. A pragmatic approach to evaluate them is in the large $y$ expansion. One way to do this is to expand the quantity 
\be
\p_m^2 \[\frac{1}{H(m)^N}\prod_{i<j} \frac{H^2(a_{ij})}{H(a_{ij}+m)H(a_{ij}-m)}\]_{m=0}
\ee
in small $a_i$. This expansion is represented by the $\ell$ sum in \eqref{insertions}. Performing the matrix integrals term-by-term in the $\ell$ sum gives an asymptotic expansion for the integrated correlators at large $y$. So computing $\p_{\tau_\m'}\p_{\bar{\tau}_\n'}\p_m^2\log\cZ$ boils down to computing the matrix integral \eqref{N4PartitionFSUN} with insertions dictated by the derivatives in \eqref{insertions} and their multi-trace generalization. For $p\leq 5$, the most general such insertion involves computing the following\footnote{For correlators involving just single-trace operators, \eqref{insertion1} is still the most general insertion for generic $p$, but when multi-trace operators are involved, there can be several more insertions of the type $\sum_{i=1}^N a_i^q$. } 
\be
\label{insertion1}
\bigg<\(\sum_{i<j}(a_i-a_j)^{2\ell} \)\(\sum_{i=1}^N a_i^p\)\(\sum_{i=1}^N a_i^q\)\bigg>
\ee
where the expectation value is computed in the Gaussian ensemble \eqref{N4PartitionFSUN} whose details we present in Section \ref{sec:EV}. Putting together various ingredients we find the following results for $C_p^{(N|i,j)}$ (in the expressions below the matrix elements $D_p^{(N|1,1)}$ were computed in \eqref{DValues})
\begin{subequations}
\label{CValues}

\ni\bul{$p=2$:}
\begin{align}
&C_2^{(N)} =  D_2^{(N)}\bigg( \frac{12 N \zeta (3)}{y\pi} -\frac{75 N^2 \zeta (5)}{y^2 \pi^2} +\frac{735 N^3 \zeta (7)}{2 y^3 \pi^3}-\frac{945 N^2 \left(7 N^2+2\right) \zeta (9)}{4 y^4 \pi^4}\no\\
&+\frac{114345 N^3 \left(N^2+1\right) \zeta (11)}{16 y^5 \pi^5}-\frac{351351 N^2 \left(11 N^4+25 N^2+4\right) \zeta (13)}{128 y^6 \pi^6}+\cdots\bigg)
\end{align}
\ni\bul{$p=3$:}
\begin{align}
&C_3^{(N)} = D_3^{(N)}\bigg(\frac{18 N \zeta (3)}{y\pi}-\frac{90 N^2 \zeta (5)}{y^2 \pi^2}+\frac{735 N^3 \zeta (7)}{2 y^3 \pi^3}-\frac{2835 N^2 \left(2 N^2+1\right) \zeta (9)}{4 y^4 \pi^4}\no\\
&+\frac{114345 N^3 \left(3 N^2+5\right) \zeta (11)}{64 y^5 \pi^5}-\frac{117117 N^2 \left(11 N^4+40 N^2+9\right) \zeta (13)}{64 y^6 \pi^6}+\cdots\bigg)
\end{align}
\ni\bul{$p=4$:}
\begin{align}\label{eq:c_p=4_1}
C_4^{(N|1,1)} &= D_4^{(N|1,1)}\bigg(\frac{24 N \zeta (3)}{y\pi} -\frac{60 N^2 \left(2 N^4-6 N^2+27\right) \zeta (5)}{\left(N^4-6 N^2+18\right) y^2 \pi^2}+\frac{105 N^3 \left(10 N^4+3 N^2+108\right) \zeta (7)}{2 \left(N^4-6 N^2+18\right) y^3 \pi^3}\no\\
& -\frac{945 N^2 \left(19 N^6+83 N^4+168 N^2+144\right) \zeta (9)}{8 \left(N^4-6 N^2+18\right) y^4 \pi^4} + \cdots \bigg)
\end{align}
\begin{align}
C_4^{(N|1,2)} &= D_4^{(N|1,2)}\bigg(\frac{24 N \zeta (3)}{y\pi} -\frac{30 N^2 \left(11 N^2-9\right) \zeta (5)}{\left(2 N^2-3\right) y^2 \pi^2} +\frac{3675 N^5 \zeta (7)}{2 \left(2 N^2-3\right) y^3 \pi^3}\no\\
& -\frac{945 N^2 \left(40 N^4+53 N^2-12\right) \zeta (9)}{4 \left(2 N^2-3\right) y^4 \pi^4} +\frac{114345 N^5 \left(13 N^2+43\right) \zeta (11)}{32 \left(2 N^2-3\right) y^5 \pi^5}+\cdots\bigg)
\end{align}
\begin{align}\label{eq:c_p=4_3}
C_4^{(N|2,2)} &= D_4^{(N|2,2)}\bigg(\frac{24 N \zeta (3)}{y\pi} -\frac{150 N^2 \left(N^2+3\right) \zeta (5)}{\left(N^2+1\right) y^2 \pi^2} +\frac{735 N^3 \left(N^2+6\right) \zeta (7)}{\left(N^2+1\right) y^3 \pi^3}\no\\
& -\frac{945 N^2 \left(7 N^4+72 N^2+20\right) \zeta (9)}{2 \left(N^2+1\right) y^4 \pi^4} +\frac{114345 N^3 \left(N^2+15\right) \zeta (11)}{8 y^5 \pi^5} + \cdots\bigg)
\end{align}
\ni\bul{$p=5$:}
\begin{align}\label{eq:c_p=5_1}
C_5^{(N|1,1)} &= D_5^{(N|1,1)}\bigg(\frac{30 N \zeta (3)}{y\pi} -\frac{150 N^2 \left(N^4+6 N^2+12\right) \zeta (5)}{\left(N^4+24\right) y^2 \pi^2}+\frac{525 N^3 \left(5 N^4+57 N^2+48\right) \zeta (7)}{4 \left(N^4+24\right) y^3 \pi^3}\no\\
& -\frac{1575 N^2 \left(7 N^6+122 N^4+126 N^2+72\right) \zeta (9)}{4 \left(N^4+24\right) y^4 \pi^4}+\cdots\bigg)
\end{align}
\begin{align}
C_5^{(N|1,2)} &= D_5^{(N|1,2)}\bigg(\frac{30 N \zeta (3)}{y\pi} -\frac{30 N^2 \left(6 N^2-5\right) \zeta (5)}{\left(N^2-2\right) y^2 \pi^2} +\frac{105 N^3 \left(17 N^2+8\right) \zeta (7)}{2 \left(N^2-2\right) y^3 \pi^3}\no\\
& -\frac{945 N^2 \left(35 N^4+87 N^2-20\right) \zeta (9)}{8 \left(N^2-2\right) y^4 \pi^4} +\frac{38115 N^3 \left(31 N^4+167 N^2+18\right) \zeta (11)}{64 \left(N^2-2\right) y^5 \pi^5} + \cdots\bigg)
\end{align}
\begin{align}\label{eq:c_p=5_3}
C_5^{(N|2,2)} &= D_5^{(N|2,2)}\bigg(\frac{30 N \zeta (3)}{y\pi} -\frac{15 N^2 \left(11 N^2+97\right) \zeta (5)}{\left(N^2+5\right) y^2 \pi^2} +\frac{735 N^3 \left(N^2+14\right) \zeta (7)}{\left(N^2+5\right) y^3 \pi^3}\no\\
&-\frac{945 N^2 \left(13 N^4+268 N^2+133\right) \zeta (9)}{4 \left(N^2+5\right) y^4 \pi^4} +\frac{114345 N^3 \left(7 N^4+200 N^2+321\right) \zeta (11)}{64 \left(N^2+5\right) y^5 \pi^5} +\cdots\bigg)
\end{align}
\end{subequations}
Having computed these results, we remark that while for $p=2$ and $3$, where the non-planar corrections start at four-loop order, for higher $p$, non-planarity sets in at lower loop orders. Specifically, for the matrix of $p=4,5$ correlators, we see that the non-planar corrections already appear at two loops which is in agreement with known results in the literature \cite{DAlessandro:2005fnh,Fleury:2019ydf}.

With these results, we now have all the ingredients to assemble the full matrix of integrated correlators $\mathcal{G}_p^{(N|i,j)}(\tau)$ defined in \eqref{eq:localisation_relation} for $p=2,3,4,5$ at finite $N$ in the zero-instanton sector in a weak-coupling expansion.

\subsection{Expectation values in the Gaussian (special) unitary ensemble}
\label{sec:EV}
Here we demonstrate how we evaluate expectation values of the type \eqref{insertion1} in the Gaussian special unitary ensemble \eqref{N4PartitionFSUN}. We found \cite{mehta, Marino:2004eq, Fiol:2013hna, Gradshteyn} as useful references for the material presented in this section.

One of the technical complications in evaluating such expectation values is to properly deal with the delta function constraint on the eigenvalues in \eqref{N4PartitionFSUN}. When the insertions are just differences of eigenvalues then such a constraint simply gives an $N$-dependent prefactor with the expectation values to be evaluated in the usual Gaussian unitary ensemble.  However, when the insertions are not differences of eigenvalues (like in \eqref{insertion1}), a systematic way to handle the constraint is to use the integral representation of the delta function, $\delta(\sum a_i) = \int \frac{d\mu}{2\pi} e^{i\mu \sum_ia_i}$.\footnote{Alternatively, one can use the physical Coulomb branch parameters $\boldsymbol{a}_i$ 
\be
\boldsymbol{a}_i = a_i -\frac{1}{N}\sum_{j=1}^N a_j\,,
\ee
which makes all insertions into functions of differences $a_{ij}$ that can then be evaluated in the $U(N)$ ensemble.} After doing a linear change of variables
\be
\label{variableChange}
a_{i}-\frac{i\mu}{2\a} = z_i
\ee
and completing the squares in the exponent, we get that \eqref{insertion1} evaluates to (where for convenience we defined $\a:=2\pi y$) 
\begin{align}
\begin{split}
\eqref{insertion1}=&\int d^Nz~  e^{-\a \sum_{i=1}^N z_i^2}   \prod_{i<j} (z_{ij})^2 \sum_{i<j}\(z_{ij}\)^{2\ell}  \int\frac{d\m}{2\pi} e^{-\frac{\mu^2N}{4\a}}  \sum_{k=1}^N \[a_k(z_k)\]^p  \sum_{k'=1}^N \[a_{k'}(z_{k'})\]^q \\
\equiv&\bigg< \sum_{i<j}\(z_{ij}\)^{2\ell}  \int\frac{d\m}{2\pi} e^{-\frac{\mu^2N}{4\a}}  \sum_{k=1}^N \[a_k(z_k)\]^p  \sum_{k'=1}^N \[a_{k'}(z_{k'})\]^q \bigg>
\label{temp0}
\end{split}
\end{align}
The expectation value in the second line is now understood to be in the usual Gaussian unitary ensemble where the eigenvalues $z_i$ are unconstrained. The sums appearing above can be opened up and simplified to the sum of following five expectation values
\begin{align}
\begin{split}
\eqref{temp0}=&~N(N-1)\bigg< z_{12}^{2\ell} \int\frac{d\m}{2\pi} e^{-\frac{\mu^2N}{4\a}} a_1^{p+q}\bigg>+N(N-1)\bigg< z_{12}^{2\ell} \int\frac{d\m}{2\pi} e^{-\frac{\mu^2N}{4\a}} a_1^pa_2^q\bigg> \\
&+\frac12N(N-1)(N-2)\bigg< z_{12}^{2\ell} \int\frac{d\m}{2\pi} e^{-\frac{\mu^2N}{4\a}} a_3^{p+q}\bigg> \\
&+ N(N-1)(N-2)\bigg< z_{12}^{2\ell} \int\frac{d\m}{2\pi} e^{-\frac{\mu^2N}{4\a}} (a_1^p a_3^q+a_1^q a_3^p) \bigg> \\
&+\frac12N(N-1)(N-2)(N-3)\bigg< z_{12}^{2\ell} \int\frac{d\m}{2\pi} e^{-\frac{\mu^2N}{4\a}} a_3^p a_4^q\bigg>
\label{tempC1}
\end{split}
\end{align}
The relation \eqref{variableChange} between the old ($a_i$) and the new ($z_i$) eigenvalues is left implicit to avoid clutter in the expression above.
Because of the $\mu$ integral and the fact that $a_i$ depends on $\mu$ through \eqref{variableChange}, it is easiest to handle these expectation values by first providing sources to them and performing the $\mu$ integral, i.e., we first do the following replacement above
\be
a ^{k} \to \partial_{\gamma }^k e^{\gamma a }\Big|_{\gamma=0}~,\qquad z_{12}^{2\ell} \to \p_\b^{2\ell} e^{\b z_{12}}\Big|_{\beta=0}
\ee
and then do the $\mu$ integrals to get
\begin{align}
\begin{split}
\label{tempC2}
\eqref{tempC1}=&~\sqrt{\frac{\a}{\pi N}}\,\p_\b^{2\ell}\bigg[N(N-1)\p_{\g_1}^{p+q}\bigg<e^{(\g_1+\b)z_1-\b z_2}\bigg>e^{-\frac{\gamma _1^2}{4 \alpha  N}} +N(N-1)\p_{\g_1}^p\p_{\g_2}^q \bigg<  e^{(\g_1+\b)z_1+(\g_2-\b)z_2} \bigg>e^{-\frac{\(\gamma _1+\gamma _2\)^2}{4 \alpha  N}}\\
&+\frac12N(N-1)(N-2)\p_{\g_3}^{p+q} \bigg<  e^{\b(z_1-z_2)+\g_3z_3} \bigg>e^{-\frac{\gamma _3^2}{4 \alpha  N}}\\
&+ N(N-1)(N-2)\p_\g^p\p_\d^q\bigg<   \(e^{(\g+\b) z_1-\b z_2 + \d z_3 }+ e^{(\d+\b) z_1-\b z_2 + \g z_3 } \) \bigg>e^{-\frac{\(\gamma _1+\gamma _3\)^2}{4 \alpha  N}}\\
&+\frac12N(N-1)(N-2)(N-3)\p_{\g_3}^p\p_{\g_4}^q\bigg<  e^{\b z_1 -\b z_2 + \g_3 z_3+\g_4 z_4}  \bigg>e^{-\frac{\(\gamma_3+\gamma_4\)^2}{4 \alpha  N}}\bigg]_{\b=\g_i=0} ~.
\end{split}
\end{align}
We see that the problem now boils down to computing expectation values of the type
\be
\label{mainEV}
\bigg< e^{\sum_{i=1}^k \g_i z_i}\bigg>
\ee
in the Gaussian unitary ensemble. 

A practical way to evaluate these expectation values is as follows. Using the orthogonal Hermite polynomials \cite{Mehta:1981xt}, it is straightforward to show that \eqref{mainEV} can be written as the following sum over a certain determinant 
\begin{align}
\label{EVsol0}
\bigg< e^{\sum_{i=1}^k \g_i z_i}\bigg> &= \frac{(N-k)!}{N!}\frac{G(N+2) \pi^{N/2}}{2^{N(N-1)/2}\a^{N^2/2}} \sum_{a_1,\cdots , a_k =0}^{N-1} \det_{i,j} ~ Q_{a_i,a_j}(\g_i)
\end{align}
where 
\begin{align}
{Q}_{m,n}(\b)&:=e^{\frac{\beta ^2}{4 \alpha }} \left(\frac{\beta }{2 \alpha }\right)^{n-m} L_m^{n-m}\left(-\frac{\beta ^2}{2 \alpha }\right)~.
\end{align}
Here $L_m^n(x)$ are the associated Laguerre polynomials. We record here the following orthogonality relation for Hermite polynomials $H_n(x)=(-1)^ne^{x^2}\(d^n/dx^n\)e^{-x^2}$
\begin{align}
\label{HermiteDef}
P_i(x) :=  \frac{1}{\(4\a\)^{i/2}} H_i(\sqrt{\alpha} x)~,~~ \int_{-\infty}^{\infty} dx ~e^{-\a x^2} P_j(x)P_k(x) = \delta_{jk}\frac{\Gamma(k+1) \sqrt{\pi}}{2^k \a^{k+\frac12}}
\end{align}
and the identity 
\begin{align}
\label{QIdentity}
\int da  e^{-\a a^{2}+\beta a}  P_m (a)  P_n (a) = \frac{\sqrt{\pi }}{ 2^{n}} m! e^{\frac{\beta ^2}{4 \alpha }} \alpha ^{-n-\frac{1}{2}} \beta ^{n-m} L_m^{n-m}\left(-\frac{\beta ^2}{2 \alpha }\right)~.
\end{align}
which are necessary to show the equality in \eqref{EVsol0}. We note that $Q$ enjoys a summation identity 
\be
\label{QSum}
\sum_{m=0}^{N-1}{Q}_{m,m}(\b) = e^{\frac{\beta ^2}{4 \alpha }}\sum_{m=0}^{N-1} L_m\left(-\frac{\beta ^2}{2 \alpha }\right) = e^{\frac{\beta ^2}{4 \alpha }} L_{N-1}^{1}\left(-\frac{\beta ^2}{2 \alpha }\right) = \frac{2 \alpha }{\beta }{Q}_{N-1,N}(\b)
\ee
which allows us to compute \eqref{mainEV} for $k=1$ in closed form as
\be
\big< e^{ \g  z_1} \big>  = \frac{2 \alpha }{\g N}{Q}_{N-1,N}(\g)\frac{G(N+2) \pi^{N/2}}{2^{N(N-1)/2}\a^{N^2/2}}~.
\ee
A convenient way to represent the sum over the determinant in \eqref{EVsol0} is to think of it as sum over products of traces of products of matrices ${Q}_{a_i,a_j}$. Working out the traces one finds that the result takes the form of an exponential times an even polynomial in $\g_i$ :
\be
\label{detSum}
\sum_{a_1,\cdots , a_k =0}^{N-1} \det_{i,j} ~ Q_{a_i,a_j}(\g_i) = e^{\sum_{i=1}^k \frac{\g_i^2}{4\a}} \sum_{i_1,i_2 \cdots i_k=0}^\infty C^{(k)}_{i_1,i_2 \cdots i_k}(N)\(\frac{\g_1 }{\sqrt{2\a}}\)^{i_1} \(\frac{\g_2 }{\sqrt{2\a}}\)^{i_2} \cdots \(\frac{\g_k }{\sqrt{2\a}}\)^{i_k}
\ee
where the coefficients $C^{(k)}_{i_1,i_2 \cdots i_k}(N)$ are functions of $N$ and completely symmetric in the indices $i_1,i_2 \cdots i_k$. We do not know them in closed form as a function of $N$ for generic indices but nevertheless can determine for a specific set of indices, which is all that we really need for calculations. 

The relation \eqref{detSum} formally determines \eqref{EVsol0} and therefore \eqref{tempC2}. Working out the relevant derivatives in \eqref{tempC2} and with the knowledge of $C^{(k)}_{i_1,i_2 \cdots i_k}(N)$ we can straightforwardly determine \eqref{insertion1} for given $\ell,p,q$.

\sec{Recursion relations for spectral overlaps involving SPO's}\label{app:spo_recursions}
One possible representation of the recursion relation for the spectral overlaps of the $p=4$ correlator with SPO's can be written in the form
\begin{align}\label{eq:recursion_2244_spo}
\begin{split}
	f_4^{(N+1)}(s) &= \frac{N-1}{4 (N+1)^2(N^2+2 N+2)^2}\bigg[P_1(N)- (N-1)(N^2-4)^2s(1-s)\bigg]\,f_2^{(N+1)}(s)\\[3pt]
	&~~~+f_4^{(N|1,1)}(s)-\frac{2 \left(2 N^2+4 N-1\right)}{(N+1) \left(N^2+2 N+2\right)}\,f_4^{(N|1,2)}(s)\\[3pt]
	&~~~+\frac{3 \left(2 N^2+2 N+3\right)}{N (N+1)}\,f_3^{(N)}(s)-\frac{P_2(N)}{2 N^2 (N+1) \left(N^2+2 N+2\right)}\,f_2^{(N)}(s)\\[3pt]
	&~~~-\frac{N \left(N^3+11 N^2+20 N-2\right)}{4 \left(N^2+2 N+2\right)}\,f_2^{(N-1)}(s)\,,
\end{split}
\end{align}
where the polynomials $P_i(N)$ are given by
\begin{align}
\begin{split}
	P_1(N)&=3 N^7+10 N^6+37 N^5+70 N^4+8 N^3-216 N^2-316 N-96\,,\\
	P_2(N)&=N^7+N^6-16 N^5+72 N^3-16 N^2-120 N-72\,.
\end{split}
\end{align}
Note that one can further manipulate the RHS of \eqref{eq:recursion_2244_spo} to repackage the terms $f_4^{(N|1,1)}(s)$ and $f_4^{(N|1,2)}(s)$ into the SPO overlap $f_4^{(N)}(s)$ by using the various relations given in the main text. However, the result of this rearrangement results in even more complicated $N$-dependent rational coefficients.

Similarly, putting together the recursion relations for $f_5^{(N|i,j)}(s)$ from the main text to obtain the $p=5$ recursion of SPO's yields
\begin{align}\label{eq:recursion_2255_spo}
\begin{split}
	f_5^{(N+1)}(s) &= \frac{2 (N-1)}{(N+1)^3(N^2+2 N+6)^2}\bigg[P_3(N)- (N-3)^2 (N^2-4)^2s(1-s)\bigg]\,f_2^{(N+1)}(s)\\[3pt]
	&~~~+f_5^{(N|1,1)}(s)-\frac{10 (N^2+2 N-1)}{(N+1)(N^2+2 N+6)}\,f_5^{(N|1,2)}(s)\\[3pt]
	&~~~+\frac{8(N^2+N+2)}{N (N+1)}\,f_4^{(N|1,1)}(s)-\frac{48 (N+2)(N^2+2 N-1)}{N (N+1) (N^2+2 N+6)}\,f_4^{(N|1,2)}(s)\\[3pt]
	&~~~-\frac{3\,P_4(N)}{N^2 (N+1)^2(N^2+2 N+6)^2}\,f_3^{(N)}(s)-\frac{2\,P_5(N)}{N^3 (N+1)^2 (N^2+2 N+6)^2}\,f_2^{(N)}(s)\\[3pt]
	&~~~+4N\,f_2^{(N-1)}(s)\,,
\end{split}
\end{align}
with polynomials
\begin{align}
	P_3(N)&=3 N^8-2 N^7+90 N^6+376 N^5+447 N^4-86 N^3+180 N^2+1152 N+432\,,\nonumber\\
	P_4(N)&=2 N^9-N^8+16 N^7-123 N^6-504 N^5-950 N^4-1356 N^3-1296 N^2-1584 N-864\,,\nonumber\\
	P_5(N)&=N^{11}-12 N^{10}-54 N^9+158 N^8+439 N^7-566 N^6\nonumber\\
	&~~~-1838 N^5-1632 N^4-312 N^3-576 N^2-3168 N-1728\,.
\end{align}

\section{One-instanton results}
\label{app:Instantons}
Here we provide results for the one-instanton contribution to the integrated correlators $\mathcal{G}_p^{(i,j)}(N;\tau)$ for $p=2,4,5$ in a weak-coupling expansion. The result for $p=3$ was already presented in the main text in Section \ref{IncludingInstantons}. 

For $p=2$ the results were previously computed in \cite{Chester:2019jas, Dorigoni:2021guq} which in our conventions are
\begin{align}
\mathcal{G}_2^{(N)}(\tau)\Big|_{\text{1-inst}}= e^{2 \pi i \tau} \frac{R_2}{4}\(-\frac{6 N \Gamma \left(N-\frac{1}{2}\right)}{\sqrt{\pi } \Gamma (N+2)} -\frac{9 (N-4) N \Gamma \left(N-\frac{3}{2}\right)}{8 \sqrt{\pi } y \pi \Gamma (N+2)} +\frac{45 N \Gamma \left(N-\frac{5}{2}\right)}{256 \sqrt{\pi } y^2\pi^2 \Gamma (N)} +\cdots \)
\end{align}
For $p=4$, the one-instanton weak-coupling expansion of the $(1,1)$ component has the following form:
\begin{align}
&\mathcal{G}_4^{(N|1,1)}(\tau)\Big|_{\text{1-inst}}=e^{2 \pi i \tau} \frac{R_4^{(1,1)}}{4} \frac{1}{ N^4-6 N^2+18 }\sum_{n=0}^\infty  \frac{\Gamma \left(N-\frac{3}{2}-n\right)}{\sqrt{\pi }  \Gamma (N+2)(\pi y)^n}h_{4,n}^{(1,1)}(N)
\end{align} 
where $h_{4,n}^{(1,1)}(N)$ are polynomials in $N$ of degree $6+n$. For first few values of $n$ these are
\begin{align}
\begin{split}
h_{4,0}^{(1,1)}(N) & =-\frac{3N}{4} \left(80 N^5-360 N^4+243 N^3+648 N^2-432 N-432\right)\\
h_{4,1}^{(1,1)}(N) & =-\frac{9N}{64}\left(80 N^6-1720 N^5+8803 N^4-14157 N^3-1392 N^2+12528 N+2880\right)\\
h_{4,2}^{(1,1)}(N) & =\frac{45}{2^{11}} N^2(N-2) \left(80 N^5+1640 N^4-32037 N^3+138231 N^2-150228 N-80136\right)\\
h_{4,3}^{(1,1)}(N) & =-\frac{315}{2^{15}} N^2 \bigg(80 N^7-40 N^6+4923 N^5-144186 N^4+936765 N^3-2308230 N^2+1935792 N+7776\bigg)
\end{split}
\end{align}
Similarly for $p=4$ and the $(1,2)$ component of the correlator, we have
\begin{align}
\mathcal{G}_4^{(N|1,2)}(\tau)\Big|_{\text{1-inst}}=&\mathcal{G}_4^{(N|2,1)}(\tau)\Big|_{\text{1-inst}}=e^{2 \pi i \tau} \frac{R_4^{(1,2)}}{4} \frac{1}{2 N^2-3}\sum_{n=0}^\infty  \frac{\Gamma \left(N-\frac{3}{2}-n\right)}{\sqrt{\pi }  \Gamma (N+2)(\pi y)^n}h_{4,n}^{(1,2)}(N)
\end{align}
where $h_{4,n}^{(1,2)}(N)$ are polynomials in $N$ of degree $4+n$. For the first few values of $n$ these are
\begin{align}
\begin{split}
h_{4,0}^{(1,2)}(N) & =-\frac{3N}{4}\(52 N^3-153 N^2+72 N+72\)\\
h_{4,1}^{(1,2)}(N) & =-\frac{9N}{64}\(52 N^4-713 N^3+2387 N^2-2088 N-480\)\\
h_{4,2}^{(1,2)}(N) & =\frac{45}{2^{11}}(N-2) N^2 \left(52 N^3+499 N^2-6357 N+13356\right)\\
h_{4,3}^{(1,2)}(N) & =-\frac{315}{2^{15}} N^2(N+1) \left(52 N^4-321 N^3+411 N^2+594 N-1296\right)
\end{split}
\end{align}
Finally, the $(2,2)$ component of the integrated correlator for $p=4$ was presented in \eqref{oneinstanton2244}. Consistently, we note that 
\begin{align}
4\mathcal{G}_4^{(2|1,1)}(\tau)\Big|_{\text{1-inst}}=2\mathcal{G}_4^{(2|1,2)}(\tau)\Big|_{\text{1-inst}}=\mathcal{G}_4^{(2|2,2)}(\tau)\Big|_{\text{1-inst}}\no\\
4\mathcal{G}_4^{(3|1,1)}(\tau)\Big|_{\text{1-inst}}=2\mathcal{G}_4^{(3|1,2)}(\tau)\Big|_{\text{1-inst}}=\mathcal{G}_4^{(3|2,2)}(\tau)\Big|_{\text{1-inst}}
\end{align}
which should be the case since (as observed for the correlator in the zero instanton sector in Section \ref{sec:spectral_overlaps}) $2T_4=T_{2,2}$ for $N=2,3$.

Likewise at $p=5$ we have the following results. For the $(1,1)$ component, we have the following  form for the one-instanton weak-coupling expansion
\begin{align}
\mathcal{G}_5^{(N|1,1)}(\tau)\Big|_{\text{1-inst}}=e^{2 \pi i \tau} \frac{R_5^{(1,1)}}{4} \frac{1}{N^4+24}\sum_{n=0}^\infty  \frac{\Gamma \left(N-\frac{3}{2}-n\right)}{\sqrt{\pi }  \Gamma (N+3)(\pi y)^n}h_{5,n}^{(1,1)}(N)
\end{align}
where $h_{5,n}^{(1,1)}(N)$ are polynomials in $N$ of degree $7+n$. For the first few values of $n$ these are
\begin{align}
\begin{split}
h_{5,0}^{(1,1)}(N) & =-\frac{15N}{2}\(16 N^6-76 N^5+42 N^4+243 N^3-48 N^2-336 N-144\)\\
h_{5,1}^{(1,1)}(N) & =-\frac{45N}{32}\(16 N^7-492 N^6+3094 N^5-5549 N^4-3333 N^3+9744 N^2+7056 N+960\)\\
h_{5,2}^{(1,1)}(N) & =\frac{225}{2^{10}}N^2 \left(16 N^7+580 N^6-17098 N^5+121349 N^4-299895 N^3+96582 N^2+338592 N+80304\right)\\
h_{5,3}^{(1,1)}(N) & =-\frac{1575}{2^{14}} N^2(N-3) \bigg(16 N^7+180 N^6+5462 N^5-145329 N^4+824655 N^3\\
&\qquad\qquad\qquad\qquad\qquad-1221822 N^2-645648 N-864\bigg)
\end{split}
\end{align}
For the $(1,2)$ component, the form of the correlator is
\begin{align}
\mathcal{G}_5^{(N|1,2)}(\tau)\Big|_{\text{1-inst}}=\mathcal{G}_5^{(N|2,1)}(\tau)\Big|_{\text{1-inst}}=e^{2 \pi i \tau} \frac{R_5^{(1,2)}}{4}  \frac{1}{ N^2-2 }\sum_{n=0}^\infty  \frac{\Gamma \left(N-\frac{3}{2}-n\right)}{\sqrt{\pi }  \Gamma (N+3)(\pi y)^n}h_{5,n}^{(1,2)}(N)
\end{align}
where $h_{5,n}^{(1,2)}(N)$ are polynomials in $N$ of degree $5+n$. For the first few values of $n$ these are
\begin{align}
\begin{split}
h_{5,0}^{(1,2)}(N) & =-\frac{3N}{2}\(32 N^4-84 N^3-31 N^2+140 N+60\)\\
h_{5,1}^{(1,2)}(N) & =-\frac{9N}{32}\(32 N^5-644 N^4+2589 N^3-1865 N^2-2940 N-400\)\\
h_{5,2}^{(1,2)}(N) & =\frac{45}{2^{10}}N^2 \left(32 N^5+684 N^4-14427 N^3+66091 N^2-78290 N-33460\right)\\
h_{5,3}^{(1,2)}(N) & =-\frac{315}{2^{14}} N^2(N-3) \left(32 N^5+156 N^4+3213 N^3-57321 N^2+161710 N+360\right)~.
\end{split}
\end{align}
And finally for the $(2,2)$ component, the form of the correlator is
\be
\mathcal{G}_5^{(N|2,2)}(\tau)\Big|_{\text{1-inst}}=e^{2 \pi i \tau} \frac{R_5^{(2,2)}}{4}\frac{1}{ N^2+5 }\sum_{n=0}^\infty  \frac{\Gamma \left(N-\frac{3}{2}-n\right)}{\sqrt{\pi }  \Gamma (N+3)(\pi y)^n}h_{5,n}^{(2,2)}(N)
\ee
where $h_{5,n}^{(2,2)}(N)$ are polynomials in $N$ of degree $5+n$. For the first few values of $n$ these are
\begin{align}
\begin{split}
h_{5,0}^{(2,2)}(N) & =-15 N (2 N+1) \left(N^3-N^2+4 N-15\right)\\
h_{5,1}^{(2,2)}(N) & =-\frac{45N}{16}\(2 N^5-19 N^4+37 N^3-180 N^2+492 N+100\)\\
h_{5,2}^{(2,2)}(N) & =\frac{225}{2^{9}}N^2 \left(2 N^5+13 N^4-154 N^3+541 N^2-2647 N+5719\right)\\
h_{5,3}^{(2,2)}(N) & =-\frac{1575}{2^{13}} N^2(N-3) (N+1) (N+2) \left(2 N^3-9 N^2+10 N+18\right)
\end{split}
\end{align}
Again, we note that the one-instanton results for the different components of the correlators are consistent with each other for low values of $N$
\begin{align}
\(\frac65\)^2\mathcal{G}_5^{(3|1,1)}(\tau)\Big|_{\text{1-inst}}=\(\frac65\)\mathcal{G}_5^{(3|1,2)}(\tau)\Big|_{\text{1-inst}}=\mathcal{G}_5^{(3|2,2)}(\tau)\Big|_{\text{1-inst}}\no\\
\(\frac65\)^2\mathcal{G}_5^{(4|1,1)}(\tau)\Big|_{\text{1-inst}}=\(\frac65\)\mathcal{G}_5^{(4|1,2)}(\tau)\Big|_{\text{1-inst}}=\mathcal{G}_5^{(4|2,2)}(\tau)\Big|_{\text{1-inst}}
\end{align}
which should be the case since $6T_5=5T_{2,3}$ for $N=3,4$.

\section{Matching to two-loop perturbation theory}\label{app:2LoopPerturbation}
In this appendix we demonstrate that the weak-coupling expansion of the integrated correlators $\mathcal{G}_p^{(N|i,j)}(\tau)$ obtained from the localisation computations matches the explicit integration of $\langle\O_2\O_2\O_p^{(i)}\O_p^{(j)}\rangle$ up to two-loop order in perturbation theory. For the $p=2$ case this has already been established in \cite{Dorigoni:2021guq},\footnote{
	See also the recent work \cite{Wen:2022oky} for a comparison up to four-loop order and interesting connections to periods of Feynman graphs.
	}
and here we extend this analysis to higher values of $p$.

In \cite{DAlessandro:2005fnh}, the family of (unintegrated) $\langle\O_2\O_2\O_p^{(i)}\O_p^{(j)}\rangle$ correlators has been computed using traditional Feynman diagram methods to order $g_{\text{YM}}^4$, with explicit data for the colour factors of different trace structures of the external operators $\O_p^{(i)}$ given up to $p=6$. To make use of the wealth of data contained in reference \cite{DAlessandro:2005fnh}, we first need to make contact with our conventions and recast their results into a manifestly $SO(6)$ covariant formulation. After applying a crossing transformation to bring the correlator into the `22pp' orientation, its full $y$-dependence is easily restored by simply uplifting the bosonic propagator factors $x_{ij}^2$ to superpropagators, i.e. by mapping $1/x_{ij}^2 \mapsto g_{ij}$. We then pull out the prefactor $g_{12}^2g_{34}^p$ and restore the factor of $\mathcal{I}$ to bring their results into the $SO(6)$ covariant form \eqref{eq:superconformal_constraint}, where the interacting part $\mathcal{H}_p^{(i,j)}$ of the correlator admits the weak-coupling expansion
\begin{align}
	\mathcal{H}_p^{(i,j)}(u,v;\tau) = g_{\text{YM}}^2\,\mathcal{H}_{p,\,\text{one-loop}}^{(i,j)}(u,v) + g_{\text{YM}}^4 \, \mathcal{H}_{p,\,\text{two-loop}}^{(i,j)}(u,v) + O(g_{\text{YM}}^6)\,.
\end{align}
The one- and two-loop contributions read
\begin{align}\label{eq:one_loop}
\begin{split}
	\mathcal{H}_{p,\,\text{one-loop}}^{(i,j)}(u,v) &= -2\,C_{1,p}^{(i,j)}(N)\cdot \frac{u}{v}\Phi^{(1)}(u,v)\,,\\
	\mathcal{H}_{p,\,\text{two-loop}}^{(i,j)}(u,v) &= N\,\frac{u}{v}\cdot\Bigg(\frac{1}{4}\Big[(1+v-u)\,C_{1,p}^{(i,j)}(N)+\frac{2u}{N}\,C_{2,p}^{(i,j)}(N)\Big]\big(\Phi^{(1)}(u,v)\big)^2\\
	&\qquad\qquad+C_{1,p}^{(i,j)}(N)\,\Big[\Phi^{(2)}(u,v)+\frac{1}{v}\Phi^{(2)}(\tfrac{1}{v},\tfrac{u}{v})\Big]\\
	&\qquad\qquad+\frac{1}{u}\Phi^{(2)}(\tfrac{1}{u},\tfrac{v}{u})\Big[C_{1,p}^{(i,j)}(N)+\frac{p-2}{N}\,C_{3,p}^{(i,j)}(N)\Big]\Bigg)\,,
\end{split}
\end{align}
where $\Phi^{(l)}(u,v)=-\frac{1}{x-\xb}\phi^{(l)}(x',\xb')$ with $\phi^{(l)}(x,\xb)$ being the well-known $l$-loop ladder integrals. The $C_{n,p}^{(i,j)}(N)$ with $n=1,2,3$ are certain colour-factors encoding the $N$-dependence of the one- and two-loop correlators. Their precise form depends on the trace structure of the external operators $\O_p^{(i)}$ and $\O_p^{(j)}$ and is given on a case-by-case basis in \cite{DAlessandro:2005fnh} (for $p=3$ see their eqs. (44)-(46), for $p=4$ see Table 2, and for $p=5,6$ see Tables 6 and 7, respectively).\footnote{\label{footnote:C}%
	In order to account for some differences in operator normalisations and conventions for Wick contractions, we simply compare their colour-factors of the free correlator and rescale their $C_{n,p}^{(i,j)}(N)$ such that the free theory answer matches ours, amounting to some factors of $4\pi^2$ and $p$.}

As a final step, we need to perform the integration over Euclidean space given by
\begin{align}
	I_2[f(u,v)] = -\frac{2}{\pi}\int_0^\infty dr\int_0^\pi d\theta\,\frac{r^3\sin^2\theta}{u^2}\,f(u,v)\vert_{u=1+r^2-2r\cos\theta,\,v=r^2}\,,
\end{align}
which defines the integrated correlator, recall \eqref{eq:integrated_correlator}. This amounts to integrating the various orientations of the one- and two-loop ladder integrals $\Phi^{(l)}(u,v)$ present in \eqref{eq:one_loop}. As shown in \cite{Dorigoni:2021guq}, one has\footnote{
	Note that we have a relative factor of 4 difference in the definition of the integral $I_2$ compared to reference \cite{Dorigoni:2021guq}, namely $I_2[f(u,v)]|_{\text{here}}=\frac{1}{4}I_2[f(u,v)]|_{\text{there}}$.}
\begin{align}\label{eq:I2_ladders}
	I_2\Big[\frac{u}{v}\Phi^{(l)}(u,v)\Big]=I_2\Big[\frac{u}{v}\Phi^{(l_1)}(u,v)\Phi^{(l_2)}(u,v)\Big]=-\frac{1}{2}\binom{2l+2}{l+1}\,\zeta(2l+1),\quad \text{for }l_1+l_2=l\,.
\end{align}
The one-loop contribution is thus straightforwardly integrated by specialising the above formula to $l=1$, which yields
\begin{align}\label{eq:I2_one-loop}
	I_2\Big[\frac{u}{v}\Phi^{(1)}(u,v)\Big] = -3\zeta(3)\,.
\end{align}
On the other hand, upon closer inspection of the two-loop correlator in \eqref{eq:one_loop} we note that six different position space structures are present, given by $\big(\Phi^{(1)}(u,v)\big)^2$ with different prefactors as well as the three orientations of $\Phi^{(2)}(u,v)$. Importantly, we find that all of the six different structures \textit{individually} integrate to the same constant obtained by setting $l=2$ in \eqref{eq:I2_ladders}, and one has
\begin{align}\label{eq:I2_two-loop}
\begin{split}
	I_2\Big[\frac{u}{v}\,\big(\Phi^{(1)}(u,v)\big)^2\Big] &= I_2\Big[u\,\big(\Phi^{(1)}(u,v)\big)^2\Big] = I_2\Big[\frac{u^2}{v}\,\big(\Phi^{(1)}(u,v)\big)^2\Big] \\
	= I_2\Big[\frac{u}{v}\,\Phi^{(2)}(u,v)\Big] &= I_2\Big[\frac{u}{v^2}\,\Phi^{(2)}(\tfrac{1}{v},\tfrac{u}{v})\Big] = I_2\Big[\frac{1}{v}\,\Phi^{(2)}(\tfrac{1}{u},\tfrac{v}{u})\Big] = -10\zeta(5)\,,
\end{split}
\end{align}
which essentially follows from the fact that these structures are related by crossing transformations and thus become indistinguishable from each other after performing the integration.

With the integrals \eqref{eq:I2_one-loop} and \eqref{eq:I2_two-loop} in place, one can easily assemble the one- and two-loop contributions to the integrated correlator (as mentioned above, to account for different conventions we have rescaled the colour-factors $C_{n,p}^{(i,j)}(N)$ from \cite{DAlessandro:2005fnh} so that the free theory correlators match, see also footnote \ref{footnote:C}). We have then compared this with the first two orders of the $1/y$ expansion obtained from the localisation integral \eqref{eq:localisation_relation}, finding perfect agreement for the cases $p=3,4,5$ for all trace configurations $(i,j)$, as well as for the $p=6$ maximal-trace correlator!

The agreement we find provides further evidence that the localisation relation \eqref{eq:localisation_relation} derived in \cite{Binder:2019jwn} not only computes integrated correlators of single-trace operators, but can be extended to multi-trace operators and thus allows us to access the entire set of integrated correlators $\mathcal{G}_p^{(N|i,j)}(\tau)$. In particular, it shows that our prescription to incorporate such higher-trace insertions in the matrix model as explained around \eqref{muindex} is indeed correct.

Furthermore, this matching also indicates that we have correctly solved the operator mixing problem on $S^4$, which we performed explicitly up to $p=5$ in Appendix \ref{denominator} and for the maximal-trace correlators for all even $p$ in Section \ref{sec:maximal_trace}, thus extending the results of \cite{Binder:2019jwn} beyond the single-trace sector and to finite $N$.

\sec{Consistency check: Ensemble average vs. genus expansion}\label{appcheck}

One may perform a consistency check of our results by demonstrating equality between the large $N$ expansion of the ensemble averages $\langle\mathcal{G}_p^{(N)}\rangle$ and the genus expansion of the spectral overlaps $f_p(s)$. This is instructive because the match involves not just the leading large $N$ limit of $\langle\mathcal{G}_p^{(N)}\rangle$ at every genus, but the subleading terms as well.

It follows from \cite{Collier:2022emf}, eq. (10.3), that the large $N$ expansion of $\langle\cG_p^{(N)} \rangle$ may be reconstructed from the genus expansion of the spectral overlaps as
\begin{align}\label{eq:large_N_average}
	\langle\cG_p^{(N)} \rangle = \frac{1}{2}\sum_{g=0}^\infty N^{p-2g}\Big[f_p^{(g)}(1)+N^{-1}f_p^{(g)}(0)\Big],
\end{align}
Note that, in terms of the quantity $\llangle\cG_p^{(g)}\rrangle$ defined earlier, $\llangle\cG_p^{(g)}\rrangle = \half f_p^{(g)}(1)$. Equation \eqr{eq:large_N_average} implies in turn that the $1/N$ expansion truncates at genus-$g^*$ if $f_p^{(g)}(1)=f_p^{(g)}(0)=0$ for all $g>g^*$. 

As one can see from the polynomials $n^{(g)}_p(s)$ given in the previous section, this truncation holds for $p=2$ and $p=3$ with thresholds $g^*=0$ and $g^*=1$, respectively. For example, specialising \eqref{eq:large_N_average} to the $p=2$ case and recalling the genus-zero overlap from \eqref{eq:genus_0_p=2}, one has
\begin{align}
	\langle \mathcal{G}_2^{(N)} \rangle = \frac{1}{2}\Big(N^2f_2^{(0)}(1)+N f_2^{(0)}(0)\Big) =\frac{N(N-1)}{4},
\end{align}
which indeed reproduces the correct $p=2$ ensemble average using only the genus-zero contribution $f_2^{(0)}(s)$. A similar check exists for $\langle \mathcal{G}_3^{(N)} \rangle$ given in \eqref{eq:ensemble_average_2233}.

On the other hand, $\langle \mathcal{G}_4^{(N)} \rangle$ and $\langle \mathcal{G}_5^{(N)} \rangle$ in \eqref{eq:ensemble_average_2244_2255_SPO} have polynomial $N$-dependent denominators, giving rise to a $1/N$ expansion which does not truncate. This is consistent with the higher-genus spectral overlaps given in the previous section (cf. \eqr{eq:higher_genus_overlaps}), which do not vanish at $s=0,1$. Nevertheless, one can still verify that \eqref{eq:large_N_average} correctly reproduces the finite $N$ averages \eqref{eq:ensemble_average_2244_2255_SPO} term by term in a $1/N$ expansion, which we have checked up to genus four.\foot{We note that the  averaged correlators $\<\cG_p^{(N|i,j)}\>$, i.e. those {\it not} projected onto the SPO basis, {\it do} admit such a truncation, as seen in equations \eqr{eq:ensemble_averages_2244}-\eqref{eq:ensemble_averages_2255}. Why these truncations exist is not clear to us.}

\section{More on the 't Hooft limit}
\label{BesselSpectralEquivalence}
In this appendix we give several details on the 't Hooft limit and its interplay with the $SL(2,\mathbb{Z})$ spectral decomposition which were skipped in the main text.
\subsection{Genus-zero spectral overlap for general $p$}
First we derive the genus-zero, generic $p$ spectral overlap \eqref{eq:genus_0_overlaps}. The idea is to make use of the known genus-zero correlator for generic $p$, recast it into the spectral form and read off the overlap. The genus-zero correlator is given by 
\begin{align}\label{eq:bessel}
	\mathcal{G}_p^{(\mathfrak{g}=0)}(\lambda) = \int_0^\infty d\omega~\omega\,\frac{J_1\big(\frac{\sqrt{\lambda}}{\pi}\omega\big)^2-J_p\big(\frac{\sqrt{\lambda}}{\pi}\omega\big)^2}{\sinh^2(\omega)}\,.
\end{align}
which was first obtained in \cite{Binder:2019jwn} through large $N$ matrix model techniques. We want to write the above expression into the spectral form in \eqref{eq:genus_0_correlator}, which we rewrite more compactly below
\begin{align}
\label{eq:genus_0_correlator_again}
	\mathcal{G}_{p,0}^{(\mathfrak{g}=0)}(\lambda) = \frac{1}{2\pi i}\int_{\text{Re}\,s=1+\epsilon} ds~\frac{\pi}{\sin(\pi s)}\,s(1-s)\,\Lambda(1-s)\,\(\frac{\lambda}{4\pi}\)^{s-1}\,f_p^{(0)}(s)~.
\end{align}
Note that the integration contour is now anchored just beyond the $s=1$ pole that comes from $\Lambda(1-s)$. This form absorbs the first term in \eqref{eq:genus_0_correlator} into the $s$ integral. Next, we make use of the following Mellin-Barnes representation for the Bessel function:
\be
\label{BesselMBRep}
J_m(x) J_n(x) = \frac{1}{2\pi i}\int_{t_0-i\infty}^{t_0+i\infty}dt \frac{\Gamma (-t) \Gamma (m+n+2 t+1)}{\Gamma (m+t+1) \Gamma (n+t+1) \Gamma (m+n+t+1)} \left(\frac{x}{2}\right)^{m+n+2 t}~,
\ee
which holds for $x>0$. The contour running parallel to the imaginary axis is anchored at a point $t_0$ such that the all poles of $\Gamma(-t)$ and all poles of $\Gamma(m+n+2t+1)$ are on opposite sides of $t_0$.  Making use of the identity 
\e{}{\int_{0}^{\infty} d x \frac{x^{\a}}{\sinh ^{2} x}=\(\frac12\)^{\a-1}\Gamma(\a+1) \zeta(\a)~,}
we have
\begin{align}
\int_{0}^{\infty} d x \frac{x^{\a}y^{\a-1}}{\sinh ^{2} x}J_m(xy) J_n(xy) = -\frac{1}{2\pi i}&\int_{t_0-i\infty}^{t_0+i\infty}dt \(\frac{\Gamma (m+n+2 t+1)}{\Gamma (t+1)\Gamma (m+t+1) \Gamma (n+t+1) \Gamma (m+n+t+1)}\)\no\\
y^{m+n+2t+\a-1}\frac{\pi }{\sin (\pi  t)}&\left(\frac{1}{2}\right)^{2m+2n+4 t+\a-1} \Gamma(m+n+2t+\a+1) \zeta(m+n+2t+\a)~.
\end{align}
Plugging this representation into \eqref{eq:bessel}, combining the two Mellin-Barnes integrals in the resulting expression into one by a change of variables and making use of the functional equation for the Riemann zeta function,
\be 
\zeta (s)= 2^s \pi ^{s-1} \zeta (1-s) \sin \left(\frac{\pi  s}{2}\right) \Gamma (1-s)\,,
\ee 
we find the spectral form \eqref{eq:genus_0_correlator_again} from which the spectral overlap $f_p^{(0)}(s)$ is extracted. 

\subsection{Genus-one correlators for various values of $p$}\label{appf2}
Following the same strategy as in the previous subsection, we record Bessel representations for genus-one SPO correlators $\mathcal{G}_{p}^{(\mathfrak{g}=1)}(\lambda)$ for $p\leq 5$, and for the maximal trace family $\widehat{\mathcal{G}}_{p}^{(\mathfrak{g}=1|\text{max})}(\lambda)$ for all $p$. 

We start with the SPOs. The result for $p=3$ was presented in \eqref{p3g1overlap}. Upon performing the $k$ sum in \eqref{eq:genus_g_correlator2} by making use of \eqref{ZetaIdentity} we find
\begin{align}
\mathcal{G}_{2}^{(\mathfrak{g}=1)}(\lambda)=&\int_{0}^{\infty} d \omega\frac{-\lambda  \omega ^3}{48 \pi ^3\sinh^2\omega} \bigg[12 \pi  J_0\big(\tfrac{\sqrt{\lambda}}{\pi }\omega\big){}^2+2 \sqrt{\lambda } \omega  J_1\big(\tfrac{\sqrt{\lambda}}{\pi }\omega\big) J_0\big(\tfrac{\sqrt{\lambda}}{\pi }\omega\big)+5 \pi  J_1\big(\tfrac{\sqrt{\lambda}}{\pi }\omega\big){}^2\bigg]
\end{align}
\begin{align}
\mathcal{G}_{4}^{(\mathfrak{g}=1)}(\lambda)=&-\int_{0}^{\infty} \frac{d \omega}{48 \pi ^3 \lambda ^2 \omega ^3\sinh^2\omega}\bigg[6 \pi  \lambda  \omega ^2 \left(7 \lambda ^2 \omega ^4+152 \pi ^2 \lambda  \omega ^2-2688 \pi ^4\right) J_0\big(\tfrac{\sqrt{\lambda}}{\pi }\omega\big){}^2\no\\
&+2 \sqrt{\lambda } \omega  \left(\lambda ^3 \omega ^6+156 \pi ^2 \lambda ^2 \omega ^4-5856 \pi ^4 \lambda  \omega ^2+32256 \pi ^6\right) J_0\big(\tfrac{\sqrt{\lambda}}{\pi }\omega\big) J_1\big(\tfrac{\sqrt{\lambda}}{\pi }\omega\big)\\
&+\pi  \left(47 \lambda ^3 \omega ^6-1080 \pi ^2 \lambda ^2 \omega ^4+19776 \pi ^4 \lambda  \omega ^2-64512 \pi ^6\right) J_1\big(\tfrac{\sqrt{\lambda}}{\pi }\omega\big){}^2\bigg]\no
\end{align}
\begin{align}
\mathcal{G}_{5}^{(\mathfrak{g}=1)}(\lambda)=&-\int_{0}^{\infty} \frac{d \omega}{4 \pi ^2 \lambda ^3 \omega ^5\sinh^2\omega}\bigg[\lambda  \omega ^2 \left(5 \lambda ^3 \omega ^6-416 \pi ^2 \lambda ^2 \omega ^4+16512 \pi ^4 \lambda  \omega ^2-147456 \pi ^6\right) J_0\big(\tfrac{\sqrt{\lambda}}{\pi }\omega\big){}^2\no\\
&+4 \pi  \sqrt{\lambda } \omega  \left(-5 \lambda ^3 \omega ^6+1712 \pi ^2 \lambda ^2 \omega ^4-34944 \pi ^4 \lambda  \omega ^2+147456 \pi ^6\right) J_0\big(\tfrac{\sqrt{\lambda}}{\pi }\omega\big) J_1\big(\tfrac{\sqrt{\lambda}}{\pi }\omega\big)\no\\
&+\left(7 \lambda ^4 \omega ^8+424 \pi ^2 \lambda ^3 \omega ^6-21248 \pi ^4 \lambda ^2 \omega ^4+213504 \pi ^6 \lambda  \omega ^2-589824 \pi ^8\right) J_1\big(\tfrac{\sqrt{\lambda}}{\pi }\omega\big){}^2\bigg]
\end{align}
Correlators at higher-genus for all $p$ can be obtained in principle in a similar way. 

Likewise, the result for the normalised genus-one maximal trace correlator $\widehat{\mathcal{G}}_{p}^{(\mathfrak{g}=1|\text{max})}(\lambda)$ is
\begin{align}
\begin{split}
	\widehat{\mathcal{G}}_{p}^{(\mathfrak{g}=1|\text{max})}(\lambda) &=\int_0^\infty d\omega~\frac{p}{48 \pi ^3 \lambda \omega \sinh^2 \omega}\bigg[12\pi\lambda \omega^2\big((p-3)\lambda w^2-4\pi ^2\big)J_0\big(\tfrac{\sqrt{\lambda}}{\pi }\omega\big)^2\\[3pt]
	&\qquad\qquad\qquad-2\sqrt{\lambda } w \big(18\pi^2(p-2)\lambda\omega^2+\lambda^2 w^4-96\pi^4\big)J_0\big(\tfrac{\sqrt{\lambda}}{\pi }\omega\big)J_1\big(\tfrac{\sqrt{\lambda}}{\pi }\omega\big)\\[3pt]
	&\qquad\qquad\qquad-\pi\big((12p-19)\lambda^2\omega^4-24\pi^2p\lambda\omega^2+192\pi^4\big)J_1\big(\tfrac{\sqrt{\lambda}}{\pi }\omega\big)^2\bigg].
\end{split}
\end{align}
This is equivalent to \eqr{eq:genus_1_correlator_normalised}. 

\subsection{Proof of \eqref{eq:genus_g_correlator1}}
\label{App:F1}
Here we give the proof of \eqref{eq:genus_g_correlator1}.

Plugging \eqref{eq:large_N_average} for the average in \eqref{eq:spectral_decomp_large_N}, the spectral decomposition of integrated correlators in a genus expansion is given by 
\begin{align}\label{eq:spectral_decomp_large_N1}
	\mathcal{G}_{p}^{(N)}(\lambda) =&~ \frac{1}{2}\sum_{g=0}^\infty N^{p-2g}\Big[f_p^{(g)}(1)+N^{-1}f_p^{(g)}(0)\Big]\\
	 &+ \frac{1}{2\pi i}\int_{\text{Re}\,s=\frac{1}{2}}ds \frac{\pi}{\sin(\pi s)}s(1-s)\sum_{g=0}^\infty N^{p-2g}\Big[\Lambda(1-s)\tilde{\lambda}^{s-1}+\Lambda(s)N^{2s-1}\tilde{\lambda}^{-s}\Big]f_p^{(g)}(s)~,\no
\end{align}
We can absorb the first line into the second at the cost of introducing two shifted contour integrals as follows:
\begin{align}
\mathcal{G}_{p}^{(N)}(\lambda) =&\frac{1}{2\pi i}\int_{\text{Re}\,s=1+\epsilon}ds \frac{\pi}{\sin(\pi s)}s(1-s)\sum_{g=0}^\infty N^{p-2g}\Lambda(1-s)\tilde{\lambda}^{s-1}f_p^{(g)}(s) \\
	 &+ \frac{1}{2\pi i}\int_{\text{Re}\,s=0-\epsilon}ds \frac{\pi}{\sin(\pi s)}s(1-s)\sum_{m=0}^\infty N^{p-2m+2s-1}\Lambda(s)\tilde{\lambda}^{-s}f_p^{(m)}(s)~,\no
\end{align}
where $\epsilon>0$ but infinitesimal. Noting that $f_p^{(m)}(s)$ has simple poles at $s=-\frac12 -n$ for $n\in\Z_{\geq0}$ and for all $m\geq 0$, commuting the $s$ integral with the genus expansion, the second line above simplifies to the following sum over residues
\begin{align}
\sum_{m=0}^\infty \sum_{n=0}^\infty N^{p-2-2(m+n)} \tilde{\lambda}^{n+\frac12} \underset{s=-\frac{1}{2}-n}{\operatorname{Res}}\left[\frac{\pi}{\sin \pi s} s(1-s) \Lambda(s) f_{{p}}^{(m)}(s)\right]
\end{align}
We now introduce the $\mg$ variable,
\be
\mg := m+n+1\,.
\ee
Note that $\mg > n$. Rewriting the previous sum in terms of $\mg$ and $n$ we get
\begin{align}
\sum_{\mathfrak{g}=1}^\infty N^{p-2\mathfrak{g}} \sum_{n=0}^{\mg-1}\tilde{\lambda}^{n+\frac12} \underset{s=-\frac{1}{2}-n}{\operatorname{Res}}\left[\frac{\pi}{\sin \pi s} s(1-s) \Lambda(s) f_{{p}}^{(\mg-n-1)}(s)\right]
\end{align}
Note that we can extend the above sum to $\mg=0$, which vanishes because of the $n$ sum. Therefore $\mathcal{G}_{p}^{(N)}(\lambda)$ now becomes
\begin{align}
\mathcal{G}_{p}^{(N)}(\lambda) =&\sum_{\mg=0}^\infty N^{p-2\mg}\bigg(\frac{1}{2\pi i}\int_{\text{Re}\,s=1+\epsilon}ds \frac{\pi}{\sin(\pi s)}s(1-s)\Lambda(1-s)\tilde{\lambda}^{s-1}f_p^{(\mg)}(s) \\
	 &+ \sum_{n=0}^{\mg-1}\tilde{\lambda}^{n+\frac12} \underset{s=-\frac{1}{2}-n}{\operatorname{Res}}\left[\frac{\pi}{\sin \pi s} s(1-s) \Lambda(s) f_{{p}}^{(\mg-n-1)}(s)\right]\bigg)~,\no
\end{align}
where we commuted the $s$ integral with the genus expansion and changed the dummy summation index from $g$ to $\mg$ in the first line. Comparing with (the perturbative part of) \eqref{genusexp} we have 
\begin{align}
\label{temp1}
\mathcal{G}_{p}^{(\mathfrak{g})}(\lambda)= &\frac{1}{2\pi i}\int_{\text{Re}\,s=1+\epsilon}ds \frac{\pi}{\sin(\pi s)}s(1-s)\Lambda(1-s)\tilde{\lambda}^{s-1}f_p^{(\mg)}(s) \no \\[5pt]
	 &+ \sum_{n=0}^{\mg-1}\tilde{\lambda}^{n+\frac12} \underset{s=-\frac{1}{2}-n}{\operatorname{Res}}\left[\frac{\pi}{\sin \pi s} s(1-s) \Lambda(s) f_{{p}}^{(\mg-n-1)}(s)\right]
\end{align}
The above equation says that a genus-$\mathfrak{g}$ correlator includes such renormalization terms from all genera $g<\mathfrak{g}$.

Next we ask, can we absorb the second line into the first by shifting the integration contour? The answer is yes, but under certain conditions on the overlaps.

First note that $f_{{p}}^{(\mg)}(s)$ has a finite number of poles at positive half integers,
\be
s=\frac32+n ~,\qquad n=0,1,2,...,\mg-1~,
\ee
as follows from \eqref{eq:higher_genus_overlaps}. For $\mg=0$ there are no such poles.  Therefore we can shift the integration contour in the first line of \eqref{temp1} towards the right over all the positive poles of $f_{{p}}^{(\mg)}(s)$ mentioned above. In doing so we encounter the positive integer poles of $\sin(\pi s)$. Therefore we need to subtract the residue of these integer $\sin$ poles. When all is said and done we land at the following expression for $\mg>0$:
\begin{align}
\label{temp2}
\mathcal{G}_{p}^{(\mathfrak{g}>0)}(\lambda)= &\frac{1}{2\pi i}\int_{\text{Re}\,s=\mg+\frac12+\epsilon}ds \frac{\pi}{\sin(\pi s)}s(1-s)\Lambda(1-s)\tilde{\lambda}^{s-1}f_p^{(\mg)}(s) \no \\[5pt]
	 &+ \sum_{n=0}^{\mg-1}\tilde{\lambda}^{n+\frac12} \underset{s=-\frac{1}{2}-n}{\operatorname{Res}}\left[\frac{\pi}{\sin \pi s} s(1-s) \Lambda(s) f_{{p}}^{(\mg-n-1)}(s)\right]\no\\[5pt]
	 &- \sum_{n=0}^{\mg-1}\tilde{\lambda}^{n+\frac12} \underset{s=\frac{3}{2}+n}{\operatorname{Res}}\left[\frac{\pi}{\sin \pi s} s(1-s) \Lambda(1-s) f_{{p}}^{(\mg)}(s)\right]\no\\[5pt]
	 &- \sum_{k=2}^{\mg}(-1)^{k+1}\tilde{\lambda}^{k-1}k(k-1) \Lambda(1-k) f_{{p}}^{(\mg)}(k)
\end{align}
Here we note the condition for consistency of the weak-coupling expansion, which must not have half-integer powers of $\lambda$. This implies 
\be
\underset{s=-\frac{1}{2}-n}{\operatorname{Res}}\left[\frac{\pi}{\sin \pi s} s(1-s) \Lambda(s) f_{{p}}^{(\mg-n-1)}(s)\right]  = \underset{s=\frac{3}{2}+n}{\operatorname{Res}}\left[\frac{\pi}{\sin \pi s} s(1-s) \Lambda(1-s) f_{{p}}^{(\mg)}(s)\right]
\ee
which is precisely eq. (7.9) of \cite{Collier:2022emf}. Hence, this consistency condition implies our final result, a spectral representation of the genus-$\mg$ correlator purely in terms of the genus-$\mg$ spectral overlap:
\begin{align}
\label{temp3}
\mathcal{G}_{p}^{(\mathfrak{g}>0)}(\lambda)= &\frac{1}{2\pi i}\int_{\text{Re}\,s=\mg+\frac12+\epsilon}ds \frac{\pi}{\sin(\pi s)}s(1-s)\Lambda(1-s)\tilde{\lambda}^{s-1}f_p^{(\mg)}(s) \no \\[5pt]
	 &- \sum_{k=2}^{\mg}(-1)^{k+1}\tilde{\lambda}^{k-1}k(k-1) \Lambda(1-k) f_{{p}}^{(\mg)}(k)
\end{align}
This is \eqr{eq:genus_g_correlator1}. At genus one, the second line is not present! However, for higher genera, $f_{{p}}^{(\mg)}(k)$ need not vanish for integer $k\geq 2$. This is the penalty we pay for having a form like \eqref{temp3} for the correlator $\mathcal{G}_{p}^{(\mathfrak{g})}(\lambda)$. Note that the above expression is valid for $\mg>0$. For $\mg=0$, we instead have
\begin{align}
\label{temp30}
\mathcal{G}_{p}^{(0)}(\lambda)= &\frac{1}{2\pi i}\int_{\text{Re}\,s=1+\epsilon}ds \frac{\pi}{\sin(\pi s)}s(1-s)\Lambda(1-s)\tilde{\lambda}^{s-1}f_p^{(0)}(s)
\end{align}

\bibliographystyle{JHEP}
\bibliography{22pp_bib}

\providecommand{\href}[2]{#2}\begingroup\raggedright\begin{thebibliography}{10}

\bibitem{Montonen:1977sn}
C.~Montonen and D.~I. Olive, {\it {Magnetic Monopoles as Gauge Particles?}},
  {\em Phys. Lett. B} {\bf 72} (1977) 117--120.

\bibitem{Witten:1978mh}
E.~Witten and D.~I. Olive, {\it {Supersymmetry Algebras That Include
  Topological Charges}},  {\em Phys. Lett. B} {\bf 78} (1978) 97--101.

\bibitem{Osborn:1979tq}
H.~Osborn, {\it {Topological Charges for N=4 Supersymmetric Gauge Theories and
  Monopoles of Spin 1}},  {\em Phys. Lett. B} {\bf 83} (1979) 321--326.

\bibitem{Argyres:2006qr}
P.~C. Argyres, A.~Kapustin, and N.~Seiberg, {\it {On S-duality for
  non-simply-laced gauge groups}},  {\em JHEP} {\bf 06} (2006) 043,
  [\href{http://arxiv.org/abs/hep-th/0603048}{{\tt hep-th/0603048}}].

\bibitem{Aharony:2013hda}
O.~Aharony, N.~Seiberg, and Y.~Tachikawa, {\it {Reading between the lines of
  four-dimensional gauge theories}},  {\em JHEP} {\bf 08} (2013) 115,
  [\href{http://arxiv.org/abs/1305.0318}{{\tt arXiv:1305.0318}}].

\bibitem{Binder:2019jwn}
D.~J. Binder, S.~M. Chester, S.~S. Pufu, and Y.~Wang, {\it {$ \mathcal{N} $ = 4
  Super-Yang-Mills correlators at strong coupling from string theory and
  localization}},  {\em JHEP} {\bf 12} (2019) 119,
  [\href{http://arxiv.org/abs/1902.06263}{{\tt arXiv:1902.06263}}].

\bibitem{Dorigoni:2021guq}
D.~Dorigoni, M.~B. Green, and C.~Wen, {\it {Exact properties of an integrated
  correlator in $ \mathcal{N} $ = 4 SU(N) SYM}},  {\em JHEP} {\bf 05} (2021)
  089, [\href{http://arxiv.org/abs/2102.09537}{{\tt arXiv:2102.09537}}].

\bibitem{Collier:2022emf}
S.~Collier and E.~Perlmutter, {\it {Harnessing S-duality in $ \mathcal{N} $ = 4
  SYM \& Supergravity as SL(2, \ensuremath{\mathbb{Z}})-Averaged Strings}},
  {\em JHEP} {\bf 08} (2022) 195, [\href{http://arxiv.org/abs/2201.05093}{{\tt
  arXiv:2201.05093}}].

\bibitem{DAlessandro:2005fnh}
M.~D'Alessandro and L.~Genovese, {\it {A Wide class of four point functions of
  BPS operators in N=4 SYM at order g**4}},  {\em Nucl. Phys. B} {\bf 732}
  (2006) 64--88, [\href{http://arxiv.org/abs/hep-th/0504061}{{\tt
  hep-th/0504061}}].

\bibitem{Drummond:2013nda}
J.~Drummond, C.~Duhr, B.~Eden, P.~Heslop, J.~Pennington, and V.~A. Smirnov,
  {\it {Leading singularities and off-shell conformal integrals}},  {\em JHEP}
  {\bf 08} (2013) 133, [\href{http://arxiv.org/abs/1303.6909}{{\tt
  arXiv:1303.6909}}].

\bibitem{Eden:2012tu}
B.~Eden, P.~Heslop, G.~P. Korchemsky, and E.~Sokatchev, {\it {Constructing the
  correlation function of four stress-tensor multiplets and the four-particle
  amplitude in N=4 SYM}},  {\em Nucl. Phys. B} {\bf 862} (2012) 450--503,
  [\href{http://arxiv.org/abs/1201.5329}{{\tt arXiv:1201.5329}}].

\bibitem{Fleury:2019ydf}
T.~Fleury and R.~Pereira, {\it {Non-planar data of $ \mathcal{N} $ = 4 SYM}},
  {\em JHEP} {\bf 03} (2020) 003, [\href{http://arxiv.org/abs/1910.09428}{{\tt
  arXiv:1910.09428}}].

\bibitem{Chicherin:2015edu}
D.~Chicherin, J.~Drummond, P.~Heslop, and E.~Sokatchev, {\it {All three-loop
  four-point correlators of half-BPS operators in planar $ \mathcal{N} $ = 4
  SYM}},  {\em JHEP} {\bf 08} (2016) 053,
  [\href{http://arxiv.org/abs/1512.02926}{{\tt arXiv:1512.02926}}].

\bibitem{Chicherin:2018avq}
D.~Chicherin, A.~Georgoudis, V.~Goncalves, and R.~Pereira, {\it {All five-loop
  planar four-point functions of half-BPS operators in $\mathcal N=4$ SYM}},
  {\em JHEP} {\bf 11} (2018) 069, [\href{http://arxiv.org/abs/1809.00551}{{\tt
  arXiv:1809.00551}}].

\bibitem{Bourjaily:2016evz}
J.~L. Bourjaily, P.~Heslop, and V.-V. Tran, {\it {Amplitudes and Correlators to
  Ten Loops Using Simple, Graphical Bootstraps}},  {\em JHEP} {\bf 11} (2016)
  125, [\href{http://arxiv.org/abs/1609.00007}{{\tt arXiv:1609.00007}}].

\bibitem{Arutyunov:2000py}
G.~Arutyunov and S.~Frolov, {\it {Four point functions of lowest weight CPOs in
  N=4 SYM(4) in supergravity approximation}},  {\em Phys. Rev. D} {\bf 62}
  (2000) 064016, [\href{http://arxiv.org/abs/hep-th/0002170}{{\tt
  hep-th/0002170}}].

\bibitem{Dolan:2006ec}
F.~A. Dolan, M.~Nirschl, and H.~Osborn, {\it {Conjectures for large N
  superconformal N=4 chiral primary four point functions}},  {\em Nucl. Phys.
  B} {\bf 749} (2006) 109--152,
  [\href{http://arxiv.org/abs/hep-th/0601148}{{\tt hep-th/0601148}}].

\bibitem{Uruchurtu:2008kp}
L.~I. Uruchurtu, {\it {Four-point correlators with higher weight superconformal
  primaries in the AdS/CFT Correspondence}},  {\em JHEP} {\bf 03} (2009) 133,
  [\href{http://arxiv.org/abs/0811.2320}{{\tt arXiv:0811.2320}}].

\bibitem{Uruchurtu:2011wh}
L.~I. Uruchurtu, {\it {Next-next-to-extremal Four Point Functions of N=4 1/2
  BPS Operators in the AdS/CFT Correspondence}},  {\em JHEP} {\bf 08} (2011)
  133, [\href{http://arxiv.org/abs/1106.0630}{{\tt arXiv:1106.0630}}].

\bibitem{Rastelli:2016nze}
L.~Rastelli and X.~Zhou, {\it {Mellin amplitudes for $AdS_5\times S^5$}},  {\em
  Phys. Rev. Lett.} {\bf 118} (2017), no.~9 091602,
  [\href{http://arxiv.org/abs/1608.06624}{{\tt arXiv:1608.06624}}].

\bibitem{Rastelli:2017udc}
L.~Rastelli and X.~Zhou, {\it {How to Succeed at Holographic Correlators
  Without Really Trying}},  {\em JHEP} {\bf 04} (2018) 014,
  [\href{http://arxiv.org/abs/1710.05923}{{\tt arXiv:1710.05923}}].

\bibitem{Arutyunov:2018tvn}
G.~Arutyunov, R.~Klabbers, and S.~Savin, {\it {Four-point functions of 1/2-BPS
  operators of any weights in the supergravity approximation}},  {\em JHEP}
  {\bf 09} (2018) 118, [\href{http://arxiv.org/abs/1808.06788}{{\tt
  arXiv:1808.06788}}].

\bibitem{Caron-Huot:2018kta}
S.~Caron-Huot and A.-K. Trinh, {\it {All tree-level correlators in
  AdS$_{5}\times$ S$_{5}$ supergravity: hidden ten-dimensional conformal
  symmetry}},  {\em JHEP} {\bf 01} (2019) 196,
  [\href{http://arxiv.org/abs/1809.09173}{{\tt arXiv:1809.09173}}].

\bibitem{Alday:2018pdi}
L.~F. Alday, A.~Bissi, and E.~Perlmutter, {\it {Genus-One String Amplitudes
  from Conformal Field Theory}},  {\em JHEP} {\bf 06} (2019) 010,
  [\href{http://arxiv.org/abs/1809.10670}{{\tt arXiv:1809.10670}}].

\bibitem{Drummond:2019odu}
J.~M. Drummond, D.~Nandan, H.~Paul, and K.~S. Rigatos, {\it {String corrections
  to AdS amplitudes and the double-trace spectrum of $ \mathcal{N} $ = 4 SYM}},
   {\em JHEP} {\bf 12} (2019) 173, [\href{http://arxiv.org/abs/1907.00992}{{\tt
  arXiv:1907.00992}}].

\bibitem{Drummond:2020dwr}
J.~M. Drummond, H.~Paul, and M.~Santagata, {\it {Bootstrapping string theory on
  AdS$_5 \times S^5$}},  \href{http://arxiv.org/abs/2004.07282}{{\tt
  arXiv:2004.07282}}.

\bibitem{Abl:2020dbx}
T.~Abl, P.~Heslop, and A.~E. Lipstein, {\it {Towards the Virasoro-Shapiro
  amplitude in AdS$_{5} \times S^{5}$}},  {\em JHEP} {\bf 04} (2021) 237,
  [\href{http://arxiv.org/abs/2012.12091}{{\tt arXiv:2012.12091}}].

\bibitem{Aprile:2020mus}
F.~Aprile, J.~M. Drummond, H.~Paul, and M.~Santagata, {\it {The
  Virasoro-Shapiro amplitude in AdS$_{5}\times$ S$^{5}$ and level splitting of
  10d conformal symmetry}},  {\em JHEP} {\bf 11} (2021) 109,
  [\href{http://arxiv.org/abs/2012.12092}{{\tt arXiv:2012.12092}}].

\bibitem{Alday:2022uxp}
L.~F. Alday, T.~Hansen, and J.~A. Silva, {\it {AdS Virasoro-Shapiro from
  dispersive sum rules}},  \href{http://arxiv.org/abs/2204.07542}{{\tt
  arXiv:2204.07542}}.

\bibitem{Alday:2017xua}
L.~F. Alday and A.~Bissi, {\it {Loop Corrections to Supergravity on $AdS_5
  \times S^5$}},  {\em Phys. Rev. Lett.} {\bf 119} (2017), no.~17 171601,
  [\href{http://arxiv.org/abs/1706.02388}{{\tt arXiv:1706.02388}}].

\bibitem{Aprile:2017bgs}
F.~Aprile, J.~M. Drummond, P.~Heslop, and H.~Paul, {\it {Quantum Gravity from
  Conformal Field Theory}},  {\em JHEP} {\bf 01} (2018) 035,
  [\href{http://arxiv.org/abs/1706.02822}{{\tt arXiv:1706.02822}}].

\bibitem{Alday:2017vkk}
L.~F. Alday and S.~Caron-Huot, {\it {Gravitational S-matrix from CFT dispersion
  relations}},  {\em JHEP} {\bf 12} (2018) 017,
  [\href{http://arxiv.org/abs/1711.02031}{{\tt arXiv:1711.02031}}].

\bibitem{Aprile:2017qoy}
F.~Aprile, J.~M. Drummond, P.~Heslop, and H.~Paul, {\it {Loop corrections for
  Kaluza-Klein AdS amplitudes}},  {\em JHEP} {\bf 05} (2018) 056,
  [\href{http://arxiv.org/abs/1711.03903}{{\tt arXiv:1711.03903}}].

\bibitem{Aprile:2019rep}
F.~Aprile, J.~Drummond, P.~Heslop, and H.~Paul, {\it {One-loop amplitudes in
  AdS$_{5}\times$S$^{5}$ supergravity from $ \mathcal{N} $ = 4 SYM at strong
  coupling}},  {\em JHEP} {\bf 03} (2020) 190,
  [\href{http://arxiv.org/abs/1912.01047}{{\tt arXiv:1912.01047}}].

\bibitem{Alday:2019nin}
L.~F. Alday and X.~Zhou, {\it {Simplicity of AdS Supergravity at One Loop}},
  {\em JHEP} {\bf 09} (2020) 008, [\href{http://arxiv.org/abs/1912.02663}{{\tt
  arXiv:1912.02663}}].

\bibitem{Drummond:2020uni}
J.~M. Drummond, R.~Glew, and H.~Paul, {\it {One-loop string corrections for AdS
  Kaluza-Klein amplitudes}},  {\em JHEP} {\bf 12} (2021) 072,
  [\href{http://arxiv.org/abs/2008.01109}{{\tt arXiv:2008.01109}}].

\bibitem{Huang:2021xws}
Z.~Huang and E.~Y. Yuan, {\it {Graviton Scattering in
  $\mathrm{AdS}_5\times\mathrm{S}^5$ at Two Loops}},
  \href{http://arxiv.org/abs/2112.15174}{{\tt arXiv:2112.15174}}.

\bibitem{Drummond:2022dxw}
J.~M. Drummond and H.~Paul, {\it {Two-loop supergravity on AdS$_{5} \times$
  S$^{5}$ from CFT}},  {\em JHEP} {\bf 08} (2022) 275,
  [\href{http://arxiv.org/abs/2204.01829}{{\tt arXiv:2204.01829}}].

\bibitem{Dorigoni:2021rdo}
D.~Dorigoni, M.~B. Green, and C.~Wen, {\it {Exact expressions for n-point
  maximal U(1)$_{Y}$-violating integrated correlators in SU(N) $ \mathcal{N} $
  = 4 SYM}},  {\em JHEP} {\bf 11} (2021) 132,
  [\href{http://arxiv.org/abs/2109.08086}{{\tt arXiv:2109.08086}}].

\bibitem{Dorigoni:2022zcr}
D.~Dorigoni, M.~B. Green, and C.~Wen, {\it {Exact results for duality-covariant
  integrated correlators in $\mathcal{N}=4$ SYM with general classical gauge
  groups}},  \href{http://arxiv.org/abs/2202.05784}{{\tt arXiv:2202.05784}}.

\bibitem{Wen:2022oky}
C.~Wen and S.-Q. Zhang, {\it {Integrated correlators in $ \mathcal{N} $ = 4
  super Yang-Mills and periods}},  {\em JHEP} {\bf 05} (2022) 126,
  [\href{http://arxiv.org/abs/2203.01890}{{\tt arXiv:2203.01890}}].

\bibitem{Hatsuda:2022enx}
Y.~Hatsuda and K.~Okuyama, {\it {Large $N$ expansion of an integrated
  correlator in $\mathcal{N}=4$ SYM}},
  \href{http://arxiv.org/abs/2208.01891}{{\tt arXiv:2208.01891}}.

\bibitem{sarnak}
P.~Sarnak, ``{Arithmetic Quantum Chaos}.''
  {http://web.math.princeton.edu/sarnak/Arithmetic\%20Quantum\%20Chaos.pdf},
  May, 1993.

\bibitem{hejrack}
D.~A. Hejhal and B.~N. Rackner, {\it On the topography of maass waveforms for
  psl(2, z)},  {\em Experimental Mathematics} {\bf 1} (1992), no.~4 275--305,
  [\href{http://arxiv.org/abs/https://doi.org/10.1080/10586458.1992.10504562}{{\tt
  https://doi.org/10.1080/10586458.1992.10504562}}].

\bibitem{sarnakk}
P.~Sarnak, {\it Spectra of hyperbolic surfaces},  {\em Contents} {\bf 40} (10,
  2003).

\bibitem{Fucito:2015ofa}
F.~Fucito, J.~F. Morales, and R.~Poghossian, {\it {Wilson loops and chiral
  correlators on squashed spheres}},  {\em JHEP} {\bf 11} (2015) 064,
  [\href{http://arxiv.org/abs/1507.05426}{{\tt arXiv:1507.05426}}].

\bibitem{Baggio:2014ioa}
M.~Baggio, V.~Niarchos, and K.~Papadodimas, {\it {tt$^{*}$ equations,
  localization and exact chiral rings in 4d $ \mathcal{N} $ =2 SCFTs}},  {\em
  JHEP} {\bf 02} (2015) 122, [\href{http://arxiv.org/abs/1409.4212}{{\tt
  arXiv:1409.4212}}].

\bibitem{Baggio:2014sna}
M.~Baggio, V.~Niarchos, and K.~Papadodimas, {\it {Exact correlation functions
  in $SU(2) \mathcal N=2$ superconformal QCD}},  {\em Phys. Rev. Lett.} {\bf
  113} (2014), no.~25 251601, [\href{http://arxiv.org/abs/1409.4217}{{\tt
  arXiv:1409.4217}}].

\bibitem{Baggio:2015vxa}
M.~Baggio, V.~Niarchos, and K.~Papadodimas, {\it {On exact correlation
  functions in SU(N) $ \mathcal{N}=2 $ superconformal QCD}},  {\em JHEP} {\bf
  11} (2015) 198, [\href{http://arxiv.org/abs/1508.03077}{{\tt
  arXiv:1508.03077}}].

\bibitem{Gerchkovitz:2016gxx}
E.~Gerchkovitz, J.~Gomis, N.~Ishtiaque, A.~Karasik, Z.~Komargodski, and S.~S.
  Pufu, {\it {Correlation Functions of Coulomb Branch Operators}},  {\em JHEP}
  {\bf 01} (2017) 103, [\href{http://arxiv.org/abs/1602.05971}{{\tt
  arXiv:1602.05971}}].

\bibitem{Hellerman:2017sur}
S.~Hellerman and S.~Maeda, {\it {On the Large $R$-charge Expansion in
  ${\mathcal N} = 2$ Superconformal Field Theories}},  {\em JHEP} {\bf 12}
  (2017) 135, [\href{http://arxiv.org/abs/1710.07336}{{\tt arXiv:1710.07336}}].

\bibitem{Grassi:2019txd}
A.~Grassi, Z.~Komargodski, and L.~Tizzano, {\it {Extremal correlators and
  random matrix theory}},  {\em JHEP} {\bf 04} (2021) 214,
  [\href{http://arxiv.org/abs/1908.10306}{{\tt arXiv:1908.10306}}].

\bibitem{Aprile:2018efk}
F.~Aprile, J.~Drummond, P.~Heslop, and H.~Paul, {\it {Double-trace spectrum of
  $N=4$ supersymmetric Yang-Mills theory at strong coupling}},  {\em Phys. Rev.
  D} {\bf 98} (2018), no.~12 126008,
  [\href{http://arxiv.org/abs/1802.06889}{{\tt arXiv:1802.06889}}].

\bibitem{Aprile:2020uxk}
F.~Aprile, J.~Drummond, P.~Heslop, H.~Paul, F.~Sanfilippo, M.~Santagata, and
  A.~Stewart, {\it {Single Particle Operators and their Correlators in Free
  $\mathcal{N}=4$ SYM}},  \href{http://arxiv.org/abs/2007.09395}{{\tt
  arXiv:2007.09395}}.

\bibitem{Eden:2000bk}
B.~Eden, A.~C. Petkou, C.~Schubert, and E.~Sokatchev, {\it {Partial
  nonrenormalization of the stress tensor four point function in N=4 SYM and
  AdS / CFT}},  {\em Nucl. Phys. B} {\bf 607} (2001) 191--212,
  [\href{http://arxiv.org/abs/hep-th/0009106}{{\tt hep-th/0009106}}].

\bibitem{Nirschl:2004pa}
M.~Nirschl and H.~Osborn, {\it {Superconformal Ward identities and their
  solution}},  {\em Nucl. Phys. B} {\bf 711} (2005) 409--479,
  [\href{http://arxiv.org/abs/hep-th/0407060}{{\tt hep-th/0407060}}].

\bibitem{Pestun:2007rz}
V.~Pestun, {\it {Localization of gauge theory on a four-sphere and
  supersymmetric Wilson loops}},  {\em Commun. Math. Phys.} {\bf 313} (2012)
  71--129, [\href{http://arxiv.org/abs/0712.2824}{{\tt arXiv:0712.2824}}].

\bibitem{Nekrasov:2002qd}
N.~A. Nekrasov, {\it {Seiberg-Witten prepotential from instanton counting}},
  {\em Adv. Theor. Math. Phys.} {\bf 7} (2003), no.~5 831--864,
  [\href{http://arxiv.org/abs/hep-th/0206161}{{\tt hep-th/0206161}}].

\bibitem{Bobev:2016nua}
N.~Bobev, H.~Elvang, U.~Kol, T.~Olson, and S.~S. Pufu, {\it {Holography for $
  \mathcal{N} $ = 1$^{*}$ on S$^{4}$}},  {\em JHEP} {\bf 10} (2016) 095,
  [\href{http://arxiv.org/abs/1605.00656}{{\tt arXiv:1605.00656}}].

\bibitem{Chester:2020vyz}
S.~M. Chester, M.~B. Green, S.~S. Pufu, Y.~Wang, and C.~Wen, {\it {New modular
  invariants in $ \mathcal{N} $ = 4 Super-Yang-Mills theory}},  {\em JHEP} {\bf
  04} (2021) 212, [\href{http://arxiv.org/abs/2008.02713}{{\tt
  arXiv:2008.02713}}].

\bibitem{Terras_2013}
A.~Terras, {\em Harmonic Analysis on Symmetric Spaces{\textemdash}Euclidean
  Space, the Sphere, and the Poincar{\'{e}} Upper Half-Plane}.
\newblock Springer New York, 2013.

\bibitem{Iwaniec2002SpectralMO}
H.~Iwaniec, {\it Spectral methods of automorphic forms},  2002.

\bibitem{DHoker:2022dxx}
E.~D'Hoker and J.~Kaidi, {\it {Lectures on modular forms and strings}},
  \href{http://arxiv.org/abs/2208.07242}{{\tt arXiv:2208.07242}}.

\bibitem{Papadodimas:2009eu}
K.~Papadodimas, {\it {Topological Anti-Topological Fusion in Four-Dimensional
  Superconformal Field Theories}},  {\em JHEP} {\bf 08} (2010) 118,
  [\href{http://arxiv.org/abs/0910.4963}{{\tt arXiv:0910.4963}}].

\bibitem{Gerchkovitz:2014gta}
E.~Gerchkovitz, J.~Gomis, and Z.~Komargodski, {\it {Sphere Partition Functions
  and the Zamolodchikov Metric}},  {\em JHEP} {\bf 11} (2014) 001,
  [\href{http://arxiv.org/abs/1405.7271}{{\tt arXiv:1405.7271}}].

\bibitem{Grassi:2014cla}
A.~Grassi, M.~Marino, and S.~Zakany, {\it {Resumming the string perturbation
  series}},  {\em JHEP} {\bf 05} (2015) 038,
  [\href{http://arxiv.org/abs/1405.4214}{{\tt arXiv:1405.4214}}].

\bibitem{Dorigoni:2017smz}
D.~Dorigoni and P.~Glass, {\it {The grin of Cheshire cat resurgence from
  supersymmetric localization}},  {\em SciPost Phys.} {\bf 4} (2018), no.~2
  012, [\href{http://arxiv.org/abs/1711.04802}{{\tt arXiv:1711.04802}}].

\bibitem{Chester:2021aun}
S.~M. Chester, R.~Dempsey, and S.~S. Pufu, {\it {Bootstrapping $\mathcal{N}=4$
  super-Yang-Mills on the conformal manifold}},
  \href{http://arxiv.org/abs/2111.07989}{{\tt arXiv:2111.07989}}.

\bibitem{Bianchi:1998nk}
M.~Bianchi, M.~B. Green, S.~Kovacs, and G.~Rossi, {\it {Instantons in
  supersymmetric Yang-Mills and D instantons in IIB superstring theory}},  {\em
  JHEP} {\bf 08} (1998) 013, [\href{http://arxiv.org/abs/hep-th/9807033}{{\tt
  hep-th/9807033}}].

\bibitem{Bianchi:1999ge}
M.~Bianchi, S.~Kovacs, G.~Rossi, and Y.~S. Stanev, {\it {On the logarithmic
  behavior in N=4 SYM theory}},  {\em JHEP} {\bf 08} (1999) 020,
  [\href{http://arxiv.org/abs/hep-th/9906188}{{\tt hep-th/9906188}}].

\bibitem{Alday:2016tll}
L.~F. Alday and G.~P. Korchemsky, {\it {Revisiting instanton corrections to the
  Konishi multiplet}},  {\em JHEP} {\bf 12} (2016) 005,
  [\href{http://arxiv.org/abs/1605.06346}{{\tt arXiv:1605.06346}}].

\bibitem{Alday:2016jeo}
L.~F. Alday and G.~P. Korchemsky, {\it {Instanton corrections to twist-two
  operators}},  {\em JHEP} {\bf 06} (2017) 008,
  [\href{http://arxiv.org/abs/1609.08164}{{\tt arXiv:1609.08164}}].

\bibitem{Alday:2016bkq}
L.~F. Alday and G.~P. Korchemsky, {\it {On instanton effects in the operator
  product expansion}},  {\em JHEP} {\bf 05} (2017) 049,
  [\href{http://arxiv.org/abs/1610.01425}{{\tt arXiv:1610.01425}}].

\bibitem{Green:2002vf}
M.~B. Green and S.~Kovacs, {\it {Instanton induced Yang-Mills correlation
  functions at large N and their AdS(5) x S**5 duals}},  {\em JHEP} {\bf 04}
  (2003) 058, [\href{http://arxiv.org/abs/hep-th/0212332}{{\tt
  hep-th/0212332}}].

\bibitem{Kovacs:2003rt}
S.~Kovacs, {\it {On instanton contributions to anomalous dimensions in N=4
  supersymmetric Yang-Mills theory}},  {\em Nucl. Phys. B} {\bf 684} (2004)
  3--74, [\href{http://arxiv.org/abs/hep-th/0310193}{{\tt hep-th/0310193}}].

\bibitem{Chester:2019jas}
S.~M. Chester, M.~B. Green, S.~S. Pufu, Y.~Wang, and C.~Wen, {\it {Modular
  invariance in superstring theory from $ \mathcal{N} $ = 4 super-Yang-Mills}},
   {\em JHEP} {\bf 11} (2020) 016, [\href{http://arxiv.org/abs/1912.13365}{{\tt
  arXiv:1912.13365}}].

\bibitem{Chester:2019pvm}
S.~M. Chester, {\it {Genus-2 holographic correlator on AdS$_{5}\times$ S$^{5}$
  from localization}},  {\em JHEP} {\bf 04} (2020) 193,
  [\href{http://arxiv.org/abs/1908.05247}{{\tt arXiv:1908.05247}}].

\bibitem{Aharony:2016dwx}
O.~Aharony, L.~F. Alday, A.~Bissi, and E.~Perlmutter, {\it {Loops in AdS from
  Conformal Field Theory}},  {\em JHEP} {\bf 07} (2017) 036,
  [\href{http://arxiv.org/abs/1612.03891}{{\tt arXiv:1612.03891}}].

\bibitem{Aprile:2020luw}
F.~Aprile and P.~Vieira, {\it {Large $p$ explorations. From SUGRA to big
  STRINGS in Mellin space}},  {\em JHEP} {\bf 12} (2020) 206,
  [\href{http://arxiv.org/abs/2007.09176}{{\tt arXiv:2007.09176}}].

\bibitem{Hellerman:2015nra}
S.~Hellerman, D.~Orlando, S.~Reffert, and M.~Watanabe, {\it {On the CFT
  Operator Spectrum at Large Global Charge}},  {\em JHEP} {\bf 12} (2015) 071,
  [\href{http://arxiv.org/abs/1505.01537}{{\tt arXiv:1505.01537}}].

\bibitem{Bourget:2018obm}
A.~Bourget, D.~Rodriguez-Gomez, and J.~G. Russo, {\it {A limit for large
  $R$-charge correlators in $\mathcal{N}=2$ theories}},  {\em JHEP} {\bf 05}
  (2018) 074, [\href{http://arxiv.org/abs/1803.00580}{{\tt arXiv:1803.00580}}].

\bibitem{Beccaria:2018xxl}
M.~Beccaria, {\it {On the large R-charge $ \mathcal{N} $ = 2 chiral correlators
  and the Toda equation}},  {\em JHEP} {\bf 02} (2019) 009,
  [\href{http://arxiv.org/abs/1809.06280}{{\tt arXiv:1809.06280}}].

\bibitem{Beccaria:2018owt}
M.~Beccaria, {\it {Double scaling limit of $N=2$ chiral correlators with
  Maldacena-Wilson loop}},  {\em JHEP} {\bf 02} (2019) 095,
  [\href{http://arxiv.org/abs/1810.10483}{{\tt arXiv:1810.10483}}].

\bibitem{Paul}
H.~Paul, E.~Perlmutter, and H.~Raj, {\it {Work in progress}}, .

\bibitem{Chester:2020dja}
S.~M. Chester and S.~S. Pufu, {\it {Far beyond the planar limit in
  strongly-coupled $ \mathcal{N} $ = 4 SYM}},  {\em JHEP} {\bf 01} (2021) 103,
  [\href{http://arxiv.org/abs/2003.08412}{{\tt arXiv:2003.08412}}].

\bibitem{Caron-Huot:2021usw}
S.~Caron-Huot and F.~Coronado, {\it {Ten dimensional symmetry of $ \mathcal{N}
  $ = 4 SYM correlators}},  {\em JHEP} {\bf 03} (2022) 151,
  [\href{http://arxiv.org/abs/2106.03892}{{\tt arXiv:2106.03892}}].

\bibitem{Cavaglia:2022qpg}
A.~Cavagli\`a, N.~Gromov, J.~Julius, and M.~Preti, {\it {Bootstrability in
  defect CFT: integrated correlators and sharper bounds}},  {\em JHEP} {\bf 05}
  (2022) 164, [\href{http://arxiv.org/abs/2203.09556}{{\tt arXiv:2203.09556}}].

\bibitem{Caron-Huot:2022sdy}
S.~Caron-Huot, F.~Coronado, A.-K. Trinh, and Z.~Zahraee, {\it {Bootstrapping
  $\mathcal{N}=4$ sYM correlators using integrability}},
  \href{http://arxiv.org/abs/2207.01615}{{\tt arXiv:2207.01615}}.

\bibitem{Benjamin:2021ygh}
N.~Benjamin, S.~Collier, A.~L. Fitzpatrick, A.~Maloney, and E.~Perlmutter, {\it
  {Harmonic analysis of 2d CFT partition functions}},  {\em JHEP} {\bf 09}
  (2021) 174, [\href{http://arxiv.org/abs/2107.10744}{{\tt arXiv:2107.10744}}].

\bibitem{Kravchuk:2021akc}
P.~Kravchuk, D.~Mazac, and S.~Pal, {\it {Automorphic Spectra and the Conformal
  Bootstrap}},  \href{http://arxiv.org/abs/2111.12716}{{\tt arXiv:2111.12716}}.

\bibitem{Bonifacio:2021aqf}
J.~Bonifacio, {\it {Bootstrapping closed hyperbolic surfaces}},  {\em JHEP}
  {\bf 03} (2022) 093, [\href{http://arxiv.org/abs/2111.13215}{{\tt
  arXiv:2111.13215}}].

\bibitem{Moore:1998zu}
G.~W. Moore, {\it {Attractors and arithmetic}},
  \href{http://arxiv.org/abs/hep-th/9807056}{{\tt hep-th/9807056}}.

\bibitem{mehta}
M.~L. Mehta, {\em {Random matrices}}.
\newblock 2nd edition, Academic Press, 1991.

\bibitem{Marino:2004eq}
M.~Marino, {\it Les houches lectures on matrix models and topological strings},
   10, 2004.
\newblock \href{http://arxiv.org/abs/hep-th/0410165}{{\tt hep-th/0410165}}.

\bibitem{Fiol:2013hna}
B.~Fiol and G.~Torrents, {\it {Exact results for Wilson loops in arbitrary
  representations}},  {\em JHEP} {\bf 01} (2014) 020,
  [\href{http://arxiv.org/abs/1311.2058}{{\tt arXiv:1311.2058}}].

\bibitem{Gradshteyn}
I.~S. Gradshteyn and I.~M. Ryzhik, {\em {Table of Integrals, Series, and
  Products}}.
\newblock Academic Press, 2007.

\bibitem{Mehta:1981xt}
M.~L. Mehta, {\it {A Method of Integration Over Matrix Variables}},  {\em
  Commun. Math. Phys.} {\bf 79} (1981) 327--340.

\end{thebibliography}\endgroup

\end{document}